\documentclass[
    aps,
    superscriptaddress,
    nofootinbib,
    notitlepage,
    twocolumn,
    prd,
    longbibliography
]{revtex4-1}

\usepackage[normalem]{ulem}

\usepackage[latin1]{inputenc}
\usepackage{graphicx}
\graphicspath{{Figures/}}
\usepackage{float}
\usepackage{latexsym}
\usepackage{graphicx}
\usepackage{amsfonts}
\usepackage{amssymb}
\usepackage{amsmath}
\usepackage{dsfont}
\usepackage[mathscr]{euscript}
\usepackage{bbold}
\usepackage[colorlinks=true,citecolor=blue,linkcolor=blue,hyperfootnotes=false]{hyperref}
\usepackage[dvipsnames]{xcolor}
\usepackage{orcidlink}
\usepackage{cool}
\usepackage{dcolumn}
\usepackage{textcomp}
\usepackage{xfrac}
\usepackage{slashed}
\usepackage{multirow}
\usepackage{tabularx} 

\def\be{\begin{equation}}
\def\ee{\end{equation}}
\def\beq{\begin{equation}}
\def\eeq{\end{equation}}
\def\figs/B{B}
\def\bea{\begin{eqnarray}}
\def\eea{\end{eqnarray}}
\def\bg{\begin{eqnarray}}
\def\nd{\end{eqnarray}}

\def\ln{{\rm ln}}

\def\mpl{M_{\rm Pl}}
\def\dd{{\rm d}}
\def\pd{\partial}

\def\calP{{\cal P}}
\def\calR{{\cal R}}
\def\calS{{\cal S}}
\def\prk{\calP_\calR}
\def\vpbh{V_{\rm PBH}}
\def\vpbha{V_{{\rm PBH},A}}
\def\vpbhb{V_{{\rm PBH},B}}

\def\fnl{f_\text{NL}}

\def\calO{{\cal O}}
\def\calM{{\cal M}}

\newcommand{\eref}[1]{Eq.~\eqref{#1}}
\newcommand{\fref}[1]{Fig.~\ref{#1}}
\newcommand{\tref}[1]{Table~\ref{#1}}
\newcommand{\sref}[1]{Sec.~\ref{#1}}

\newcommand{\rref}[1]{Ref.~\cite{#1}}

\begin{document}

\title{Primordial Black Holes from Inflation with a Spectator Field}

\author{Dario L.~Lorenzoni \orcidlink{0000-0001-9832-8183}}
\email{lorenzod@myumanitoba.ca}
\affiliation{Department of Physics \& Astronomy, University of Manitoba, Winnipeg, MB R3T 2N2, Canada}
\affiliation{Department of Physics, University of Winnipeg, Winnipeg, MB R3B 2E9, Canada}

\author{Sarah R.~Geller \orcidlink{0000-0003-4126-4662}}
\email{sarah.geller@yale.edu}
\affiliation{Santa Cruz Institute for Particle Physics, Santa Cruz, CA 95064, USA}
\affiliation{Department of Physics, University of California, Santa Cruz, Santa Cruz, CA 95064, USA}
\affiliation{Department of Astronomy, Yale University, Kline Tower, 266 Whitney Avenue, New Haven, CT 06511, USA}

\author{David I.~Kaiser \orcidlink{0000-0002-5054-6744}}
\email{dikaiser@mit.edu}
\affiliation{Department of Physics, Massachusetts Institute of Technology, Cambridge, MA 02139, USA}

\author{Evan McDonough \orcidlink{0000-0002-2130-3903}}
\email{e.mcdonough@uwinnipeg.ca}
\affiliation{Department of Physics, University of Winnipeg, Winnipeg, MB R3B 2E9, Canada}

\begin{abstract}
    How is the production of primordial black holes (PBHs) in single-field models of inflation impacted by the presence of additional scalar fields? We consider the effect of a spectator field---a free scalar field with sub-Hubble mass and no direct coupling to the inflaton and which makes a subdominant contribution to the total energy density---in the context of single-field models of inflation featuring a transient phase of ultra-slow-roll evolution. Despite the modest title, a spectator field can have a dramatic impact: the slow-roll evolution of the spectator prevents the combined inflaton-and-spectator system from entering into ultra-slow-roll, which naively might be expected to preclude the production of PBHs. However, we demonstrate that the growth of perturbations is maintained or enhanced by the spectator, through the rich interplay of curvature and isocurvature perturbations.  We show in a model-independent way that the single-field phase of ultra-slow-roll is replaced by two turns in field space encompassing a phase of tachyonic instability for the isocurvature perturbations and a transfer of power from isocurvature to curvature modes. Furthermore, we highlight a degeneracy between the fine-tuning of the feature in the inflaton potential and the parameters of the spectator, leading to an overall {\it resilience} of model predictions to parameter variations. This makes it easier for the underlying PBH model to accommodate both high-precision cosmic microwave background constraints and production of PBHs in the asteroid-mass range.
\end{abstract}

\maketitle

\section{Introduction}
\label{sec:intro}

Primordial black holes (PBHs) \cite{Zeldovich:1967lct,Hawking:1971ei,Carr:1974nx,Meszaros:1974tb,Carr:1975qj,Khlopov:1985jw,Niemeyer:1999ak} have a long and storied history, predating the modern era of precision cosmology.  In recent years, PBHs  have emerged as a leading candidate to explain some or all of the observed dark matter~\cite{Khlopov:2008qy,Carr:2009jm,Sasaki:2018dmp,Carr:2020gox,Carr:2020xqk,Green:2020jor,Escriva:2021aeh,Villanueva-Domingo:2021spv,Escriva:2022duf,Gorton:2024cdm}. Primordial black holes may also explain the JWST observations of high-redshift supermassive black holes \cite{Volonteri:2021sfo,Dayal:2024zwq,Hai-LongHuang:2024vvz,Hai-LongHuang:2024gtx,Huang:2024aog}, and a variety of observations, from gravitational waves at LIGO \cite{Bird:2016dcv}, NANOGrav \cite{DeLuca:2020agl,DeLorenci:2025wbn}, and future detectors \cite{Qin:2023lgo}, to ultrahigh-energy neutrinos \cite{Klipfel:2025jql,Baker:2025cff} and other cosmic rays \cite{Klipfel:2025bvh}. In all these cases the question that remains is the origin of the large density fluctuations necessary for the direct collapse formation of PBHs. 

The origin of PBHs finds a natural home in models of cosmic inflation.%
\footnote{
    Alternative production mechanisms include bubble collision \cite{Hawking:1982ga} and collapse \cite{Deng:2017uwc}, collapse of cosmic strings \cite{Hawking:1987bn,Polnarev:1988dh} or domain walls \cite{Rubin:2001yw}, quark confinement \cite{Dvali:2021byy}, and others. These mechanisms are beyond the scope of our work; for a review of these processes, see Refs.~\cite{Escriva:2022duf,Khlopov:2008qy}.
}
Inflation provides a causal mechanism for structure formation, with primordial perturbations observed in the cosmic microwave background (CMB) and large-scale structure originating from quantum fluctuations. It is possible these same quantum fluctuations, on much shorter length scales than those directly probed by the CMB, could seed the formation of primordial black holes. 

A significant body of work has been dedicated to this possibility. For example, single-field inflation models with a potential $V (\varphi)$ that is tuned to yield a brief period of ultra-slow-roll (USR) evolution can lead to an enhancement of perturbations on a characteristic scale, resulting in a spike in the primordial power spectrum of curvature perturbations \cite{Kinney:2005vj,Martin:2012pe,Ezquiaga:2017fvi,Garcia-Bellido:2017mdw,Germani:2017bcs,Kannike:2017bxn,Motohashi:2017kbs,Di:2017ndc,Ballesteros:2017fsr,Pattison:2017mbe,Passaglia:2018ixg,Biagetti:2018pjj,Mishra:2019pzq,Figueroa:2020jkf,Karam:2022nym,Ozsoy:2023ryl,Cole:2023wyx,Cicoli:2018asa,Cicoli:2022sih,Cai:2022erk,Inomata:2021uqj,Inomata:2021tpx,Bhaumik:2019tvl,Pi:2022ysn,Choudhury:2024one}. However, the demand that the model both satisfy CMB constraints \cite{Planck:2019kim,Planck:2018vyg,Planck:2018jri,BICEP:2021xfz} and produce a population of PBHs consistent with observations \cite{Khlopov:2008qy,Carr:2009jm,Sasaki:2018dmp,Carr:2020gox,Carr:2020xqk,Green:2020jor,Escriva:2021aeh,Villanueva-Domingo:2021spv,Escriva:2022duf,Gorton:2024cdm}, places a significant restriction on model building. In the context of single-field USR models, this manifests as a significant parameter fine-tuning \cite{Cole:2023wyx}, in the sense that the predictions of the model are exponentially sensitive to small variations in one or more model parameters.

On the other hand, it is not {\it a priori} the case that the early Universe can be described by a single scalar field. The realm of particle physics is far from single-field: the Higgs doublet of the Standard Model comprises four real scalars, and particle physics theories beyond the Standard Model (such as the two-Higgs-doublet model \cite{Branco:2011iw}) typically add yet more scalars \cite{Lyth:1998xn,Mazumdar:2010sa}. String theory provides another example, with potentially many moduli fields at low energies, including the string theory axiverse, which could add up to {\it tens of thousands} of axion-like particles to the spectrum \cite{Fallon:2025lvn}. Field theory models can also yield an axiverse-like spectrum \cite{McDonough:2020gmn,Maleknejad:2022gyf,Alexander:2023wgk,Alexander:2024nvi}. It stands to reason that the dynamics of cosmic inflation could rely on multiple scalar fields, usually termed ``multifield inflation.''

Multifield inflation offers ample opportunities for PBH production \cite{Randall:1995dj,Garcia-Bellido:1996mdl,Lyth:2010zq,Bugaev:2011wy,Halpern:2014mca,Clesse:2015wea,Kawasaki:2015ppx,Braglia:2022phb,Fumagalli:2020adf,Braglia:2020eai,Palma:2020ejf,Geller:2022nkr,Qin:2023lgo,Ferraz:2024bvd,Bhattacharya:2022fze,Cai:2021wzd,Pi:2017gih}.
Such models typically consider a direct coupling between the different fields, either in the kinetic sector or in their potential or both, and may include multiple distinct phases of inflation driven by the different scalar fields of the model.

A minimal realization of multifield inflation is given by the seemingly trivial case of a {\it spectator field}, which is subdominant in energy density to the inflaton and, in its minimal realization, has no direct coupling to the inflaton. The spectator ostensibly plays no role in the background evolution, but can have important phenomenological implications: spectator fields can source CMB non-Gaussianity \cite{Chen:2018uul}, gravitational waves \cite{Dimastrogiovanni:2016fuu,McDonough:2018xzh,Holland:2020jdh}, and jointly the matter-antimatter asymmetry and production of dark matter \cite{Alexander:2018fjp}, among other signatures. 

In this work, and in the accompanying paper \cite{Lorenzoni:2025gni}, we consider the influence of spectator fields in the context of single-field models of PBH production. The spectator field is defined as a free scalar field with no direct coupling to the inflaton, with a mass that is small compared to the Hubble scale during inflation, and which makes a subdominant contribution to the energy density at all times. In this companion paper we expand upon the origin and phenomenological implications of the spectator scenario presented in Ref.~\cite{Lorenzoni:2025gni}, in which one of many available single-field USR PBH models is supplemented with a simple spectator field.

Our main finding is that the inclusion of a spectator field as defined above can maintain or enhance the production of primordial black holes in single-field models, despite significantly altering the dynamics of both the background evolution and perturbations.
Notably, {\it the spectator can prevent the system from ever entering a phase of ultra-slow-roll, and yet still realize the enhancement of perturbations required to seed PBHs}. The would-be USR enhancement is supplanted by a tachyonic growth of isocurvature perturbations, which is converted to curvature perturbations during a turn in field space.

With the requirement of USR removed, so too is the extreme fine-tuning of the feature in the potential. The spectator therefore imbues the single-field PBH model with a \textit{resilience} to small variations in the inflaton model parameters. While the feature in the potential is still a required ingredient, the degree of precision with which the single-field model parameters must be specified is significantly reduced. As we detail in this companion paper and the accompanying paper \cite{Lorenzoni:2025gni}, this is realized through a compensation between inflaton model parameters and $\calO(1)$ variations in the spectator parameters.
This indicates that this setup may alleviate the well-known issue of fine-tuning in single-field models of inflation that produce PBHs, by considering the physically motivated presence of additional fields.

This paper is structured as a comparison between our spectator PBH scenario and the single-field models upon which it is built. 
We begin by introducing the background dynamics of both setups in \sref{sec:background}, in which we review the physics of ultra-slow-roll inflation before delving into the multifield dynamics. Section~\ref{sec:perturbations} describes the formation of inflationary perturbations and highlights the differences and similarities between the single-field and spectator PBH scenarios. The shape of the curvature power spectrum is analyzed in \sref{sec:spectrum}, in which we explicitly show consistency with CMB constraints and with the production of PBHs in the so-called asteroid-mass range, with $10^{17}~{\rm g} \leq M_{\rm PBH} \leq 10^{23}~{\rm g}$. 
The dependence of the model (and its observables) on both the inflaton and spectator parameters is the subject of \sref{sec:parameters}.
While we consider, for simplicity, a particular form of the inflaton potential throughout the previous sections, we show in \sref{sec:models} that this mechanism is generic to single-field inflation models that produce PBHs through a phase of USR evolution.
Finally, we summarize and discuss our results in \sref{sec:discussion}, in which we highlight future research directions.

\section{The spectator PBH Model: Background Evolution}
\label{sec:background}

Single-field models of inflation can produce PBHs through a transient phase of ultra-slow-roll, during which the evolution of the inflaton field, $\varphi$, slows down. As we will see in \sref{sec:perturbations}, this 
amplifies curvature perturbations, enhancing the inflationary power spectrum on appropriate scales and allowing for the formation of PBHs.

We begin by asking a simple question: \textit{what would happen to PBH production in these single-field models of inflation if another scalar field were to simply exist in the Universe?}
Such a field would be \textit{spectating} inflation, having no direct coupling to the inflaton and a subdominant contribution to the energy density of the Universe. Including the presence of this spectator field, $\chi$, the action of the system becomes
\begin{equation}\label{eq:action}
    S= \int \dd^4 x \sqrt{ -g} \left[ \frac{M_\text{Pl}^2}{2} R - \frac{1}{2} (\partial \varphi)^2 - \frac{1}{2}(\partial \chi)^2 - V(\varphi,\chi)\right]\,,
\end{equation}
where $M_{\text{Pl}}\equiv1/\sqrt{8\pi G}$ is the reduced Planck mass, and $g$ and $R$ are the determinant of the spacetime metric and its Ricci scalar, respectively. The potential of the system is simply given by the sum of the inflaton and spectator contributions:
\begin{equation}\label{eq:potential-generic}
    V(\varphi,\chi)=\vpbh(\varphi) + V_S(\chi)\,,
\end{equation}
with the inflaton potential being of the PBH-producing kind, supporting a phase of USR.

There are therefore \textit{no direct couplings} between the inflaton and the spectator field in this setup, neither in the kinetic sector nor in the potential.
The mechanism 
on which we focus here requires neither a curved field-space manifold with its associated noncanonical kinetic sector 
nor multiple phases of inflation.
This setup is thus starkly different from the ones considered in the literature, which explicitly depend on some kind of field coupling \cite{Randall:1995dj,Garcia-Bellido:1996mdl,Lyth:2010zq,Bugaev:2011wy,Halpern:2014mca,Clesse:2015wea,Kawasaki:2015ppx,Braglia:2022phb,Fumagalli:2020adf,Braglia:2020eai,Palma:2020ejf,Geller:2022nkr,Qin:2023lgo,Bhattacharya:2022fze}.

\subsection{Single-Field Inflation}
\label{sec:background:SF}

\begin{figure}
    \centering
    \includegraphics[width=1\linewidth]{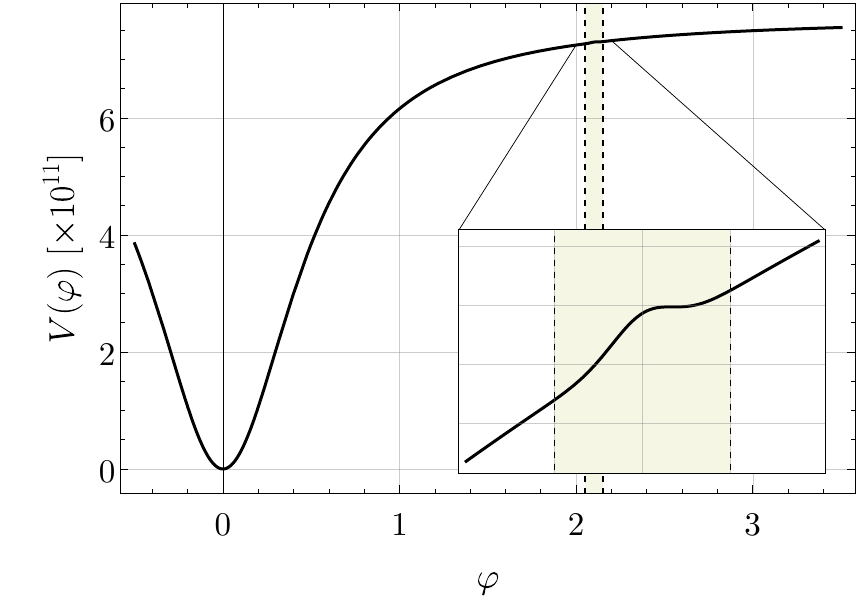}
    \caption{
    The inflationary potential $\vpbha$ of \eref{eq:VPBH-A}, in units of $M_{\rm Pl}$. The inset highlights the presence of a local inflection point that causes the inflaton field to slow down as it rolls down the potential, yielding a temporary phase of USR evolution. The parameters used in this figure are delineated in the ``Reference'' row of \tref{tab:PBHA-params}.}
    \label{fig:PBHA-V}
\end{figure}

Let us first review the dynamics of PBH-producing single-field models in the absence of spectator fields, i.e. setting $V_S(\chi)=0$ and $\chi=0$ in \eref{eq:action}:
\begin{equation}\label{eq:action-SF}
    S = \int \dd^4 x \sqrt{-g} \left[ \frac{\mpl^2}{2}R-\frac{1}{2}(\pd\varphi)^2-V_\text{PBH}(\varphi) \right]\,.
\end{equation}
In order to support a phase of ultra-slow-roll evolution, the inflaton potential $\vpbh(\varphi)$ needs to exhibit a local feature that will cause the field to temporarily slow down its evolution. This requires the potential to be nearly flat, such as in the vicinity of a (quasi-)inflection point.
As a concrete example, consider the potential of \rref{Mishra:2019pzq},
\begin{equation}\label{eq:VPBH-A}
    V_{{\rm PBH},A} (\varphi) = V_0 \frac{ \varphi^2}{\varphi^2 + M^2} \left( 1 + A e^{- (\varphi - \varphi_d)^2/2\sigma^2} \right)\,,
\end{equation}
which superimposes a Gaussian ``bump'' (of width $\sigma$, at position $\varphi_d$) on the well-known Kachru-Kallosh-Linde-Trivedi (KKLT) potential \cite{Kachru:2003aw}. 
The shape of $\vpbha$ is illustrated in \fref{fig:PBHA-V}, for the reference model parameters of \tref{tab:PBHA-params}: a local minimum appears, followed by a local maximum. (This feature could also be a plateau region.)
We will adopt this potential in the following discussion, and will show in \sref{sec:models} that the mechanism described hereafter is independent of the particular form of $\vpbh$.

\begin{figure}
    \centering
    \includegraphics[width=1\linewidth]{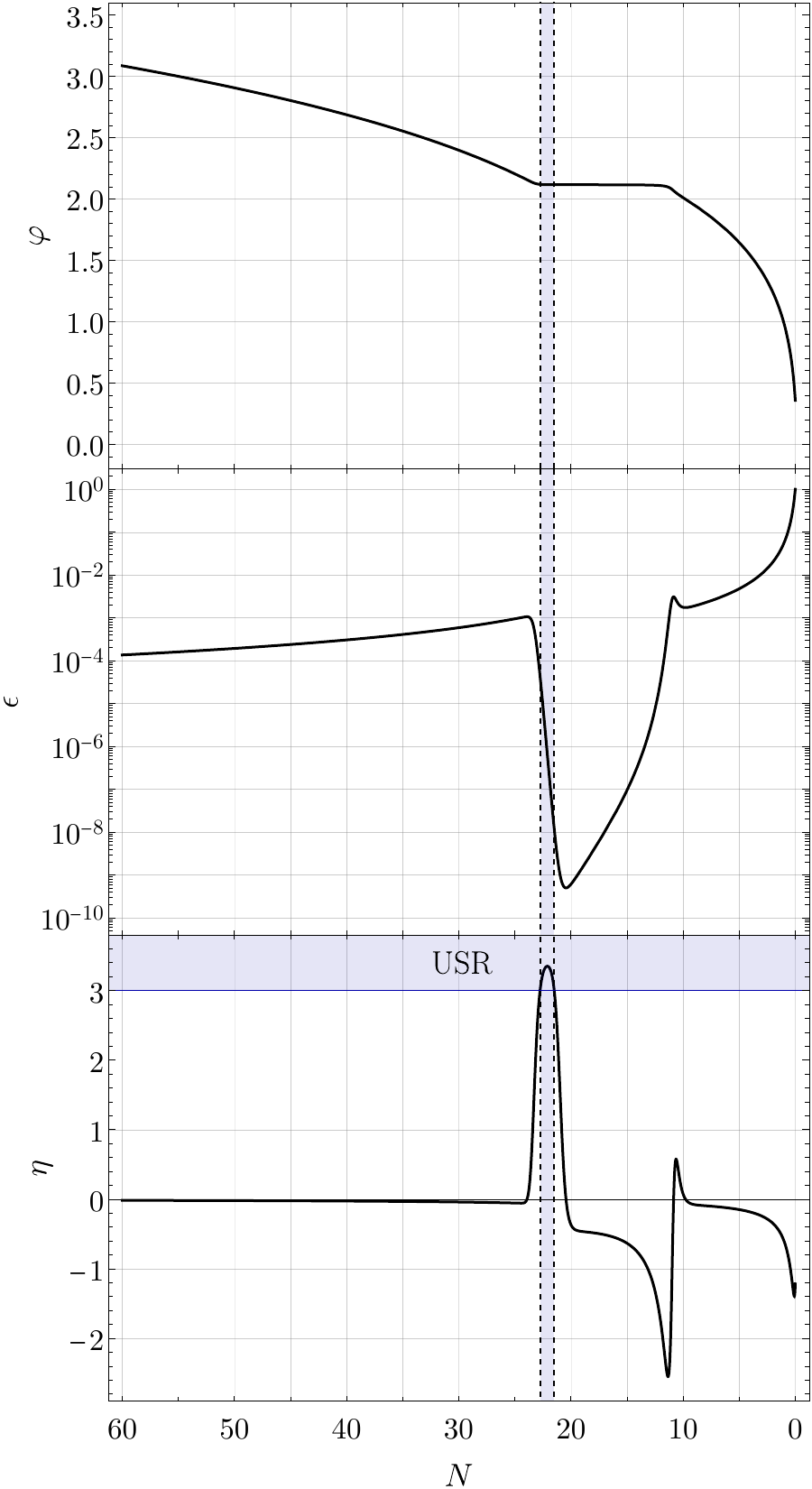}
    \caption{Evolution of the background field $\varphi$ and of the slow-roll parameters $\epsilon$ and $\eta$ in the single-field regime of \eref{eq:action-SF}. This is shown for the $\vpbha$ potential of \eref{eq:VPBH-A}, using the ``Reference'' model parameters of \tref{tab:PBHA-params}. The time evolution is shown in terms of the number of e-folds $N$ before the end of inflation, see \eref{eq:N-efolds}.
    \textit{Top panel:} The inflaton $\varphi$ rolls slowly down its potential at the beginning and at the end of its evolution. When it reaches the feature in $\vpbha$, $\varphi$'s evolution is considerably slowed down ($\varphi$ barely evolves during $20\lesssim N\lesssim 10$) until it evolves beyond the feature. 
    \textit{Middle panel:} As the inflaton is slowed down by the feature in $\vpbha$, the slow-roll parameter $\epsilon$ quickly decreases from $\calO(10^{-4})$ to $\calO(10^{-9})$. It increases again as $\varphi$ picks up speed after the local feature, eventually reaching the value $\epsilon=1$ and ending the inflationary epoch.
    \textit{Bottom panel:} The significant slowing down of the field causes a temporary phase of ultra-slow-roll: as $\epsilon$ decreases, $\eta$ becomes large, crossing into the USR regime for $1.4$ e-folds.
    \textit{All panels:} The shaded region highlights the phase of ultra-slow-roll $\eta\geq 3$.
    }
    \label{fig:PBHA-background-SF}
\end{figure}

The background evolution of the inflaton field is governed by its equation of motion and by the Friedmann equation:
\begin{gather}
    \ddot{\varphi}+3 H \dot{\varphi} = - \pd_\varphi V_{\rm PBH}\,, 
    \label{eq:EoM-bckgr-SF}
    \\
    H^2=\frac{1}{3\mpl^2}\left( \frac{\dot{\varphi}^2}{2} + V_{\rm PBH}\right)\,.
    \label{eq:Friedmann-SF}
\end{gather}
The dot notation indicates time derivatives, while $\pd_\varphi=\pd/\pd\varphi$ denotes differentiation with respect to the field $\varphi$.
The Friedmann equation~\eqref{eq:Friedmann-SF} relates the field dynamics to the expansion of space, encoded in the Hubble parameter $H=\dot{a}/a$, where $a$ is the scale factor of the Universe. The inflationary dynamics can be described by the slow-roll parameters
\begin{eqnarray}
    \epsilon&&\equiv -\frac{\dot{H}}{H^2} = \frac{1}{2}\frac{\dot{\varphi}^2}{H^2\mpl^2}\,,\label{eq:epsilon-SF}
    \\
    \eta&&\equiv 2\epsilon-\frac{\dot{\epsilon}}{2H\epsilon}\,.
    \label{eq:eta}
\end{eqnarray}
Inflation (namely, accelerated expansion of space) requires $\epsilon <1$, while {\it slow-roll} inflation refers to the regime $\epsilon \ll 1$ and $|\eta|\ll 1$. {\it Ultra}-slow-roll (USR) inflation occurs when $\epsilon \ll 1$ but $\eta  \geq 3$, corresponding to the condition that $\pd_\varphi \vpbh$ is a subdominant contribution to the field's equation of motion, Eq.~\eqref{eq:EoM-bckgr-SF}.
In summary,
\begin{align}
    \text{SR:}\quad& \epsilon\ll1\,, \quad |\eta|\ll1\,,\label{eq:SR-condition}
    \\
    \text{USR:}\quad& \epsilon\rightarrow 0^+\,,\quad \eta\geq 3\,.\label{eq:USR-condition}
\end{align}
In the vicinity of a (quasi-)inflection point, the potential flattens and its derivative $\pd_\varphi \vpbh\sim0$. From \eref{eq:EoM-bckgr-SF}, under these conditions the inflaton will evolve as $\ddot{\varphi}\approx-3H\dot{\varphi}$ and enter a phase of USR.
This is illustrated in \fref{fig:PBHA-background-SF}: when $\varphi$ encounters the local feature in the potential, it slows down significantly; $\epsilon$ correspondingly becomes small ($\calO(10^{-9})$), while $\eta$ temporarily grows, surpassing the USR threshold of $\eta=3$.
In order for inflation to end, eventually seeding the conditions that could form today's Universe, $\varphi$ needs to exit the USR phase and accelerate until $\epsilon=1$.

For completeness we consider an alternative definition of the second slow-roll parameter, defined by
\begin{equation}\label{eq:epsilon2}
    \epsilon_2\equiv\frac{\dd\,\ln\,\epsilon}{\dd N} =\frac{\dot{\epsilon}}{H\epsilon}\,.
\end{equation}
This is related to our definition of $\eta$ by $\epsilon_2=-2\eta+4\epsilon$; note that in the USR regime, $\epsilon_2\leq-6$. 
We introduce the number of e-folds of expansion before the end of inflation, 
\begin{equation}\label{eq:N-efolds}
    N\equiv-\ln\,a\,,\quad \dd N = -\dd\,\ln\,a = H\,\dd t ,
\end{equation}
where $N=0$ at the end of inflation. In \fref{fig:PBHA-background-SF} and following, we adopt $N$ as our time coordinate.

\begin{table*}[htb!]
    \centering
    \begin{tabular}{@{\extracolsep{10pt}}c|*{5}c}
        \hline\hline
        Model 
        & $V_0~[\mpl^4]$ & $\varphi_d~[\mpl]$ 
        & $m_{\chi}~[\mpl]$ 
        & $\chi_i~[\mpl]$
        \\ 
        \hline 
        SF Reference 
        & $7.7\times 10^{-11}$ 
        & $2.1$ 
        & $-$ & $-$
        \\
        SF Variation
        & $6.8\times 10^{-11}$ 
        & $2.1\times(1-4\times 10^{-4})$
        & $-$ & $-$
        \\
        PBHspec
        & $9.7\times 10^{-11}$ 
        & $2.1\times(1-4\times 10^{-4})$
        & $8\times10^{-7}$ & $3.3$
        \\
        \hline\hline
    \end{tabular}
    \caption{Parameter values for the PBH$A$ models considered in this work. We fix the values $M=1/2$, $A=1.421\times 10^{-3}$, and $\sigma=1.7138\times10^{-2}$ (in units of $\mpl$) for each set of parameters, and always consider the inflaton's initial value to be $\varphi_i=3.5 \mpl$. (The dynamics are nearly independent of $\varphi_i$, so long as it produces sufficient e-folds of inflationary expansion.) We set $\dot{\varphi}_i = \dot{\chi}_i = 0$.
    The single-field ``Reference'' model is defined so that it produces an enhancement of the curvature power spectrum consistent with PBH production, similarly to the baseline models considered in Refs.~\cite{Cole:2023wyx,Mishra:2019pzq}. The single-field ``Variation'' model shifts the parameter $\varphi_d\rightarrow\varphi_{d,\text{ref}}\times(1-4\times 10^{-4})$; this excludes PBH production. The ``PBHspec'' model then adds a spectator field to the Variation model, keeping the $\vpbha$ parameters fixed (apart from the overall normalization scale $V_0$). By including the spectator field, PBHspec allows for PBH production in the asteroid-mass range.
    }
    \label{tab:PBHA-params}
\end{table*}

\subsection{Enter the Spectator}
\label{sec:background:PBHspec}

We now turn back to the full form of \eref{eq:action}; i.e., we add a spectator field $\chi$ to the single-field model of Eq.~\eqref{eq:action-SF}.
For concreteness, we continue to consider the potential $\vpbha$ of \eref{eq:VPBH-A} in what follows. We shall assume the spectator is a free scalar field with mass $m_{\chi}$ and potential given by $V_S(\chi)=\frac{1}{2}m_\chi\chi^2$.

\begin{figure}[htb!]
    \centering
    \includegraphics[width=\linewidth]{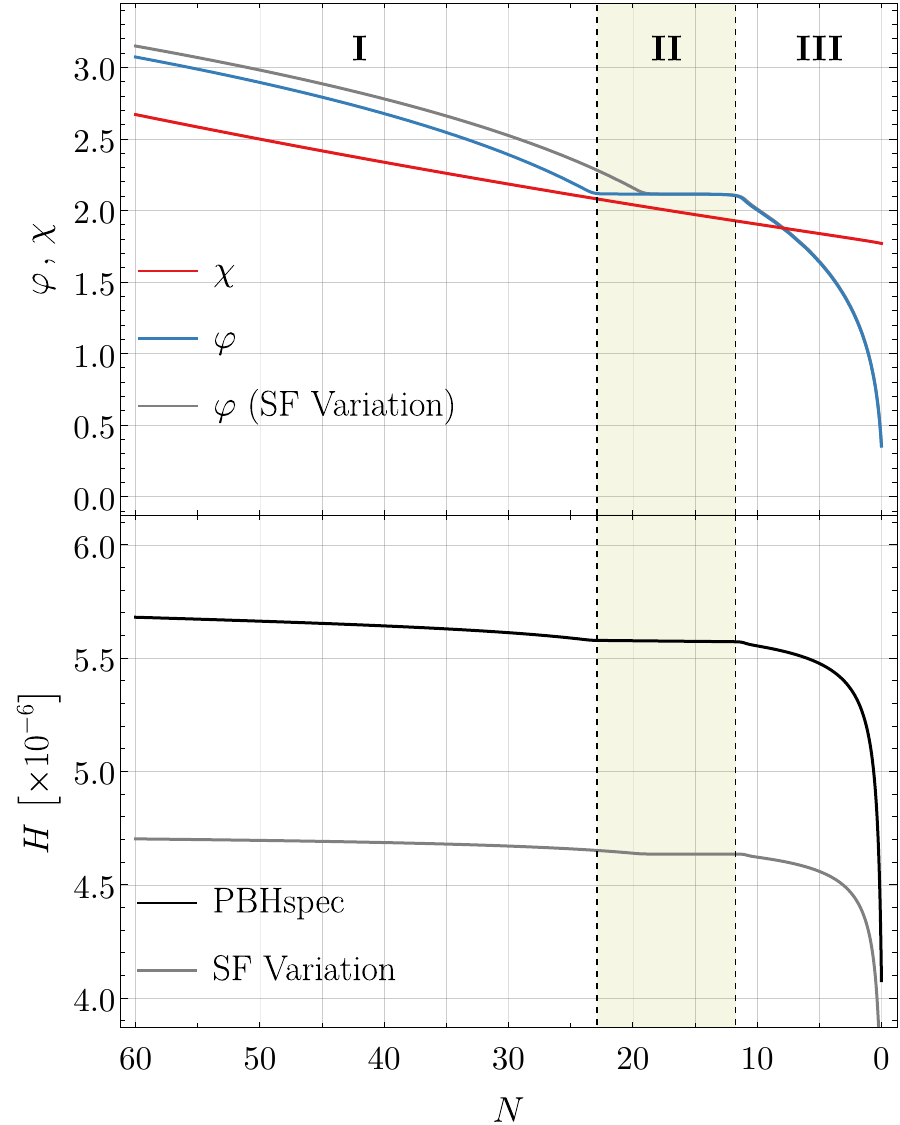}
    \caption{
    Evolution of the background fields and the Hubble parameter (in units of $M_{\rm Pl}$) for the PBHspec model as compared to the single-field case, for fixed $\vpbha$ parameters (reported in Table~\ref{tab:PBHA-params}: ``spectator'' refers to the PBHspec model, ``no spectator'' refers to the Variation model).
    \emph{Top panel:} The evolution of $\varphi$ is similar to its single-field behavior, slowing down when it hits the feature in $\vpbh$ and slow-rolling elsewhere. The spectator field $\chi$ slow-rolls at near-constant velocity, since $V_\text{S}(\chi)$ does not exhibit any particular feature.
    \emph{Bottom panel:} The Hubble parameter is increased with respect to its single-field value by the presence of $\chi$. This causes additional Hubble friction, which further slows down $\varphi$.
    \emph{All panels:} The highlighted ``phase II'' corresponds to the period of slower evolution of $\varphi$, encapsulating the single-field USR phase. As seen in \fref{fig:PBHA-background-eps-eta-omega}, it is delimited by the turns in field space. 
    }
    \label{fig:PBHA-background-fields-H}
\end{figure}

We denote the inflaton-plus-spectator setup as ``PBHspec,'' in contrast to the single-field or ``SF'' scenario. For the PBHspec case,
the background evolution of the fields is governed by their decoupled equations of motion,
\begin{equation}\label{eq:EoM-bckgr}
    \ddot{\varphi}+3 H \dot{\varphi} = - \pd_\varphi V_{\rm PBH} \, ,\quad 
    \ddot{\chi}+3 H \dot{\chi} = -\pd_\chi V_{\rm S}\,, 
\end{equation}
with $\pd_\chi=\pd/\pd\chi$, as well as one common Friedmann equation,
\begin{equation}\label{eq:Friedmann}
    H^2=\frac{1}{3M_\text{Pl}^2}\left( \frac{\dot{\varphi}^2}{2} + \frac{\dot{\chi}^2}{2} + V_{\rm PBH} + V_\text{S}\right)\,.
\end{equation}
An \textit{indirect coupling} between the two fields therefore emerges through their contributions to the Hubble parameter. Being subdominant, $\chi$ causes a slight increase of $H$, therefore augmenting the Hubble friction of $\varphi$ in \eref{eq:EoM-bckgr}.

The evolution of the background fields and of the Hubble parameter, found by solving Eqs.~\eqref{eq:EoM-bckgr} and~\eqref{eq:Friedmann}, is shown in \fref{fig:PBHA-background-fields-H}.
The PBHspec scenario is compared to the single-field PBH model, for the same $\vpbha$ parameters (see ``Variation'' model in \tref{tab:PBHA-params}; note this is not the same as the ``Reference'' parameters illustrated in \fref{fig:PBHA-background-SF}).
First, we note from the top panel that the evolution of $\varphi$ (blue curve) is similar to its single-field behavior (gray curve), insofar as it substantially slows down when $\varphi$ encounters the feature in $\vpbh(\varphi)$: this corresponds to the single-field USR phase. The inflaton $\varphi$ is slowed down for a longer period in the PBHspec case. Meanwhile, since $V_S(\chi)$ has no special features, the spectator field $\chi$ slow-rolls down its potential at a near-constant rate throughout the inflationary evolution. 
Correspondingly, the value of $H$ is slightly increased throughout inflation by the presence of $\chi$; most notably, the phase during which $H$ ``plateaus'' is extended. This augments the Hubble friction that $\varphi$ is subjected to, through \eref{eq:EoM-bckgr}, further slowing it down; this also explains why $\varphi$ plateaus for a more extended period of time.

\begin{figure}[tb!]
    \centering
    \includegraphics[width=\linewidth]{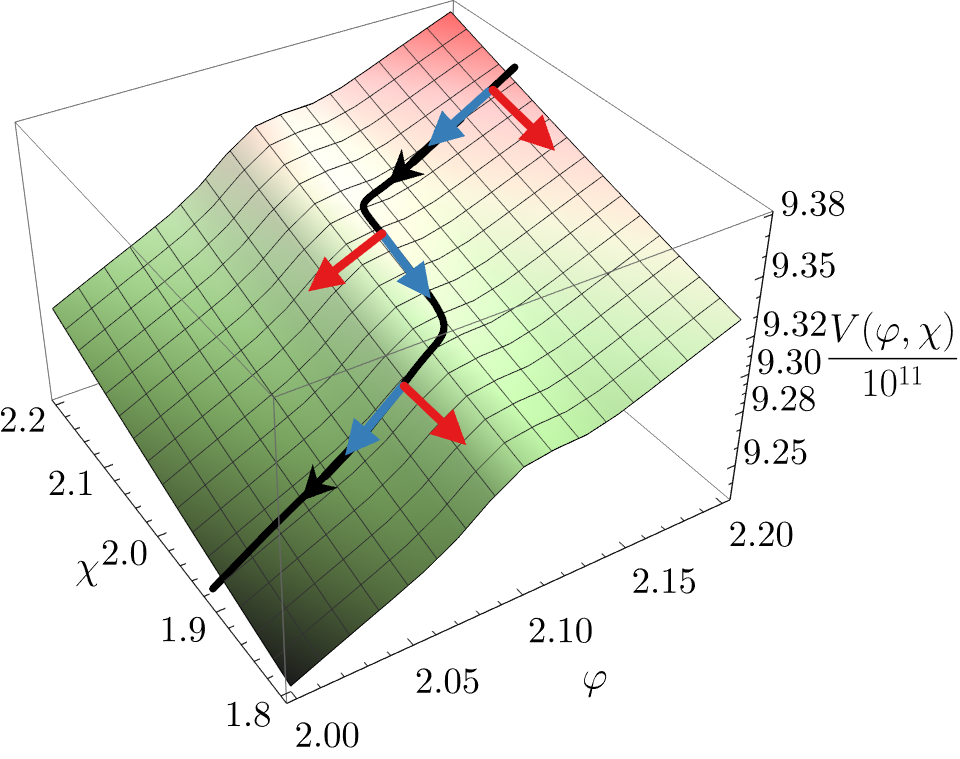}
    \caption{
    Inflationary potential $V(\varphi,\chi)=\vpbha(\varphi)+V_\text{S}(\chi)$ (in units of $M_{\rm Pl}$), enlarged on the region around the $\vpbha$ feature (corresponding to the inset in \fref{fig:PBHA-V}). The trajectory of the background field system is superimposed (black curve) as well as the direction of $\hat{\sigma}^I$ (blue) and $\hat{s}^I$ (red) at three points of the inflationary evolution corresponding to phases I--III. The ``bump'' along the $\varphi$ direction causes the trajectory to temporarily turn from the $\varphi$ direction to the $\chi$ direction; during this phase, $\hat{\sigma}^I$ and $\hat{s}^I$ switch roles. The PBHspec parameter values are listed in \tref{tab:PBHA-params}.
    }
    \label{fig:PBHA-V3d}
\end{figure}

We introduce the usual multifield notation $\phi^I=\{\varphi,\chi\}$, where the index $I$ spans over the number of fields; in the present scenario, $I=1,2$. (For a review of the formalism of multifield inflation see, e.g., Refs.~\cite{Sasaki:1995aw,Gordon:2000hv,Wands:2002bn,Langlois:2008mn,Peterson:2010np,Gong:2011uw,Kaiser:2012ak,Gong:2016qmq}.) The magnitude of the velocity of the background fields is given by 
\begin{equation}\label{eq:sigma-dot}
    \dot{\sigma}\equiv |\dot{\phi}^I| =\sqrt{\dot{\varphi}^2 + \dot{\chi}^2}\,,
\end{equation}
where the second equality is a consequence of the decoupling of the fields in our scenarios. Because of this decoupling, the first slow-roll parameter $\epsilon$ of \eref{eq:epsilon-SF}, which in a multifield scenario is given by $\epsilon=\dot{\sigma}^2/(2H^2\mpl^2)$, becomes sum separable:
\begin{equation}\label{eq:epsilon}
    \epsilon =\frac{1}{2}\frac{\dot{\varphi}^2}{H^2 M_\text{Pl}^2} + \frac{1}{2}\frac{\dot{\chi}^2}{H^2 M_\text{Pl}^2} \equiv \epsilon_\varphi+\epsilon_\chi\,.
\end{equation}

\begin{figure}[htb!]
    \centering
    \includegraphics[width=\linewidth]{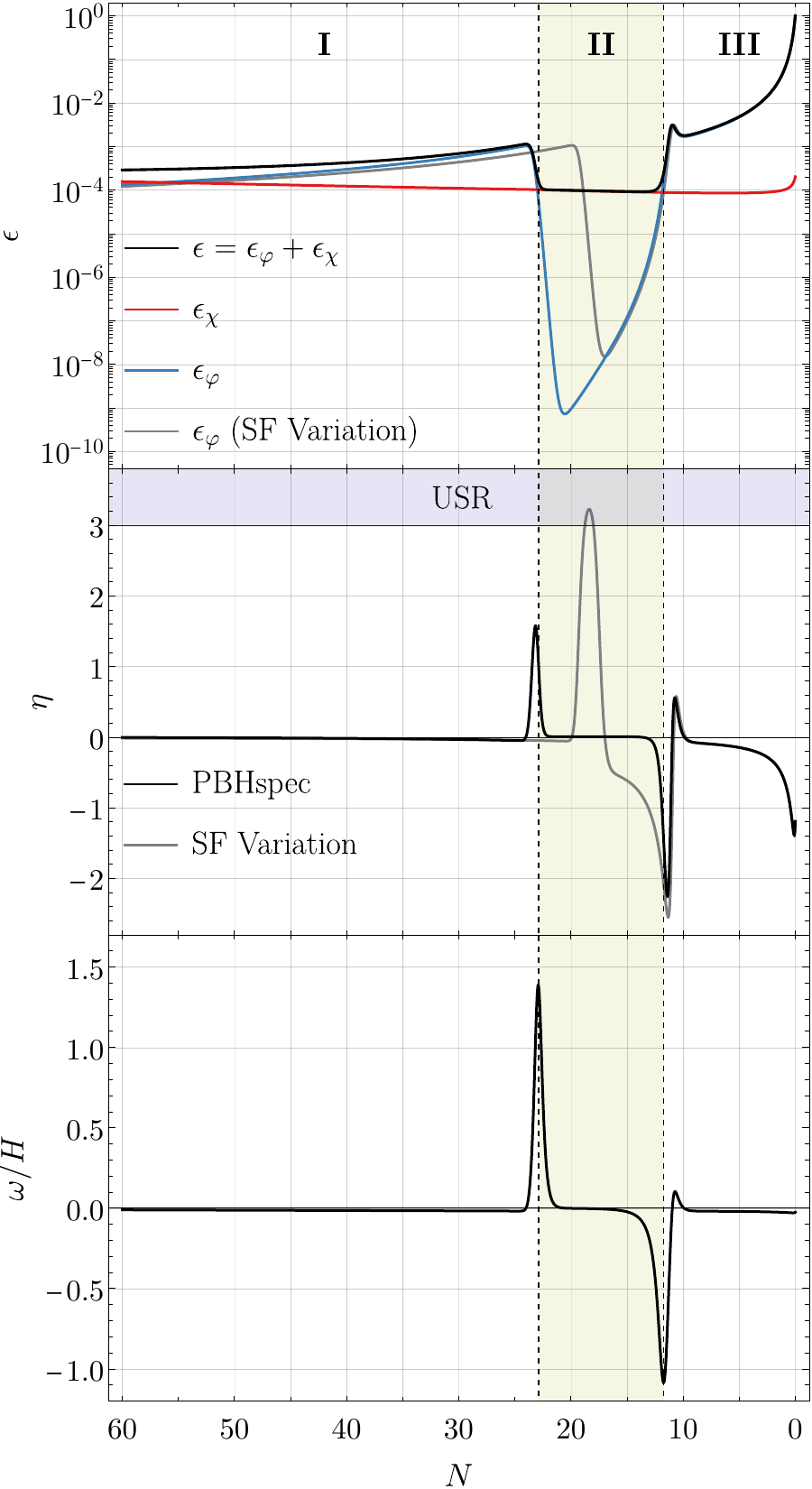}
    \caption{
    Evolution of the slow-roll parameters and the turn rate for the PBHspec and single-field models considered in \fref{fig:PBHA-background-fields-H} (parameter values reported in Table~\ref{tab:PBHA-params}).
    \emph{Top panel:} The presence of $\chi$ sets a floor on $\epsilon\geq\epsilon_\chi$: while $\epsilon_\varphi$ becomes small (tracing the single-field case), $\epsilon_\chi \sim 10^{-4}$ and we never see $\epsilon\rightarrow0$.
    \emph{Middle panel:} As $\epsilon$ never becomes unusually small, $\eta$ never enters the USR regime (unlike the single-field case).
    \emph{Bottom panel:} The trajectory of the background fields exhibits two turns, corresponding to the transitions between $\epsilon\approx\epsilon_\varphi$ and $\epsilon\approx\epsilon_\chi$ (and vice versa).
    \emph{All panels:} The highlighted phase II is delimited by the turns in field space (the dashed vertical lines correspond to the extrema of $\omega$).
    During this phase, $\epsilon_\chi\gg \epsilon_\varphi$.
    }
    \label{fig:PBHA-background-eps-eta-omega}
\end{figure}

Being an intrinsically multifield scenario, one must account for the possibility of turning field trajectories in field space.
We describe the direction of the background field system by defining the field-space unit vectors
\begin{equation}\label{eq:field-space-vectors}
    \hat{\sigma}^I \equiv \frac{\dot{\phi}^I}{\dot{\sigma}} = \left\{ \frac{\sqrt{\epsilon_\varphi}}{\sqrt{\epsilon}}, \frac{\sqrt{\epsilon_\chi}}{\sqrt{\epsilon}} \right\}\,,
    \quad
    \hat{s}^J\equiv\varepsilon^{IJ} \hat{\sigma}_I\,.
\end{equation}
The former points along the direction of the fields' motion (called the \textit{adiabatic} or \textit{curvature} direction), while the latter is orthogonal to it (the \textit{entropy} or \textit{isocurvature} direction). In \eref{eq:field-space-vectors}, $\varepsilon^{IJ}$ is the two-dimensional, antisymmetric Levi-Civita pseudotensor.
Lastly, we define the turn rate pseudovector and pseudoscalar,
\begin{equation}\label{eq:turn-rate}
    \omega^I\equiv\pd_t\hat{\sigma}^I  
    \,,\quad
    \omega\equiv \varepsilon_{IJ}\hat{\sigma}^I\omega^J\,,
\end{equation}
which describe the time evolution of the fields' trajectory in field space. In the setup of \eref{eq:action}, the turn rate simplifies to $\omega=\dot{\theta}$, where $\theta$ is the angle between $\hat{\sigma}^I$ and the $\varphi$ axis.

The field-space trajectory of the $\vpbha$ model is illustrated in \fref{fig:PBHA-V3d}, superimposed onto the two-field potential of \eref{eq:potential-generic}. This trajectory exhibits two turns, taking place when the field system reaches the feature in $\vpbha$: when the inflaton $\varphi$ slows down, the small but steady evolution of the spectator field $\chi$ becomes important, inducing the trajectory to temporarily turn along the $\chi$ direction until $\varphi$ accelerates again. The unit vector $\hat{\sigma}^I$ (in blue, at different stages of the inflationary evolution) traces the direction of the trajectory, while $\hat{s}^I$ (in red) is orthogonal to it.

\begin{figure}[tb!]
    \centering
    \includegraphics[width=\linewidth]{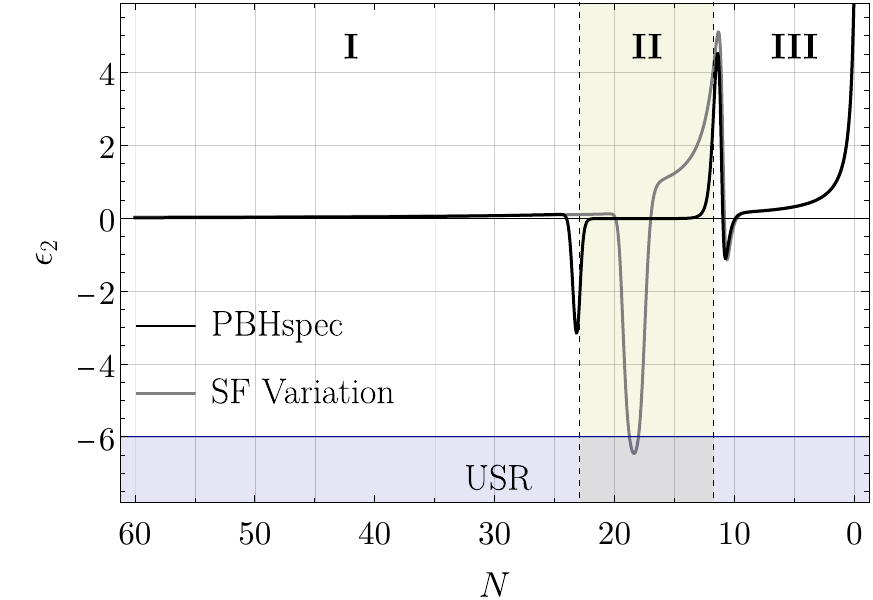}
    \caption{
    Evolution of the $\epsilon_2$ slow-roll parameter, Eq.~\eqref{eq:epsilon2}, for the same models considered in Figs.~\ref{fig:PBHA-background-fields-H}--\ref{fig:PBHA-background-eps-eta-omega}. Consistent with the middle panel of the latter figure, we can see here that the PBHspec model never enters ultra-slow-roll, whereas the single-field model spends $\sim {\cal O}(1)$ e-fold within USR.
    }
    \label{fig:PBHA-background-eps2}
\end{figure}

Figure~\ref{fig:PBHA-background-eps-eta-omega} shows the evolution of the slow-roll parameters and of the turn rate, for the PBHspec scenario as compared to the single-field PBH model. The same parameters as in \fref{fig:PBHA-background-fields-H} are considered here.

From the evolution of $\epsilon$ (top panel), we note that the sum separability of \eref{eq:epsilon} means that the presence of $\chi$ sets a \textit{floor} on the total $\epsilon$, which can never be smaller than $\epsilon\approx\epsilon_\chi$. While $\epsilon_\varphi$ reaches values of $\calO(10^{-9})$, similarly to the single-field case (although it dips lower than the gray curve, due to the additional Hubble friction provided by $\chi$), $\epsilon_\chi$ never departs from slow-roll and is near constant at $\sim\calO(10^{-4})$. The first slow-roll parameter is therefore bounded by $\epsilon\geq\epsilon_\chi$. As shown in the middle panel of Fig.~\ref{fig:PBHA-background-eps-eta-omega}, by incorporating the decoupled spectator field, the system {\it never enters USR}: even when $\varphi$ slows significantly, the system's velocity along the $\chi$ direction prevents the system from entering ultra-slow-roll, with $\eta < 3$ at all times.

Lastly, we confirm from the plot of $\omega$ (bottom panel of Fig.~\ref{fig:PBHA-background-eps-eta-omega}) that two turns take place. Interestingly, these correspond to the beginning and end of the phase in which $\epsilon_\varphi$ becomes very small and $\epsilon$ is bounded from below by $\epsilon_\chi$, denoted by ``phase II.'' 
This can be understood by looking at the form of the adiabatic unit vector $\hat{\sigma}^I$ of \eref{eq:field-space-vectors}. During the first phase of inflation, $\epsilon_\varphi>\epsilon_\chi$ and $\hat{\sigma}^I$ predominantly points along the $\varphi$ direction. When $\varphi$ slows down and $\epsilon_\varphi$ decreases, we rapidly find $\epsilon_\chi\gg\epsilon_\varphi$ and $\hat{\sigma}^I$ turns to point along the $\chi$ direction. Once $\varphi$ exits the feature in $\vpbh$, it quickly accelerates and we return to the regime $\epsilon_\varphi>\epsilon_\chi$: $\hat{\sigma}^I$ thus turns back to point along the $\varphi$ direction. Note that $\chi$ never accelerates: its near-constant velocity, subdominant during phases I and III, becomes relevant when $\varphi$ decelerates sufficiently. Only during the brief phase II does $\chi$ temporarily dominate the evolution of the system.

For completeness, we also show in \fref{fig:PBHA-background-eps2} the evolution of the $\epsilon_2$ slow-roll parameter of \eref{eq:epsilon2}, for the same parameter sets considered above. This clearly matches the behavior of $\eta$: the single-field model enters a short phase of ultra-slow-roll (defined here as $\epsilon_2\leq-6$), while the PBHspec scenario is prevented from entering this regime. We also point out that the spectator field's relative contribution to the energy density of the Universe is subdominant throughout the inflationary evolution, $\rho_\chi/\rho_\text{tot}\lesssim0.02$.

\subsection{Resilience to Parameter Variations}
\label{sec:background:resilience}

\begin{figure}[tb!]
    \centering
    \includegraphics[width=\linewidth]{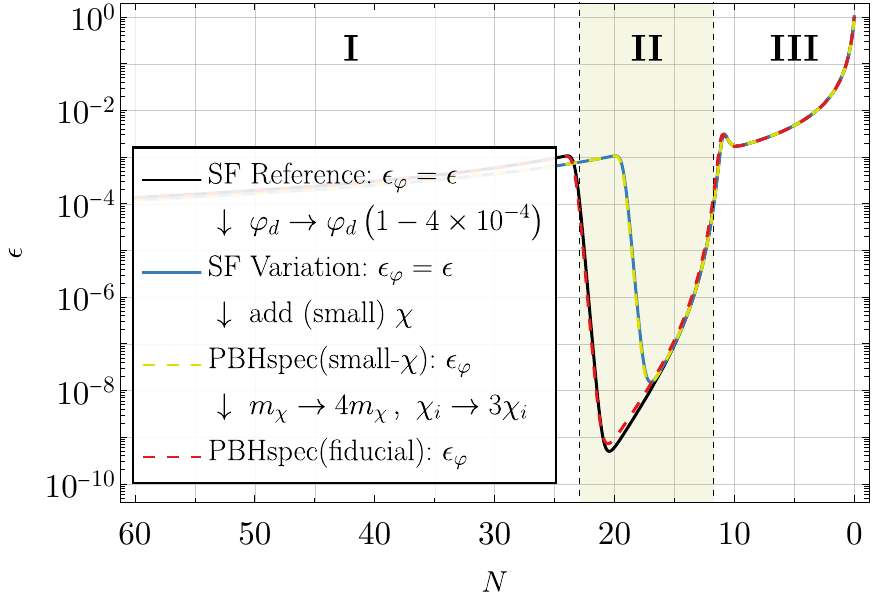}
    \caption{
    Evolution of the inflaton's slow-roll parameter $\epsilon_\varphi$ in the single-field and PBHspec models, and parameter compensations. In the single-field model, a small variation $\varphi_{d}\rightarrow (1-4\times 10^{-4})\varphi_{d}$ drastically modifies the dynamics (black to blue curve). When a spectator field is present, this shift can be compensated by $\calO(1)$ variations in the spectator field's parameters (yellow-dashed to red-dashed).
   }
    \label{fig:epsilon-compensations}
\end{figure}

The addition of the spectator field endows the PBH model with a \textit{resilience} to parameter variations. This can be observed already at the background level, from the evolution of $\epsilon_\varphi$. In Fig.~\ref{fig:epsilon-compensations} we compare $\epsilon$ in the single-field case to $\epsilon_{\varphi}$ in the PBHspec model. The black curve depicts $\epsilon$ for the single-field ``Reference'' model of \tref{tab:PBHA-params}, protagonist of \sref{sec:background:SF} and consistent with PBH production: as noted previously, $\epsilon$ decreases to $\calO(10^{-9})$. Through a tiny variation in one of the $\vpbha$ parameters, $\varphi_d\rightarrow\varphi_d (1-4\times 10^{-4})$, the black curve is shifted to the blue curve (``Variation'' model of \tref{tab:PBHA-params}). The evolution of $\epsilon$ is changed dramatically, still supporting a phase of USR (gray curve in \fref{fig:PBHA-background-eps-eta-omega}) but not sufficient to produce PBHs.

Adding a spectator field to the single-field Variation model may not change the situation (blue to yellow-dashed curve), but making the spectator more massive can restore $\epsilon_\varphi$ to its original form (yellow-dashed to red-dashed curve: $m_\chi\rightarrow6m_\chi$, $\chi_i\rightarrow3\chi_i$) \textit{without} requiring any fine-tuning beyond $\calO(1)$ changes. The evolution of $\epsilon_\varphi$ is therefore resilient to small variations in the $\vpbh$ parameters, due to compensations by the spectator field.

\section{Perturbations}
\label{sec:perturbations}

We now consider perturbations to the background system, using the familiar multifield formalism of Refs.~\cite{Sasaki:1995aw,Gordon:2000hv,Wands:2002bn,Langlois:2008mn,Peterson:2010np,Gong:2011uw,Kaiser:2012ak,Gong:2016qmq}. Throughout the classical evolution of the background fields described in the previous section, quantum fluctuations will be sourced. At any point in space and time, each field can be expanded around its homogeneous background value,
\begin{equation}\label{eq:field-expansion-pert}
    \Phi^I(x^\mu)=\phi^I(t) + \delta \phi^I(x^\mu)\,,
\end{equation}
where $\phi^I=\{\varphi,\chi\}$ obey \eref{eq:EoM-bckgr} and $\delta\phi^I$ represent small departures from the background value.
Considering perturbations around flat Friedmann-Lema\^itre-Robertson-Walker spacetimes, one may construct the Mukhanov-Sasaki variables, which remain gauge invariant to first order in the perturbations,
\begin{equation}\label{eq:MS-var-def}
    Q^I \equiv \delta\phi^I + \frac{\dot{\phi}^I}{H}\psi\,,
\end{equation}
where $\psi(x^\mu)$ is the comoving scalar metric perturbation on equal-time hypersurfaces. 

The PBHspec setup discussed in \sref{sec:background:PBHspec} is intrinsically multifield. Having multiple possible directions in field space, quantum fluctuations will be produced both \textit{along} the direction of the background fields' motion, and \textit{perpendicular} to it. These correspond to the adiabatic and isocurvature directions identified in \eref{eq:field-space-vectors}. 
The Mukhanov-Sasaki variables of Eq.~\eqref{eq:MS-var-def} can then be projected along the field-space unit vectors as
\begin{equation}\label{eq:MS-var-decomp}
    Q^I = \hat{\sigma}^I Q_\sigma + \hat{s}^I Q_s\,,
\end{equation}
thus defining the scalar adiabatic and isocurvature perturbations $Q_\sigma$ and $Q_s$, respectively.  
These can be recast as the canonically normalized comoving perturbations
\begin{gather}
    \calR = \frac{H}{\dot{\sigma}}Q_\sigma\,,\label{eq:R-def}
    \\
    \calS = \frac{H}{\dot{\sigma}}Q_s\,.\label{eq:S-def}
\end{gather}
The adiabatic (or curvature) perturbation $\calR$ is also produced in single-field inflationary models, while the isocurvature perturbation $\calS$ is unique to multifield systems. Whereas the former 
reflects variations in the total energy density of the Universe, the latter leaves the total energy density unchanged while modifying the ratio of energy density associated with distinct particle species at various spacetime locations.

The power spectra for the curvature and isocurvature perturbations are defined as the two-point correlators
\begin{equation}\label{eq:PRk-def}
    \calP_{\cal X}(k,N) \equiv \frac{k^3}{2\pi^2}|{\cal X}_k(N)|^2~,
\end{equation}
for ${\cal X}=\{\calR,\calS\}$. 
When the time dependence is not explicitly stated, the power spectrum will be evaluated at the end of inflation, $\calP_{\cal X}(k)\equiv\calP_{\cal X}(k,N_\text{end})$.
This is the principal quantity of interest for studying predictions from inflation, as its shape can be tested against observational data, as discussed further in \sref{sec:spectrum:CMB}.

\subsection{Single-Field Ultra-Slow-Roll}
\label{sec:perturbations:USR}

In the single-field model described in \sref{sec:background:SF}, the Mukhanov-Sasaki variable of Eq.~\eqref{eq:MS-var-decomp} reduces to the scalar perturbation along the trajectory of the inflaton, $Q_\sigma$. The curvature perturbation satisfies
\begin{equation}\label{eq:R-def-SF}
    \calR \equiv \frac{H}{\dot{\varphi}}Q_\sigma = \frac{H}{\dot{\varphi}}\delta\varphi + \psi\,.
\end{equation}
It can be decomposed into Fourier modes $\calR_k$, 
which cross the Hubble radius $(aH)^{-1}$ at a time $N_*$ where the comoving wavenumber $k=a(N_*)H(N_*)$.
To linear order in the perturbations, these modes obey the equation of motion 
\begin{equation}\label{eq:EoM-R-SF}
    \ddot{\calR}_k + (3+\delta)H\dot{\calR}_k + \frac{k^2}{a^2}\calR_k = 0\, ,
\end{equation}
where we use the notation $\delta=\epsilon_2=4\epsilon-2\eta$ for consistency with the literature.
In the limit $k \ll aH$, curvature perturbations will be enhanced whenever the Hubble damping term $(3+\delta)H\dot{\calR}_k$ becomes negative. A phase of ultra-slow-roll, as defined by~\eref{eq:USR-condition}, provides exactly these conditions.
In particular, whenever $\epsilon\ll1$ and $\eta\geq3/2$, long-wavelength modes for which $k^2/(aH)^2<|\dot{\calR}_k/(H\calR_k)|$ will be \textit{antidamped}, growing quasiexponentially.

\begin{figure}[tb!]
    \centering
    \includegraphics[width=\linewidth]{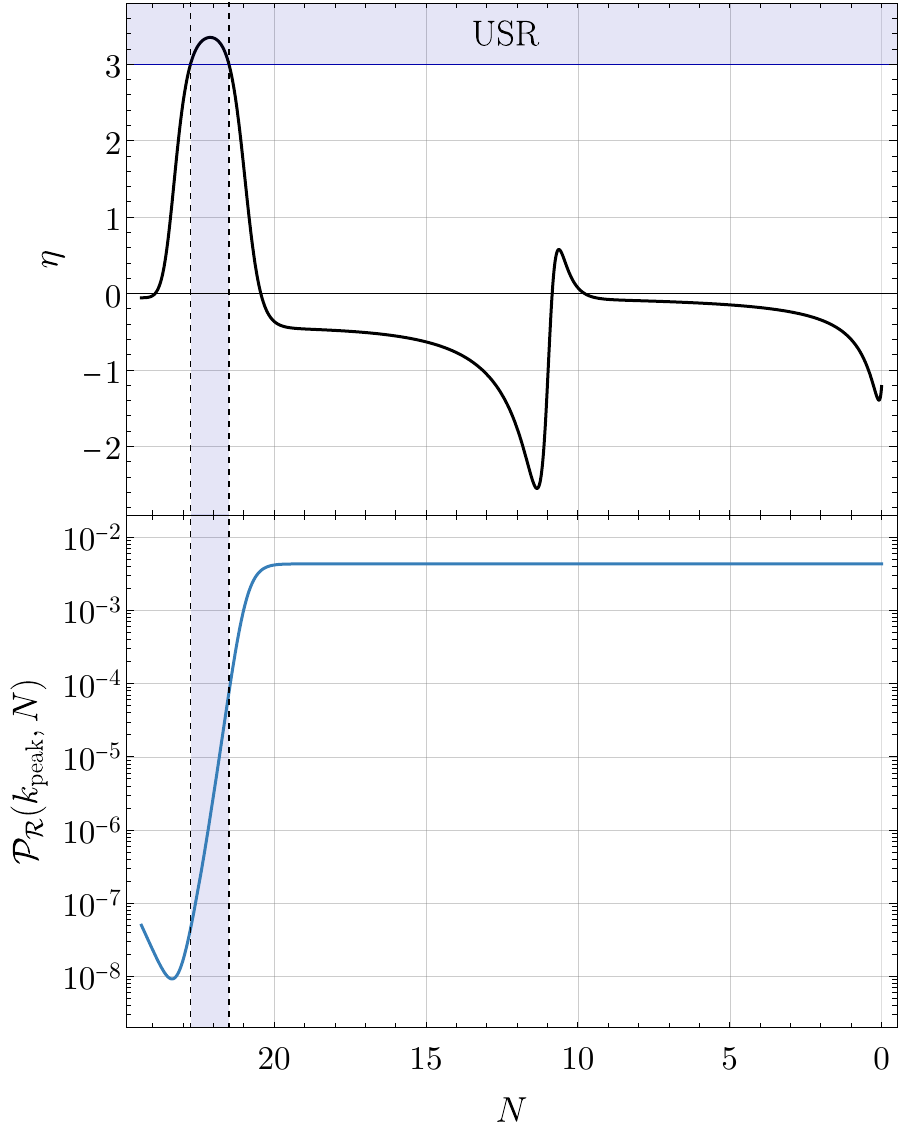}
    \caption{
    Evolution of the second slow-roll parameter $\eta$ and corresponding amplification of the curvature mode $\calR_{k_\text{peak}}$, in the single-field ``Reference'' model of \tref{tab:PBHA-params}.
    \textit{Top panel:} As seen previously in \fref{fig:PBHA-background-SF}, $\eta$ crosses the USR threshold for $\sim1.4$ e-folds, inducing a phase of ultra-slow-roll evolution.
    \textit{Bottom panel:} We depict the time-dependent power spectrum for a curvature mode $\calR_{k_\text{peak}}$ that exits the Hubble radius shortly before the onset of the USR phase (at $N_*\approx23$). While $\eta\geq3/2$, the curvature perturbation grows, with $\prk(k_\text{peak},N)$ being enhanced by $\sim10^{6}$. After USR, the curvature mode remains frozen at this enhanced value until the end of inflation.
    }
    \label{fig:PBHA-perturbations-SR}
\end{figure}

This behaviour can be seen in \fref{fig:PBHA-perturbations-SR} for the single-field $\vpbha$ potential of \eref{eq:VPBH-A}: when the system enters the phase of ultra-slow-roll (top panel, $\eta\geq3$), the curvature perturbation grows quasiexponentially (bottom panel), resulting in an enhancement of $\sim 10^{6}$ with respect to its pre-USR values. Notice that the figure portrays the evolution of the time-dependent power spectrum for a fixed $k$-mode.
The curvature mode shown here crosses the Hubble radius shortly before the onset of the USR phase, thus simultaneously satisfying the condition for antidamping and being subjected to the entirety of the USR-induced growth. It thus corresponds to the mode $\calR_{k_\text{peak}}$, which is maximally enhanced.

\subsection{Influence of the Spectator}
\label{sec:perturbations:PBHspec}

Curvature perturbations are always produced during the inflationary expansion. Additionally, the presence of the spectator field makes it possible for the isocurvature fluctuations of \eref{eq:S-def} to be produced as well. The equations of motion for the Fourier modes of these two types of perturbations are coupled. To linear order, they obey
\begin{gather}
    \frac{d}{dt}  \left(\dot {\cal R}_k - 2 \omega {\cal S}_k \right)
    + (3 + \delta) H \left( \dot  {\cal R}_k - 2 \omega {\cal S}_k \right)  + \frac{k^2}{a^2}{\cal R}_k = 0\,,
    \label{eq:EoM-R}
    \\
    \ddot{\cal S}_k+ (3+ \delta) H \dot{\cal S}_k + \left(\frac{k^2}{a^2} + \mu^2_s\right){\cal S}_k = - 2 \omega \dot{\cal R}_k\,.
    \label{eq:EoM-S}
\end{gather}
We define the effective isocurvature mass as
\begin{equation}\label{eq:muS}
    \mu_s ^2 = {\cal M}_{ss} - {\cal M}_{\sigma\sigma} + 2 H^2 \epsilon\left(3+\delta-\epsilon\right)\, ,
\end{equation}
with $\calM_{\sigma\sigma}$ and $\calM_{ss}$ being the projections of the mass-matrix $\calM^I_{\,J}=\pd^I\pd_J V$ along the adiabatic and isocurvature directions, respectively%
\footnote{%
    Given our simple setup, with strictly canonical kinetic terms for both fields and hence a trivial field-space structure, the effective masses ${\cal M}_{\sigma \sigma}$ and ${\cal M}_{ss}$ do not involve any terms related to the (here vanishing) Riemann tensor of the field-space manifold. For the more general case, see, e.g., Ref.~\cite{Kaiser:2012ak}.
}%
:
\begin{gather}
    \calM_{\sigma\sigma} = \hat{\sigma}^I\hat{\sigma}^J\pd_I\pd_J V\,,
    \\
    \calM_{ss} = \hat{s}^I\hat{s}^J\pd_I\pd_J V\,,
\end{gather}
with $\pd_I V\equiv\pd V/\pd\phi^I$.

In the long-wavelength limit, that is, on super-Hubble scales ($k/a\ll H$), the equations of motion of Eqs.~\eqref{eq:EoM-R} and~\eqref{eq:EoM-S} simplify to
\begin{gather}
    \dot {\cal R}_k \simeq
    2\omega \calS_k\,,
    \label{eq:EoM-R-superHubble}
    \\
    \ddot{\cal S}_k+ (3+ \delta) H \dot{\cal S}_k + \tilde{\mu}^2_s{\cal S}_k \simeq 0\,,
    \label{eq:EoM-S-superHubble}
\end{gather}
with the effective isocurvature mass defined as $\tilde{\mu}_s^2 \equiv \mu_s^2 + 4\omega^2$.
The equation of motion for $\calS_k$ indicates that isocurvature modes will be enhanced whenever $\tilde{\mu}_s^2<0$, corresponding to a phase of tachyonic instability, and they will decay whenever $\tilde{\mu}_s^2>0$. The evolution of $\calR_k$, meanwhile, indicates that there can be a super-Hubble transfer of power from isocurvature to curvature perturbations whenever the background fields' trajectory experiences a turn in field space, with $\omega \neq 0$.

\begin{figure}[h!]
    \centering
    \includegraphics[width=\linewidth]{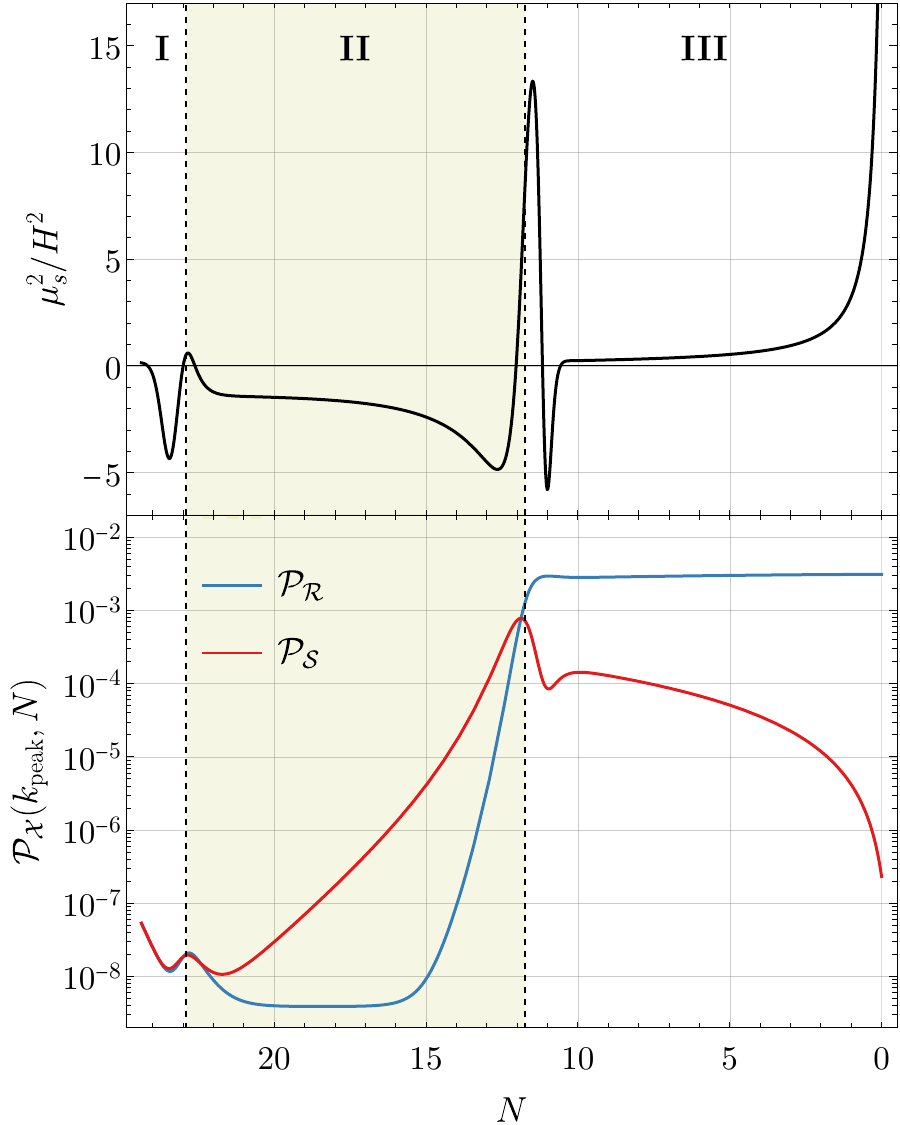}
    \caption{
    Evolution of the effective isocurvature mass $\mu_s^2$ and interaction of the curvature and isocurvature perturbations $\calR_{k_\text{peak}}$ and $\calS_{k_\text{peak}}$ for the PBHspec model. 
    \textit{Top panel:} During phase II (when $\epsilon_\chi\gg \epsilon_\varphi$ and the system evolves along the $\chi$ direction), $\mu_s^2$ becomes negative: the isocurvature perturbations experience a tachyonic instability. Toward the end of inflation, $\mu_s^2$ becomes large and positive.
    \textit{Bottom panel:} The time-dependent power spectra for curvature $\calR_{k_\text{peak}}$ and isocurvature $\calS_{k_\text{peak}}$ modes that exit the Hubble radius shortly before the beginning of phase II, that is, just before the first turn in field space (at $N_*\approx23$). The tachyonic phase causes a near-exponential growth of $\calS_k$, which then transfers its power to $\calR_k$ at the end of phase II, when the second turn takes place. During phase III, the positive $\mu_s^2$ causes $\calS_k$ to decay.
    }
    \label{fig:PBHA-perturbations}
\end{figure}

In the PBHspec model of \eref{eq:action}, the USR enhancement of curvature modes is supplanted by a transient period of tachyonic isocurvature growth and a subsequent transfer of power to curvature modes.
After the first turn in field space, when $\epsilon_\chi\gg\epsilon_\varphi$, the adiabatic unit vector $\hat{\sigma}^I$ will temporarily point along the $\chi$ direction and, vice versa, the isocurvature vector $\hat{s}^I$ will trace the $\varphi$ direction. For this reason, within phase II, the mass-matrix projections become $\calM_{\sigma\sigma}\approx\pd_\chi^2 V_S(\chi)$ and $\calM_{ss}\approx\pd_\varphi^2 \vpbh(\varphi)$. The effective isocurvature mass then simplifies to
\begin{equation}
    \mu_s^2\approx \pd_\varphi^2 \vpbh(\varphi) - \pd_\chi^2 V_S(\chi)\, ,
\label{musapprox}
\end{equation}
since the last term in \eref{eq:muS} is suppressed by $\epsilon$. The second term in Eq.~(\ref{musapprox}) is negative definite since the spectator field's mass must be a real quantity. As mentioned at the end of \sref{sec:background:PBHspec}, the first term necessarily becomes temporarily negative in order for $\varphi$ to evolve beyond the local feature in $\vpbh$, and for inflation to end. This indicates that during most of phase II, $\mu_s^2$ is negative, indicating a tachyonic instability of the isocurvature perturbations.

We can indeed observe a tachyonic instability when examining the evolution of $\mu_s^2$ for $\vpbha$ of \eref{eq:VPBH-A}, as shown in the top panel of \fref{fig:PBHA-perturbations}.\footnote{Note that we consider the evolution of $\mu_s^2$ instead of the super-Hubble mass $\tilde{\mu}_s^2$ introduced in \eref{eq:EoM-S-superHubble}, because the $k$-modes analyzed here only exit the Hubble radius at the time of the first turn. Given that $k/a\sim H$ at those times, the effective mass $\mu_s^2$ of Eq.~(\ref{eq:muS}) remains most relevant. Shortly thereafter, though, the regime $k/a\ll H$ becomes valid, and for most of phases II and III, these modes evolve with $\tilde{\mu}_s^2 \simeq \mu_s^2$, as shown in \fref{fig:muS}.}

Correspondingly, isocurvature modes $\calS_k$ experience a rapid growth during phase II (bottom panel, red curve). They then transfer their power to $\calR_k$ (blue curve) during the second turn in field space, at the interface between phases II and III. Afterward, $\calS_k$ modes quickly decay due to $\mu_s^2$ becoming large and positive, while modes $\calR_k$ instead become frozen at their amplified value.

\subsection{Common Ground}
\label{sec:perturbations:ddV}

Both the single-field and PBHspec models show an enhancement of perturbations, though by means of totally different mechanisms. It is helpful to pause and consider the relation between the two models.
 
To understand the mapping between the single-field and PBHspec model, consider the equations of motion for the field fluctuations, namely the gauge-invariant Mukhanov-Sasaki variables of \eref{eq:MS-var-decomp}. Denoting the perturbations $Q^I=\{Q_\varphi,Q_\chi\}$, we note that these are related to $Q_\sigma, Q_s$ by a time-dependent rotation defined by the direction of $\hat{\sigma}^I$. 
Let us compare the single-field USR phase with the PBHspec phase II, when $\omega=0$ and the crucial growth of isocurvature occurs.

In the single-field case, $Q_{\sigma}=Q_\varphi$, which evolves following
\cite{Kaiser:2012ak}
\begin{equation}\label{eq:EoM-Qphi-SF}
    \ddot{Q}_\varphi  + 3 H \dot{Q}_\varphi + \left[ \frac{k^2}{a^2} 
    + \calM_{\sigma\sigma} - \frac{1}{\mpl^2 a^3} \frac{\dd}{\dd t} \left( \frac{a^3 \dot{\varphi}^2}{H} \right) \right] Q_\varphi =0\,.
\end{equation}
The third term in the brackets is suppressed by $\epsilon$, which is exponentially small during the single-field USR phase. Since $\calM_{\sigma\sigma}=\pd^2_\varphi \vpbh$ in this scenario, we can therefore approximate \eref{eq:EoM-Qphi-SF} as
\begin{equation}\label{eq:EoM-Qphi-SF-approx}
    \ddot{Q}_\varphi  + 3 H \dot{Q}_\varphi + \left[ \frac{k^2}{a^2} 
    + \pd^2_\varphi \vpbh(\varphi) \right] Q_\varphi \approx 0\,.
\end{equation}

\begin{figure}[tb!]
    \centering
    \includegraphics[width=\linewidth]{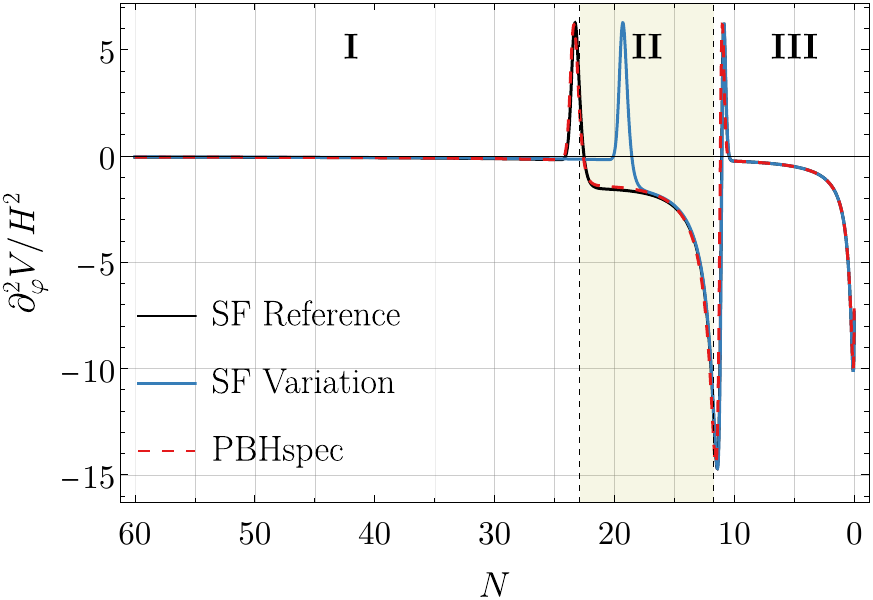}
    \caption{    
    Evolution of $\pd_\varphi^2V/H^2$. This quantity determines the USR growth of the single-field PBH model and the tachyonic growth of isocurvature perturbations in the PBHspec model. The former relies on a finely tuned feature of the potential, whereas the latter does not.
    }
    \label{fig:PBHA-ddV}
\end{figure}

In the PBHspec case, one has both $Q_{\sigma}$ and $Q_s$. We are especially interested in the latter, which evolves as \cite{Kaiser:2012ak,Kaiser:2013sna,Schutz:2013fua}
\begin{equation}\label{eq:EoM-Qs}
    \ddot{Q}_s + 3 H \dot{Q}_s + \left[ \frac{k^2}{a^2} + {\cal M}_{ss} + 3 \omega^2  \right] Q_s = 4 M_{\rm pl}^2 \frac{\omega}{\dot{\sigma}} \frac{k^2}{a^2} \psi\,.
\end{equation}
During phase II, while $\hat{\sigma}^I$ is aligned with the $\chi$ direction, one has $Q_s \approx Q_{\varphi}$ and ${\cal M}_{ss} \approx \pd^2_\varphi \vpbh$. Since $\omega=0$ during this phase, \eref{eq:EoM-Qs} takes the simple form
\begin{equation}\label{eq:EoM-Qphi-PBHspec-approx}
   \ddot{Q}_\varphi + 3 H \dot{Q}_\varphi + \left[ \frac{k^2}{a^2} +  \pd^2_\varphi \vpbh(\varphi) \right] Q_\varphi 
    \approx 0 \,,
\end{equation}
which exactly matches the equation for $Q_\varphi$ in the single-field model, \eref{eq:EoM-Qphi-SF-approx}. Both are completely determined by $H$ and $\pd^2_\varphi \vpbh$, and in particular, we expect the evolution to be completely described by the evolution of the dimensionless ratio $\pd^2_\varphi \vpbh/H^2$. To make this more explicit, we rewrite \eref{eq:EoM-Qphi-PBHspec-approx} by changing variables from cosmic time to number of e-folds (see \eref{eq:N-efolds}; primes denote $N$-derivatives) and dividing by $H^2$:
\begin{equation}\label{eq:EoM-Qphi-PBHspec-approx-Ne}
   Q_\varphi'' + (3-\epsilon)\, Q_\varphi' + \left[ \frac{k^2}{(aH)^2} +  \frac{\pd^2_\varphi \vpbh(\varphi)}{H^2} \right] Q_\varphi 
    \approx 0 \,.
\end{equation}
The quantity $\pd^2_\varphi \vpbh/H^2$ will therefore act as a mass term for the evolution of $Q_\varphi$. On super-Hubble scales, when $k/(aH)\ll1$, this equation will display the same behavior as \eref{eq:EoM-S-superHubble}: we expect $Q_\varphi$ to grow when $\pd^2_\varphi \vpbh/H^2<0$, and to decay when $\pd^2_\varphi \vpbh/H^2>0$. 
In \fref{fig:PBHA-ddV} we compare the evolution of $\pd^2_\varphi \vpbh/H^2$ in the single-field model and in the PBHspec model.

This establishes a correspondence between the growth of perturbations in the two models. In the single-field case, the growth is attributed to ultra-slow-roll evolution while the field traverses a fine-tuned feature of its potential. In the PBHspec scenario, this is realized as a tachyonic growth of isocurvature perturbations, which can then transfer power to the curvature perturbations during a field-space turn. Importantly, the latter is realized {\it without the need to fine-tune the potential}.

\section{Power Spectrum and CMB Observables}
\label{sec:spectrum}

We numerically solve the background equations of Eqs.~\eqref{eq:EoM-bckgr} and~\eqref{eq:Friedmann} and perturbation equations of Eqs.~\eqref{eq:EoM-R} and~\eqref{eq:EoM-S}, with Bunch-Davies initial conditions imposed on the perturbations. We perform the calculations of both background quantities and perturbations using two independent codes written in Python and Mathematica, respectively, and find excellent agreement between the two; versions of each of these codes are available at the public GitHub repository \texttt{PBHspec}~\cite{CODEnote}. In Python, we use the publicly available code \texttt{PyTransport}~\cite{Dias:2016rjq,Mulryne:2016mzv}, which uses the transport method \cite{Dias:2015rca} to directly compute the time evolution of correlation functions. In Mathematica, we solve the perturbation equations in the form presented in \rref{Achucarro:2010da} and, from this, construct the power spectra.

\begin{figure}[tb!]
    \centering
    \includegraphics[width=\linewidth]{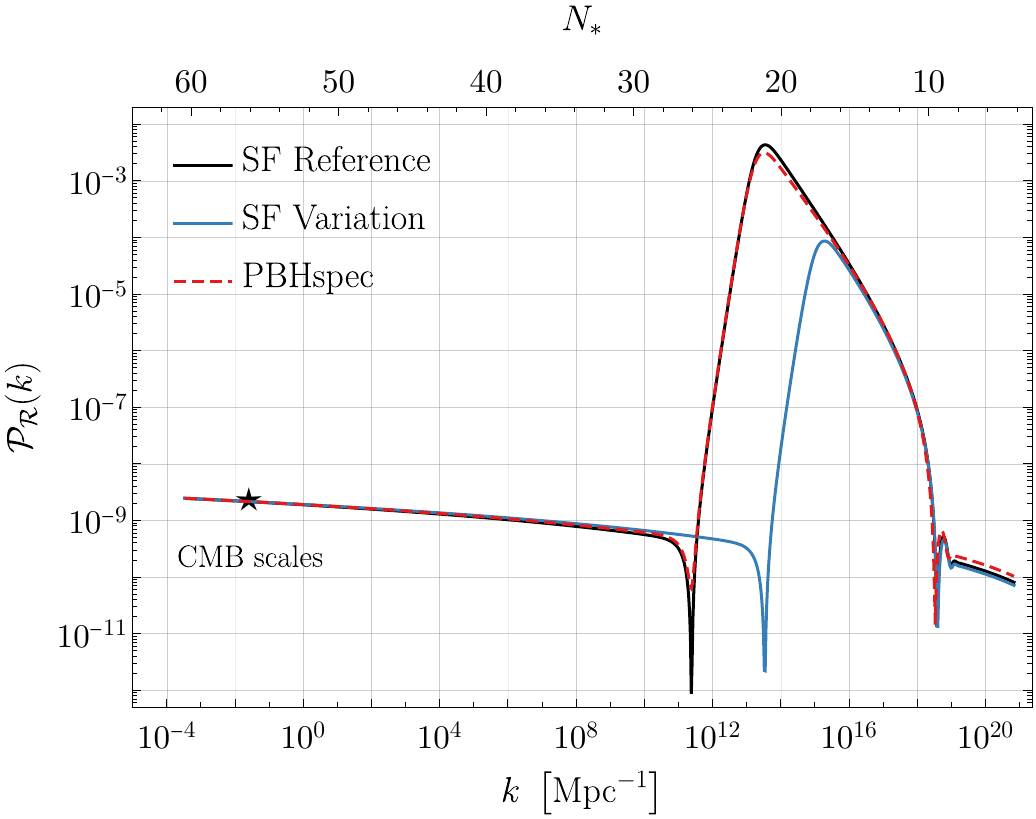}
    \caption{
    Power spectrum of curvature perturbations for the $\vpbha$ potential of \eref{eq:VPBH-A}, in both the single-field (black and blue curves) and PBHspec (red curve) scenarios. The parameters used in the figure are summarized in \tref{tab:PBHA-params}. The black and red curves exhibit a large (and similar) enhancement of $\prk$, albeit due to different mechanisms; as the peak of these spectra surpasses the threshold of $10^{-3}$, these models allow for the production of primordial black holes. These models satisfy the ``steepest growth'' constraints \cite{Byrnes:2018txb}, as $\prk\propto k^4$ to the left of the peak, and the CMB parametrization $\prk =A_s(k/k_{\rm CMB})^{n_s-1}$ at the CMB scales highlighted by the black star.
    }
    \label{fig:PBHA-PRk}
\end{figure}

The power spectrum of curvature perturbations $\prk(k)$ for the PBHspec model is shown in \fref{fig:PBHA-PRk}. In this section we analyze this power spectrum in detail, separating our discussion into the modes relevant to PBH formation and those relevant to CMB observations.

\subsection{Shape of the Power Spectrum}
\label{sec:spectrum:shape}

As we have seen in \sref{sec:perturbations}, the dynamics of the single-field and PBHspec models produce an amplification of the curvature perturbations for certain ``peak'' $k$-modes, albeit through very different mechanisms. 
This leads to an enhancement of the curvature power spectrum on a characteristic scale $k$ set by the timing of the onset of the USR (single-field) or tachyonic (PBHspec) phase. 
This is illustrated in \fref{fig:PBHA-PRk}, in which we show the curvature power spectrum for varying $k$ evaluated at the end of inflation. 

One may appreciate that the characteristic peak of $\prk(k)$ in the PBHspec scenario (red curve) closely mirrors the single-field result (black and blue curves), even though the multifield PBHspec system never enters a phase of USR. To the left of the peak, the spectrum realizes the well-studied ``steepest growth,'' with $\prk(k)\propto k^4$ \cite{Byrnes:2018txb}. To the right of the peak, the spectrum features a sharp decay (which is a power law for constant-$\eta$ models \cite{Ozsoy:2023ryl}). The peak of the power spectrum is realized by the modes which exit the Hubble radius shortly before or during phase II, analogous to the single-field scenario in which the peak is due to modes that exit at the onset of the USR phase.

\begin{figure}
    \centering
    \includegraphics[width=1\linewidth]{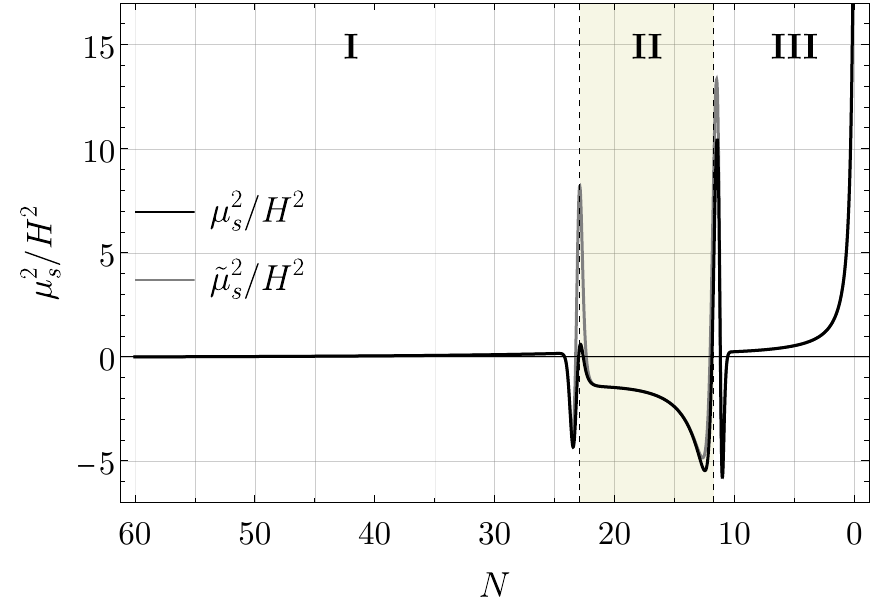}
    \caption{Evolution of the effective isocurvature mass for sub-Hubble ($\mu_s^2$) and super-Hubble ($\tilde{\mu}_s^2$) modes, for the PBHspec model of $\vpbha$. The two quantities differ by $4\omega^2$, which becomes significant during the transitions between different phases. Most remarkably, at the transition between phases I and II, $\tilde{\mu}_s^2$ gains a substantial contribution from the turn. As $\tilde{\mu}_s^2$ becomes large and positive, it causes isocurvature modes that are already super-Hubble at this time to be largely suppressed. The tachyonic instability is near identical for the two quantities, meaning all $\calS_k$ modes experience a similar tachyonic growth during phase II, and eventually decay during phase III.
    }
    \label{fig:muS}
\end{figure}

\begin{figure}[tb!]
    \centering
    \includegraphics[width=\linewidth]{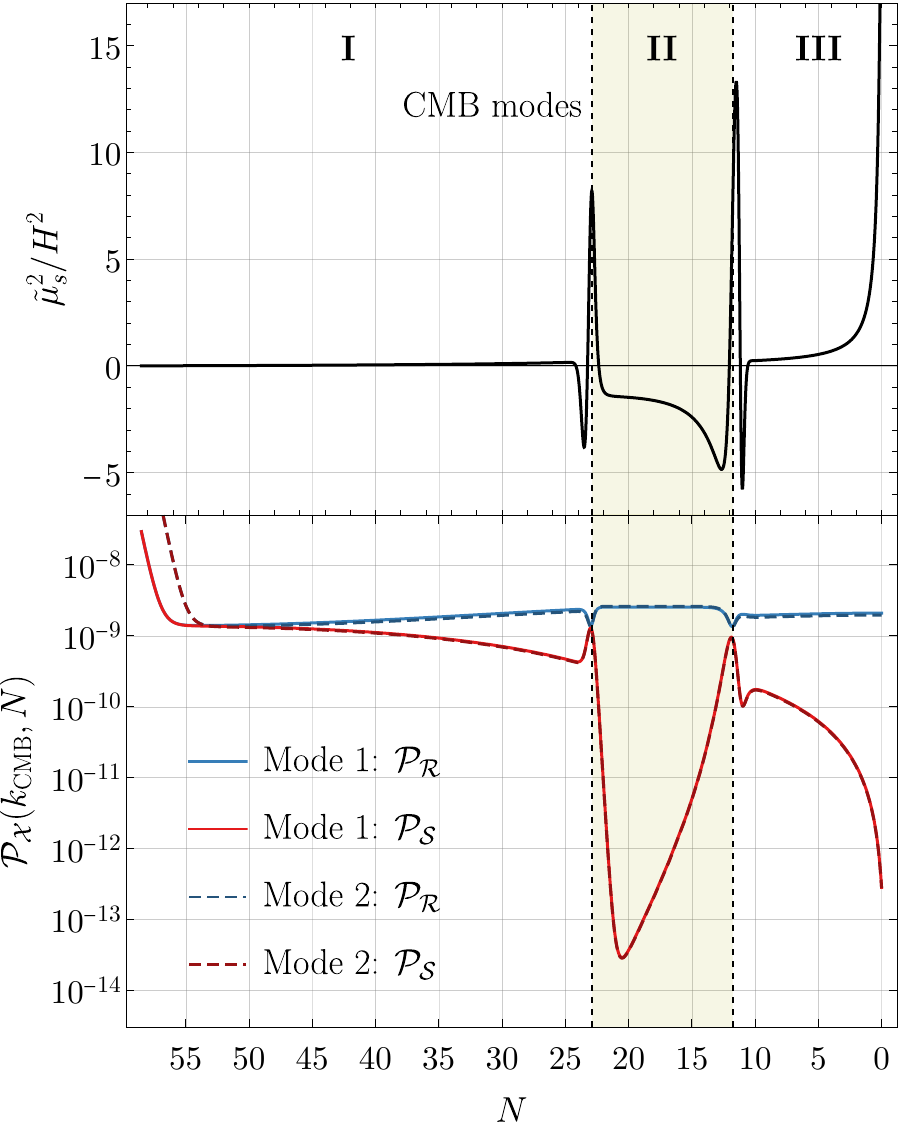}
    \caption{
    Evolution of the super-Hubble effective isocurvature mass $\tilde{\mu}_s^2$ and the curvature and isocurvature perturbations $\calR_{k}$ and $\calS_{k}$, for two $k$-modes that cross the Hubble radius around CMB scales, near $k_{\rm CMB}=0.05~{\rm Mpc}^{-1}$. 
    \textit{Top panel:} At the boundary between phases I and II, $\tilde{\mu}_s^2$ briefly becomes large and positive due to the presence of a turn in field space. During phase II (when $\epsilon_\chi\gg \epsilon_\varphi$ and the system evolves along the $\chi$ direction), $\tilde{\mu}_s^2$ drives a tachyonic instability for modes ${\cal S}_k$, while in phase III it grows to become large and positive.
    \textit{Bottom panel:} The time-dependent power spectra of \eref{eq:PRk-def}, for the curvature and isocurvature perturbations of two modes $k_\text{CMB,1}$ and $k_\text{CMB,2}$ that cross the Hubble radius two e-folds apart, at CMB scales. Before exhibiting the tachyonic growth of phase II (similarly to the ``peak'' modes of \fref{fig:PBHA-perturbations}) and the late-time decay of phase III,  CMB modes are suppressed by $\sim\calO(10^{-4})$ between phases I and II. This is due to the positive peak of $\tilde{\mu}_s^2$, which affects the evolution of modes near $k_{\rm CMB}$ but which occurs while modes with $k_{\rm peak} \gg k_{\rm CMB}$ remain deep inside the Hubble radius. The sharp peak of $\tilde{\mu}_s^2$ therefore compensates the tachyonic enhancement for modes near $k_{\rm CMB}$ but not for those near $k_{\rm peak}$.  
    Furthermore, the two opposing field-space turns that occur while the modes with $k_{\rm CMB}$ are outside the Hubble radius 
    cause two transfers of power from $\calS_k$ to $\calR_k$: the first one is positive, while the second is negative. The value of $\prk(k_\text{CMB})$ at the end of inflation is therefore near identical to its single-field value when CMB modes first cross the Hubble radius (explaining the similarity in $A_s$).
    Furthermore, the evolution of $k_\text{CMB,1}$ and $k_\text{CMB,2}$ is self-similar, explaining why the values of $n_s$ (and $\alpha_s$) are similar to the single-field model.    
    }
    \label{fig:PBHA-perturbations-CMB}
\end{figure}

To understand the emergence of the peak in the spectrum, it is useful to contrast the peak modes with modes that exit the horizon much earlier than the first turn in field space. Here there is a key consideration: the effective isocurvature mass is different for sub-Hubble and super-Hubble modes. In particular, the super-Hubble mass $\tilde{\mu}_s^2$ gains an additional term $4\omega^2$. The effective isocurvature mass is shown in Fig.~\ref{fig:muS}, where black and gray correspond to sub- and super-Hubble modes, respectively. One can appreciate from this that modes which are already super-Hubble at the onset of phase II will experience a brief period of superheavy mass: $\tilde{\mu}_s^2 / H^2$ spikes up to ${\cal O}(10)$.  This causes isocurvature modes that are already super-Hubble at that time to be suppressed, before the start of the phase II tachyonic instability. This behaviour is illustrated in \fref{fig:PBHA-perturbations-CMB} for modes which exit the Hubble radius at CMB scales. Similarly, modes that become super-Hubble in phase III, after the tachyonic instability, have predominantly heavy isocurvature mass, resulting in negligible growth (and then decay) of isocurvature perturbations.

Additionally, the peak modes---as opposed to modes on scales relevant to CMB measurements---only satisfy $k < aH$ during  
one turn in field space (the second one). This causes a positive transfer of power from the enhanced $\calS_k$ to the $\calR_k$ modes at the end of phase II, as was previously observed in \fref{fig:PBHA-perturbations}. On the other hand, CMB modes (and any $k$-mode that is super-Hubble at the onset of phase II) satisfy $k \ll aH$ during  
two opposing turns in field space. As can be appreciated from \fref{fig:PBHA-perturbations-CMB}, the first turn induces a positive transfer of power from $\calS_k$ to $\calR_k$, while the second one---being of opposite sign---induces a \textit{negative} transfer from $\calS_k$ to $\calR_k$. The curvature perturbations ${\cal R}_k$ on such scales are therefore decreased by the second turn, reducing them near exactly to their phase I values.

In sum, these effects lead to the characteristic peak structure of the curvature power spectrum, in which the enhanced modes 
become amplified by the transfer of power from the tachyonic growth of isocurvature modes identified in the previous section.

\subsection{Primordial Black Holes}
\label{sec:spectrum:PBH}

Primordial black holes are formed through the gravitational collapse of large overdensities after the end of inflation. For such overdensities to be produced, the inflationary curvature power spectrum needs to be enhanced by $\calO(10^6)$ with respect to its CMB scales while remaining within the perturbative regime. A power spectrum which exhibits a peak with $10^{-3}\leq\prk(k_\text{peak})\ll 1$ will therefore support PBH production \cite{Young:2019yug,Kehagias:2019eil,Escriva:2019phb,DeLuca:2020ioi,Musco:2020jjb,Escriva:2021aeh,Kehagias:2019eil,Escriva:2019phb,DeLuca:2020ioi,Musco:2020jjb,Escriva:2021aeh}. 

In the PBHspec model, the peak of the power spectrum occurs at the mode $k_{\rm peak}$ which exits the horizon shortly before the onset of phase II. This maps onto the typical mass of the produced primordial black holes (that is, the mass at the peak of the PBH mass distribution \cite{Klipfel:2025bvh}) in an identical manner to the single-field case, as
\begin{equation}
    \frac{M_{\rm PBH,f}}{30~M_{\odot}} \approx \left(\frac{\gamma}{0.2}\right)\left(\frac{g_{*}\left(T_{\rm f}\right)}{106.75}\right)^{-\frac{1}{6}}\left(\frac{k_{\rm peak}}{3.2 \times 10^{5}~\mathrm{Mpc}^{-1}}\right)^{-2}\,,
\end{equation}
where $M_{\rm PBH,f}$ is the PBH mass at the time of its formation ~\cite{Ozsoy:2023ryl}.
We take the efficiency factory $\gamma=0.2$ \cite{Carr:1975qj} and consider the effective number of degrees of freedom to be $g_*(T_{\rm f})=106.75$ at the time of formation; the solar mass is $M_\odot\approx 2.0\times10^{33}~{\rm g}$. 
From this, we find that the power spectrum in Fig.~\ref{fig:PBHA-PRk} yields primordial black holes with typical mass $5.7 \times 10^{18}~{\rm g}$, which is within the asteroid mass range, $10^{17}~{\rm g} \leq M_{\rm PBH} \leq 10^{23}~{\rm g}$ \cite{Khlopov:2008qy,Carr:2009jm,Sasaki:2018dmp,Carr:2020gox,Carr:2020xqk,Green:2020jor,Escriva:2021aeh,Villanueva-Domingo:2021spv,Escriva:2022duf,Gorton:2024cdm}.

\subsection{CMB Observables}
\label{sec:spectrum:CMB}

\begin{table*}[htb!]
    \centering
    \makebox[\textwidth][c]{
    \begin{tabular}{@{\extracolsep{10pt}}c|*{7}c|*{2}c}
        \hline\hline
        Model
        & $A_s$ & $n_s$ & $\alpha_s$ & $r$ & $\beta_{\rm iso}$ & $f_{\rm NL}^{\rm local}$ & $M_{\rm PBH}~
        [{\rm g}]$
        & $V_\text{S}/V_\text{PBH}$ & $m_\chi/H$
        \\ 
        \hline 
        SF Reference  
        & $2.09\times 10^{-9}$ & $0.9661$ & $-7.6\times 10^{-4}$ & $0.002$ & $-$ & $0.02$ & $5.0 \times 10^{18}$
        & $-$ & $-$
        \\
        SF Variation 
        & $2.10\times 10^{-9}$ & $0.9689$ & $-6.4\times 10^{-4}$ & $0.002$ & $-$ & $0.02$ & $-$
        & $-$ & $-$
        \\
        PBHspec 
        & $2.11\times 10^{-9}$ & $0.9670$ & $-6.9\times 10^{-4}$ & $0.003$ & $1.4\times 10^{-4}$ & $0.02$ & $5.7 \times 10^{18}$
        & $0.02$ & $0.14$
        \\
        \hline\hline
    \end{tabular}
    }
    \caption{Observables for the parameter sets considered for the PBH$A$ model (as defined in \tref{tab:PBHA-params}). CMB scales exit at $N_{\rm CMB} = 56$. The non-Gaussianities are mildly peaked in the local configuration, and $\beta_\text{iso}$ is taken at the scale $k=0.05~{\rm Mpc}^{-1}$ (i.e., $k_\text{mid}$ in Ref.~\cite{Planck:2018jri}). We evaluate the ``spectator-ness'' measures $V_\text{S}/V_\text{PBH}$ and $m_\chi/H$ at $N_{\rm CMB}$. Observables are computed using \texttt{PyTransport}~\cite{Dias:2016rjq,Mulryne:2016mzv}.
    }
    \label{tab:PBHA-obs}
\end{table*}

A remarkable feature of the PBHspec scenario is that it preserves the CMB predictions of the single-field model. The CMB observables for all the models considered in this work are summarized in \tref{tab:PBHA-obs}.

Modes that correspond to the comoving CMB pivot scale $k_{\rm CMB}=0.05~{\rm Mpc}^{-1}$ exit the Hubble radius at a time \cite{Dodelson:2003vq,Liddle:2003as}
\begin{align}\label{eq:Ncmb}
    N_{\rm CMB} &\simeq 62 + \frac{1}{4} {\rm ln} \left( \frac{ \rho_{\rm CMB}^2}{3 \mpl^6 H^2_{e} } \right) + \frac{1-3 w_{\rm reh}}{12(1+ w_{\rm reh})}   \ln\left(\frac{\rho_{\rm rad}}{\rho_{e}}\right) \nonumber \\
    &\simeq 56 \pm 5 \,
\end{align}
before the end of inflation. Here $\rho_{\rm CMB}$ is the energy density of the Universe at the time $N_{\rm CMB}$, while $\rho_{\rm rad}$ is the value at the onset of radiation domination~\cite{Amin:2014eta,Cook:2015vqa,Martin:2016oyk,Allahverdi:2020bys}; the quantities $H_e$ and $\rho_e$ are calculated at the end of inflation. The equation of state during reheating falls within the range $\omega_{\rm reh}\in \{-1/3,+1\}$: the result $N_{\rm CMB}=56$ in \eref{eq:Ncmb} assumes instantaneous reheating ($\omega_{\rm reh}=1/3$), and the range $\pm 5$ comes from the uncertainty in the duration of the phase of reheating \cite{Amin:2014eta}.

We parametrize the primordial power spectrum of curvature perturbations $\prk(k)$  on CMB scales via the conventional form \cite{Planck:2015sxf} 
\begin{equation}
    \prk(k)\equiv A_{\rm s} \left(\frac{k}{k_\text{CMB}}\right)^{n_{\rm s}-1+\frac{1}{2}\alpha_{\rm s}\ln\left(\frac{k}{k_\text{CMB}}\right)}\,,
\end{equation}
where $A_{\rm s}=\prk(k_\text{CMB})$ is the amplitude at the CMB pivot scale, $n_{\rm s}$ is the spectral index, and $\alpha_s$ is the running of the spectral index.  The latter are defined as
\begin{align}
    n_{\rm s}&\equiv 1+\left.\frac{\dd\,\ln\prk(k)}{\dd\,\ln k}\right|_{k=k_\text{CMB}}\,,\\
    \alpha_{\rm s}&\equiv\left.\frac{\dd\,n_{\rm s}}{\dd\,\ln k}\right|_{k=k_\text{CMB}}\,.
\end{align}
We also define the tensor-to-scalar ratio $r$ as the ratio between the amplitudes of tensor and scalar (curvature plus isocurvature) perturbations at the CMB pivot scale:
\begin{equation}
    r\equiv \left. \frac{\calP_t(k)}{\prk(k)+\calP_\calS(k)} \right|_{k=k_\text{CMB}}\,.
\end{equation}
The primordial isocurvature fraction is defined as~\cite{Planck:2018jri}
\begin{equation}
    \beta_\text{iso}(k)\equiv\frac{{\cal P}_{\cal S}(k)}{\prk(k)+{\cal P}_{\cal S}(k)}\,,
\end{equation}
which we evaluate at the CMB pivot scale, corresponding to the scale $k_\text{mid}$ in Ref.~\cite{Planck:2018jri}. Constraints on these quantities are summarized in~\tref{tab:CMB-constraints}.

\begin{table}[htb!]
    \centering
    \begin{tabular}{c|c}
        \hline\hline
        Parameter & Constraint
        \\ 
        \hline
        $A_{\rm s}~[\times10^{-9}]$ & $2.10\pm0.03$
        \\
        $n_{\rm s}$ & $0.9665\pm0.0038$
        \\
        $\alpha_{\rm s}$ & $-0.0045\pm0.0067$
        \\
        $r$ & $<0.037$
        \\
        $\beta_{\rm iso}$ & $<0.001$
        \\
        \hline\hline
    \end{tabular}
    \caption{Parameter constraints for CMB observables from Planck 2018 CMB data~\cite{Planck:2018vyg,Planck:2018jri,BICEP:2021xfz}. For $\beta_{\rm iso}$, we adopt the most conservative constraint on the isocurvature fraction at the CMB pivot scale from Table~14 of \cite{Planck:2018jri}.
    }
    \label{tab:CMB-constraints}
\end{table}

As discussed in the previous section, the CMB-scale modes do not experience a significant enhancement in the PBHspec scenario. This results in an overall preservation of the single-field predictions for the CMB.

The presence of the spectator field may rescale the amplitude $A_s$ by $\calO(0.1-1)$ with respect to its single-field values. This effect is due to a partial compensation between the turns into and out of phase II and is completely degenerate with the normalization of the potential. An $\calO(1)$ change in the $\vpbha$ parameter $V_0$ can thus keep the PBHspec model consistent with CMB constraints.

Crucially, nearby $k$-modes on CMB scales experience a near-identical change in amplitude, resulting in a negligible change to $n_s$. To demonstrate this explicitly, we show in \fref{fig:PBHA-perturbations-CMB} the evolution of two neighboring $k$-modes exiting the Hubble radius around CMB scales (two e-folds apart). One may see that the two $k$-modes experience identical evolution during phase II.
In particular, the small difference between the amplitudes of $\calR_k$ for the two modes is maintained to subpercent precision between early times ($N\sim 40$, long before the phase II mechanism has an impact on the dynamics) and the end of inflation ($N=0$, after the phase II dynamics). The same reasoning also explains the small change in $\alpha_s$.

The isocurvature fraction $\beta_{\rm iso}$ may also be read off from \fref{fig:PBHA-perturbations-CMB}, from the relative size of ${\cal P}_{\cal S}$ and ${\cal P}_{\cal R}$ at the end of inflation. From this we find $\beta_{\rm iso} = 1.5\times 10^{-4}$, well below the CMB constraint.

Meanwhile, the tensor modes are not impacted at all by the spectator dynamics. The spectrum of tensor modes is given by ${\cal P}_t =H^2 (N_{\rm CMB})/(8\pi^2 \mpl^2)$, from which we infer $r = 8 \pi^2 A_s/H^2 (N_{\rm CMB}) = 3\times 10^{-3}$, essentially unchanged from the single-field model.

\subsection{Non-Gaussianity}
\label{sec:spectrum:NG}

\begin{figure*}[htb!]
    \centering
    \includegraphics[width=\linewidth]{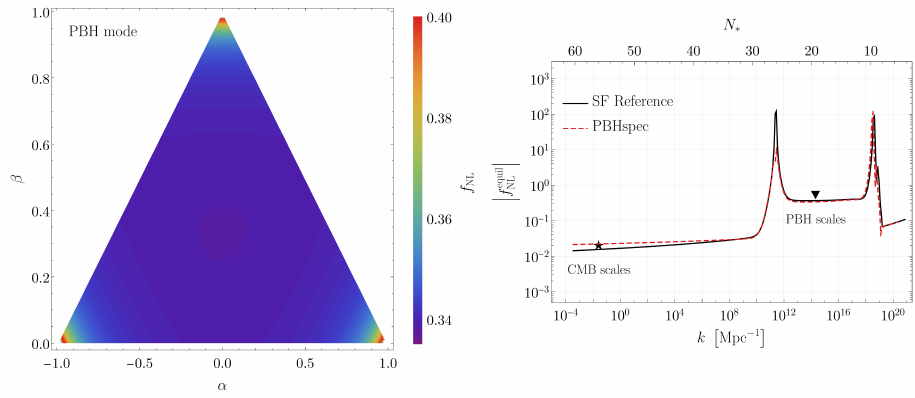}
    \caption{Reduced bispectrum of curvature perturbations, $\fnl$.
    \textit{Left panel:} Projection of $\fnl$ onto the $\alpha-\beta$ plane for a fixed $k$-mode exiting the Hubble radius at PBH scales ($k=k_{\rm peak}$), evaluated at the end of inflation. We find the non-Gaussianity is near constant across all shapes, being weakly peaked in the local configuration with $\fnl^{\rm local}=0.4$. The presence of significant non-Gaussianity at these scales may assist the formation of primordial black holes.
    \textit{Right-panel:} Amplitude of the equilateral configuration of the non-Gaussianity, $\fnl^{\rm equil}$, for modes that exit the Hubble radius at different times during inflation. The star and triangle indicate CMB and PBH scales, respectively. The single-field and PBHspec models exhibit small and similar non-Gaussianities at CMB scales, which are significantly enhanced on PBH scales.
    }
    \label{fig:PBHA-NG}
\end{figure*}

Finally, we examine the non-Gaussianity of the PBHspec model. That is, we investigate how the curvature perturbations produced in this setup depart from a Gaussian distribution, as probed by the three-point correlator,
\begin{equation}
    \langle \calR_{{\bf k}_1} \calR_{{\bf k}_2} \calR_{{\bf k}_3} \rangle \equiv (2\pi)^3 \delta^3({{\bf k}_1} + {{\bf k}_2} + {{\bf k}_3}) {B}_\calR(k_1,k_2,k_3)\,,
\end{equation}
which defines the bispectrum $B_\calR(k_1,k_2,k_3)$ for three $k$-modes which form a closed triangle in Fourier space, ${\bf k}_1+{\bf k}_2+{\bf k}_3=0$. It is useful to define the reduced bispectrum, or non-Gaussianity parameter \cite{Komatsu:2001rj},
\begin{equation}
    \fnl\equiv\frac{5}{6} \frac{B_\calR(k_1, k_2, k_3)}{P_\calR(k_1)P_\calR(k_2)+ P_\calR(k_2)P_\calR(k_3) + P_\calR(k_1)P_\calR(k_3)}
\end{equation}
where $P_\calR(k)=2\pi^2\prk(k)/k^3$ is the dimensionful power spectrum.

To determine the shape of the non-Gaussianity and in which configuration it is peaked, we adopt the $\alpha-\beta$ formalism of Refs.~\cite{Fergusson:2006pr,Rigopoulos:2004ba}. This defines the variables
\begin{equation}
    \alpha=\frac{k_2-k_3}{k}\,,\quad \beta=\frac{k-k_1}{k}\,,\quad\text{with}\>\>
    k=\frac{k_1+k_2+k_3}{2} ,
\end{equation}
where $0\leq\beta\leq1$ and $-(1-\beta)\leq\alpha\leq1-\beta$. As illustrated in \fref{fig:PBHA-NG}, the bispectrum can be projected onto the $\alpha-\beta$ plane, forming a triangle. The center of the triangle corresponds to the equilateral configuration ($k_1 = k_2 = k_3$), its corners represent the squeezed (or local) limit ($k_1\ll k_2 = k_3$), and the middle of the sides indicate the orthogonal configuration ($k_1=2k_2=2k_3$).

We first calculate the non-Gaussianity
for modes that cross the Hubble radius at CMB scales. The projection of $\fnl$ onto the $\alpha-\beta$ plane yields nearly equal values for each of the three typical configurations, with $f_{\rm NL}^{\rm local} \simeq f_{\rm NL}^{\rm equil} \simeq f_{\rm NL}^{\rm ortho} \simeq 0.02$. The local non-Gaussianity on CMB scales is very weakly peaked in the local configuration, with each component sitting well below the CMB bounds summarized in \tref{tab:NG-constraints}. 
Interestingly, these values are very similar to those in the single-field model. This can be understood by the fact that $k$-modes exiting around CMB scales are only modestly impacted by the tachyonic instability of the isocurvature modes, thus reproducing the single-field results.

\begin{table}[htb!]
    \centering
    \begin{tabular}{c|c}
        \hline\hline
        Parameter & Constraint
        \\ 
        \hline
        $f_{\rm NL}^{\rm local}$ & $-0.9 \pm 5.1$
        \\
        $f_{\rm NL}^{\rm equil}$ & $-26 \pm 47$
        \\
        $f_{\rm NL}^{\rm ortho}$ & $-38 \pm 24$
        \\
        \hline\hline
    \end{tabular}
    \caption{Parameter constraints for the non-Gaussianity parameter $\fnl$ from Planck 2018 CMB data~\cite{Planck:2019kim}.
    }
    \label{tab:NG-constraints}
\end{table}

\begin{figure*}[htb!]
    \centering
    \includegraphics[width=0.49\linewidth]{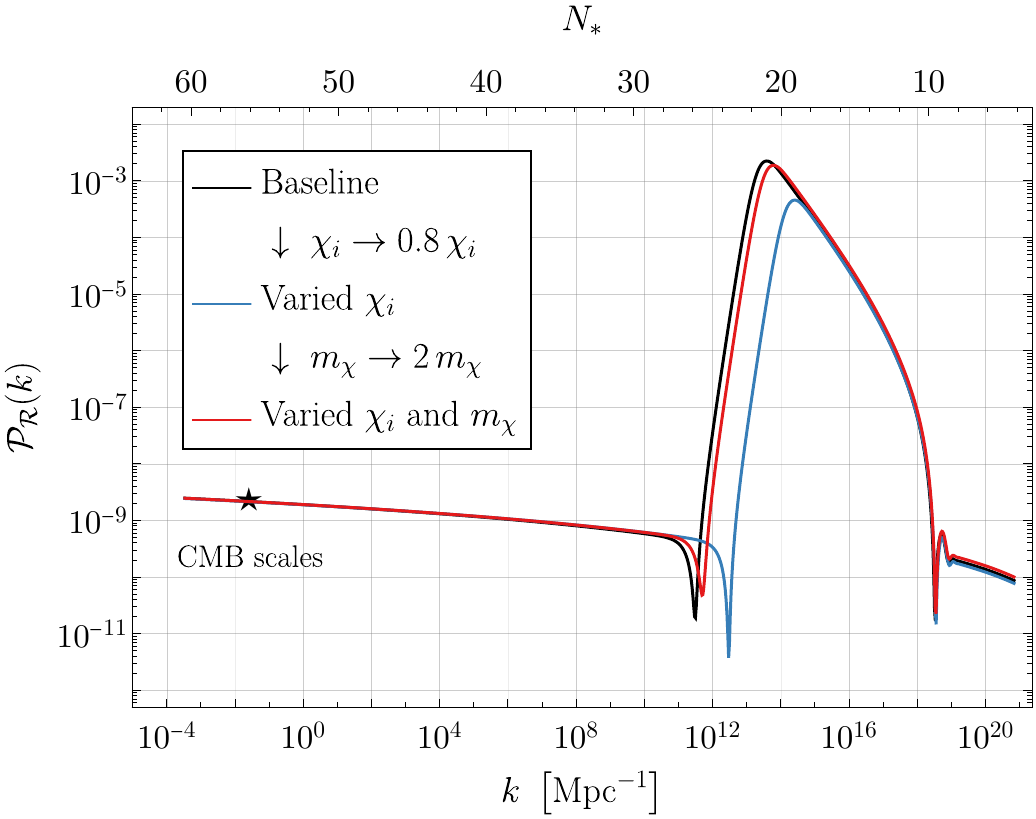}
    \includegraphics[width=0.49\linewidth]{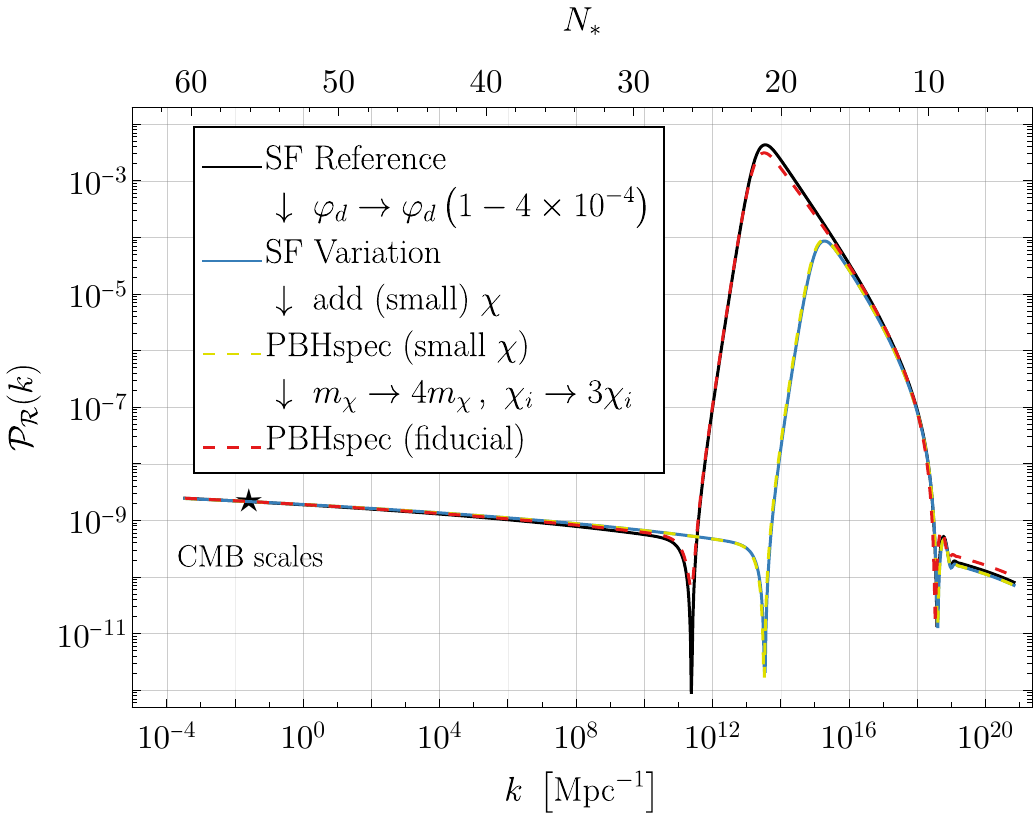}
    \caption{Parameter dependence of the curvature power spectrum for the PBHspec model. The base model PBH$A$ of Eq.~\eqref{eq:VPBH-A} is adopted for illustrative purposes.
    \textit{Left panel:} Dependence of the model on ${\cal O}(1)$ variations in the spectator parameters, producing $\calO(10)$ shifts in the amplitude of the peak of $\prk$ (black to blue and blue to red), and partial compensation between the shifts in $m_\chi$ and $\chi_i$ (black to red). The PBH parameters for all the curves are fixed to their fiducial values as reported in the PBHspec row of \tref{tab:PBHA-params}, with the ``baseline'' model having $m_\chi=3.5\times 10^{-7} \mpl$ and $\chi_i=4 \mpl$; the corresponding observables are consistent with those described in \tref{tab:PBHA-obs}.     
    \textit{Right panel:} Dependence of the single-field model on a $\calO(10^{-4})$ variation in the inflaton parameter $\varphi_d$, producing an $\calO(10)$ shift in the amplitude of the peak of $\prk$ (black to blue). Null effect of adding a small spectator field to the system (blue to yellow-dashed) and compensation of the previous shift by $\calO(1)$ variations in the spectator parameters (yellow-dashed to red-dashed). The model parameters used here are reported in \tref{tab:PBHA-params}.
    }
    \label{fig:parameter_compensations}
\end{figure*}

The left panel of \fref{fig:PBHA-NG} shows the $\alpha-\beta$ projection of the non-Gaussianity for modes exiting the Hubble radius at PBH scales. One may appreciate an enhancement of the non-Gaussianity for these modes, leading to $f_{\rm NL} \sim 0.4$. This is mildly peaked in the local configuration, and is similar to the single-field model. 

To understand the scale dependence of the non-Gaussianity, in the right panel of \fref{fig:PBHA-NG} we plot $\fnl^\text{equil}$ as a function of wavenumber $k$. From this, one may appreciate the dramatic enhancement of non-Gaussianity on small scales around the PBH scale. In this case, we also include the single-field PBH model for reference.

\section{Parameter Dependence}
\label{sec:parameters}

The spectator sector of the model of Eq.~\eqref{eq:action} depends only on two parameters: the field's mass, $m_\chi$, and its initial value, $\chi_i$. An ${\cal O}(1)$ variation in either parameter can produce $\sim{\cal O}(10)$ shifts in the peak amplitude of $\prk$, which is significant but modest in comparison to the sensitivity of the single-field model to variations in the parameters of $V_{\rm PBH}$. 
An example of this dependence is illustrated in the left panel of \fref{fig:parameter_compensations}: varying $\chi_i\rightarrow0.7\chi_i$ can decrease the peak of $\prk(k)$ by $\calO(10)$ (black to blue curve), and shifting $m_\chi\rightarrow2m_\chi$ can increase it by a similar amount (blue to red curve). Since these shifts compensate each other, one may also appreciate a partial degeneracy in $m_{\chi}$ and $\chi_i$. This can be physically understood in terms of their combined contribution to the Hubble parameter and to the slow-roll parameter $\epsilon_\chi$.

On the other hand, the inflaton sector of the model is highly dependent on parameter variations. It is well known that the issue of fine-tuning plagues single-field inflationary models that produce PBHs through a phase of USR \cite{Cole:2023wyx}. Indeed, predictions for observables in such models are exponentially sensitive to small variations of their model parameters: a delicate balance needs to be struck to produce an appropriate abundance of PBHs (of the correct mass) while also satisfying high-precision CMB constraints. This is shown in the right panel of \fref{fig:parameter_compensations}: a small variation (of ${\cal O}(10^{-4})$) in the PBH$A$ model parameter $\varphi_d$ shifts the peak amplitude of the curvature power spectrum by ${\cal O}(10)$, from $\prk\sim\text{few}\times 10^{-3}$ (black curve; consistent with PBH production) to $\prk\lesssim 10^{-4}$ (blue curve; too low to produce PBHs).

These two contrasting trends indicate that the spectator-plus-inflaton system retains a certain \textit{resilience} to small variations in the PBH model parameters.  
In the illustrative example of \fref{fig:parameter_compensations} (right panel), the above-mentioned ${\cal O}(10)$ shift in the peak amplitude of $\prk$ due to the small PBH parameter variation (black to blue curve) can be compensated by a ${\cal O}(1)$ variation in the spectator parameters (yellow-dashed to red-dashed curve). The compensated model (red) is consistent with the initial one (black) and compatible with PBH production in the asteroid-mass range as well as with all CMB constraints.

This significant resilience of the model to small parameter variations suggests that fine-tuning of the single-field model may be ameliorated.  To quantify this statement, we perform a Markov chain Monte Carlo analysis of the single-field and spectator models fit to CMB data under the prior of asteroid-mass PBH dark matter, and compute the Bayesian evidence. This will be presented elsewhere \cite{to-appear}.

\section{Model (In)dependence}
\label{sec:models}

\begin{figure}[htb!]
    \centering
    \includegraphics[width=\linewidth]{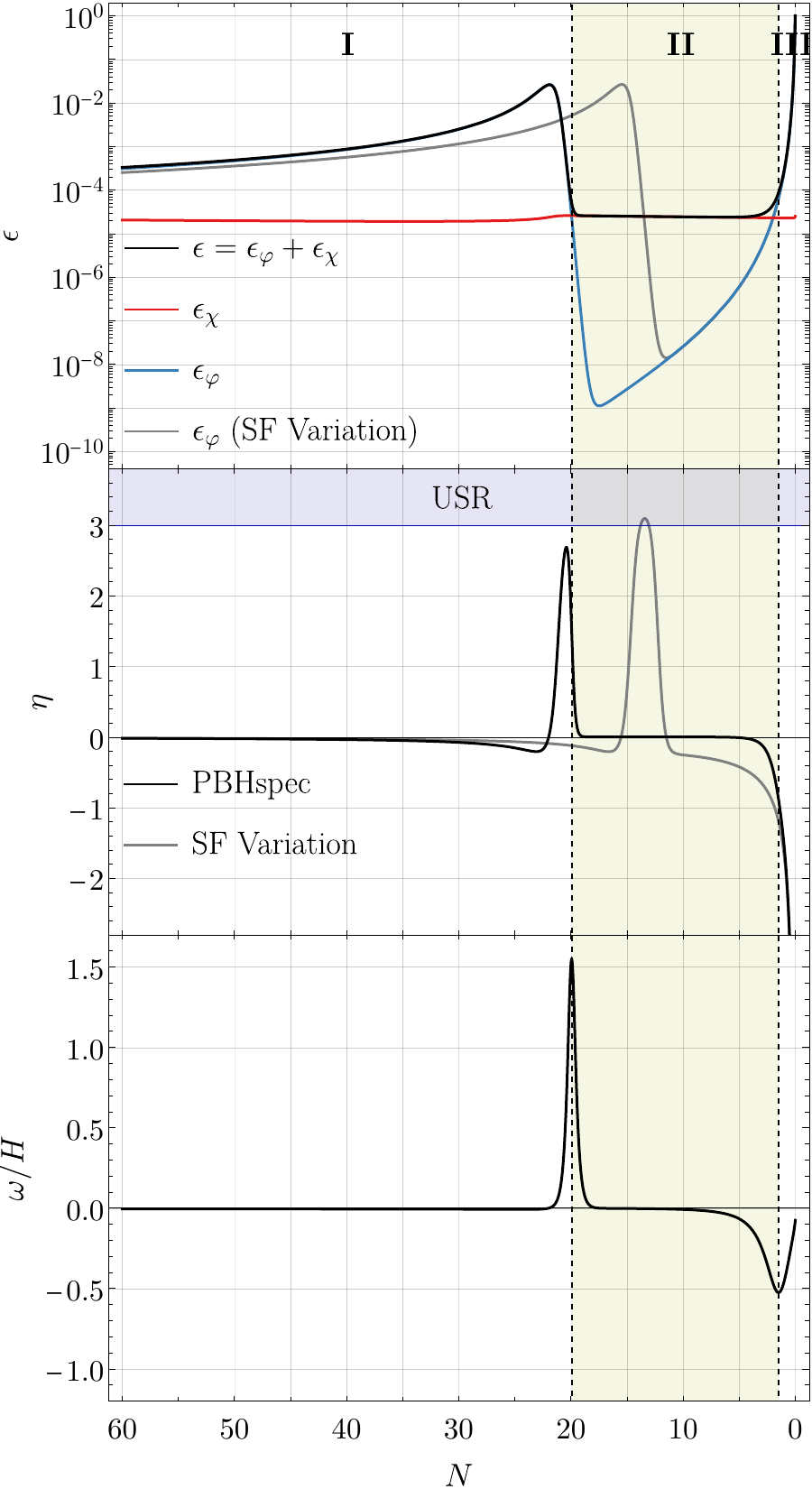}
    \caption{
    Evolution of $\epsilon$, $\eta$, and $\omega$ in the PBHspec scenario, for the $\vpbhb$ potential of \eref{eq:VPBH-B}. The parameters used in the figure are summarized in \tref{tab:PBHB-params}.
    }
    \label{fig:PBHB-background}
\end{figure}

\begin{figure}[htb!]
    \centering
    \includegraphics[width=\linewidth]{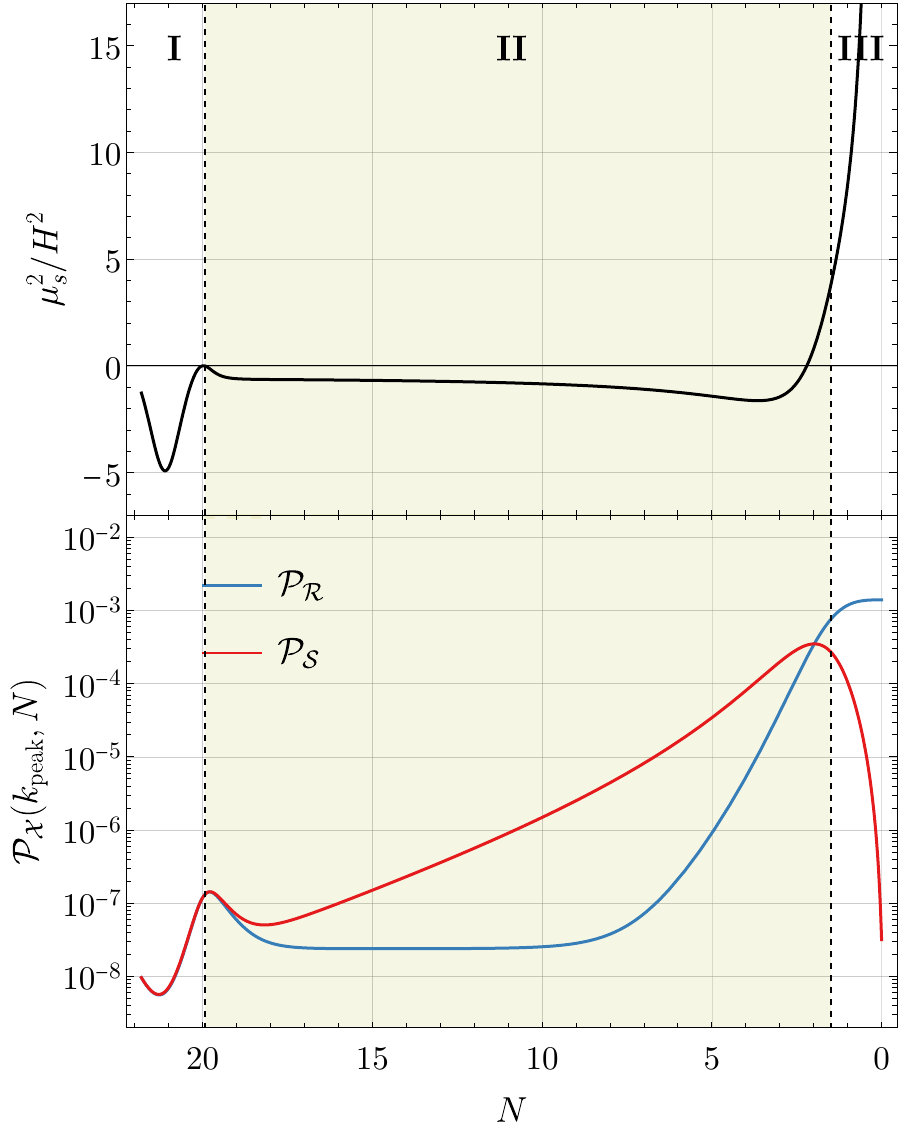}
    \caption{
    Evolution of $\mu_s^2$ and of the curvature and isocurvature perturbations $\calR$ and $\calS$, for the $\vpbhb$ potential of \eref{eq:VPBH-B}. The parameters used in the figure are summarized in \tref{tab:PBHB-params}.
    }
    \label{fig:PBHB-perturbations}
\end{figure}

In the previous sections, we took the inflaton potential to follow the particular form of \eref{eq:VPBH-A} for the sake of clarity. However, the mechanism we present is strongly model independent: any potential $\vpbh(\varphi)$ that produces PBHs through a transient phase of ultra-slow-roll (when treated as a single-field model) would be similarly affected by the presence of a spectator.
This is due to two reasons.

In any such model, a light spectator field can easily realize an $\epsilon_\chi$ larger than $\epsilon_\varphi$ when $\varphi$ slows down (causing $\epsilon_\varphi$ to become small). This will produce two opposite turns in field space, delimiting a phase wherein the field trajectory is aligned with the $\chi$ direction (our phase II).

Additionally, the presence of a local USR-inducing feature in $\vpbh$ requires $\pd_\varphi \vpbh \sim0$. In order for $\varphi$ to eventually escape the USR feature and terminate inflation, the potential must (briefly) obey $\pd_\varphi^2\vpbh<0$ during phase II. The latter condition {\it necessarily} causes $\mu_s^2$ to temporarily become negative when the field trajectory is aligned with the $\chi$ direction, sourcing a tachyonic growth of isocurvature perturbations. These will enhance the curvature modes when the second turn occurs, at the end of phase II.

To demonstrate this point, we calculate the evolution of the background fields and their perturbations for a second single-field model that can produce PBHs:
\begin{equation}\label{eq:VPBH-B}
    V_{{\rm PBH},B} (\varphi) = \frac{ M_{\rm Pl}^4 \lambda v^4}{12} \frac{ x^2 \left( 6 - 4 a x + 3 x^2 \right)}{\left( M_{\rm Pl}^2 + b x^2 \right)^2}\,, 
\end{equation}
with $x \equiv \varphi / v$. This is the model developed in Refs.~\cite{Garcia-Bellido:2017mdw,Germani:2017bcs}, in which a cubic term is added to a Higgs-like potential \cite{Bezrukov:2007ep} to engineer a local USR-inducing quasi-inflection point.

Note that, apart from exhibiting a local feature that slows down the evolution of $\varphi$, the potentials $\vpbha$ of Eq.~\eqref{eq:VPBH-A} and $\vpbhb$ of Eq.~\eqref{eq:VPBH-B} do not have much in common. They are in fact derived from different considerations and use different sets of parameters.

\begin{table*}[htb!]
    \centering
    \begin{tabular}{@{\extracolsep{10pt}}c|*{7}c}
        \hline\hline
        Model 
        & $\lambda$ & $v~[\mpl]$ 
        & $m_{\chi}~[\mpl]$ 
        & $\chi_{i}~[\mpl]$
        \\ 
        \hline 
        SF Reference  
        & $1.16\times 10^{-6}$ & $0.19669$ 
        & $-$ & $-$ 
        \\
        SF Variation
        & $9.10\times 10^{-7}$ &
        $0.19669\times (1-4\times10^{-3})$
        & $-$ & $-$ 
        \\
        PBHspec  
        & $1.19\times 10^{-6}$ &
        $0.19669\times (1-4\times10^{-3})$
        & $7\times10^{-7}$ & $2.5$ 
        \\
        \hline\hline
    \end{tabular}
    \caption{Parameter values for the PBH$B$ models, with $a=0.719527$, $b=1.500016$, and $\varphi_i=3 \mpl$ fixed throughout (we take $\dot{\varphi}_i = \dot{\chi}_i = 0$).
    The single-field ``Reference'' model reproduces the results introduced in Ref.~\cite{Garcia-Bellido:2017mdw}, modified to align with the latest CMB constraints and produce PBHs in the asteroid-mass range. The ``Variation'' model shifts the parameter $v\rightarrow v_{\text{ref}}\times(1-4\times10^{-3})$; this excludes PBH production. The ``PBHspec'' model includes the effect of a spectator field to the ``Variation'' $\vpbhb$ parameters; it is again consistent with PBH production in the asteroid-mass range. 
    }
    \label{tab:PBHB-params}
\end{table*}

\begin{table*}[htb!]
    \centering
    \makebox[\textwidth][c]{
    \begin{tabular}{@{\extracolsep{10pt}}c|*{7}c|*{2}c}
        \hline\hline
        Model 
        & $A_s$ & $n_s$ & $\alpha_s$ & $r$ & $\beta_{\rm iso}$ & $f_{\rm NL}^{\rm local}$ & $M_{\rm PBH}~[{\rm g}]$
        & $R_V$ & $m_\chi/H$
        \\ 
        \hline 
        Reference 
        & $2.10\times 10^{-9}$ & $0.9599$ & $-1.8\times 10^{-3}$ & $0.005$ & $-$ & $0.02$ & $1.1 \times 10^{17}$
        & $-$ & $-$
        \\
        Variation
        & $2.10\times 10^{-9}$ & $0.9665$ & $-8.2\times 10^{-4}$ & $0.004$ & $-$ & $0.01$ & $-$ 
        & $-$ & $-$
        \\
        PBHspec
        & $2.09\times 10^{-9}$ & $0.9640$ & $-1.0\times 10^{-3}$ & $0.005$ & $2.9\times10^{-4}$ & $0.02$ & $4.0 \times 10^{16}$
        & $0.008$ & $0.10$
        \\
        \hline\hline
    \end{tabular}
    }
    \caption{Observables for the parameter sets considered for the PBH$B$ model (as defined in \tref{tab:PBHB-params}). CMB scales exit at $N_{\rm CMB} = 56$. The non-Gaussianities are mildly peaked in the local configuration, and $\beta_\text{iso}$ is taken at the scale $k=0.05~\text{Mpc}^{-1}$ (i.e., $k_\text{mid}$ in Ref~\cite{Planck:2018jri}). We evaluate the spectatorness measures $V_\text{S}/V_\text{PBH}$ and $m_\chi/H$ at $N_\text{CMB}$. Observables are computed using \texttt{PyTransport}~\cite{Dias:2016rjq,Mulryne:2016mzv}.
    }
    \label{tab:PBHB-obs}
\end{table*}

The evolution of the background quantities $\epsilon$, $\eta$, and $\omega$ for the potential $\vpbhb$ of \eref{eq:VPBH-B} is shown in \fref{fig:PBHB-background}. The same behavior previously encountered in \fref{fig:PBHA-background-eps-eta-omega} can be found here: the presence of the spectator field $\chi$ sets a floor on $\epsilon$ and causes two turns in field space while eliminating the phase of ultra-slow-roll.

\begin{figure}[tb!]
    \centering
    \includegraphics[width=\linewidth]{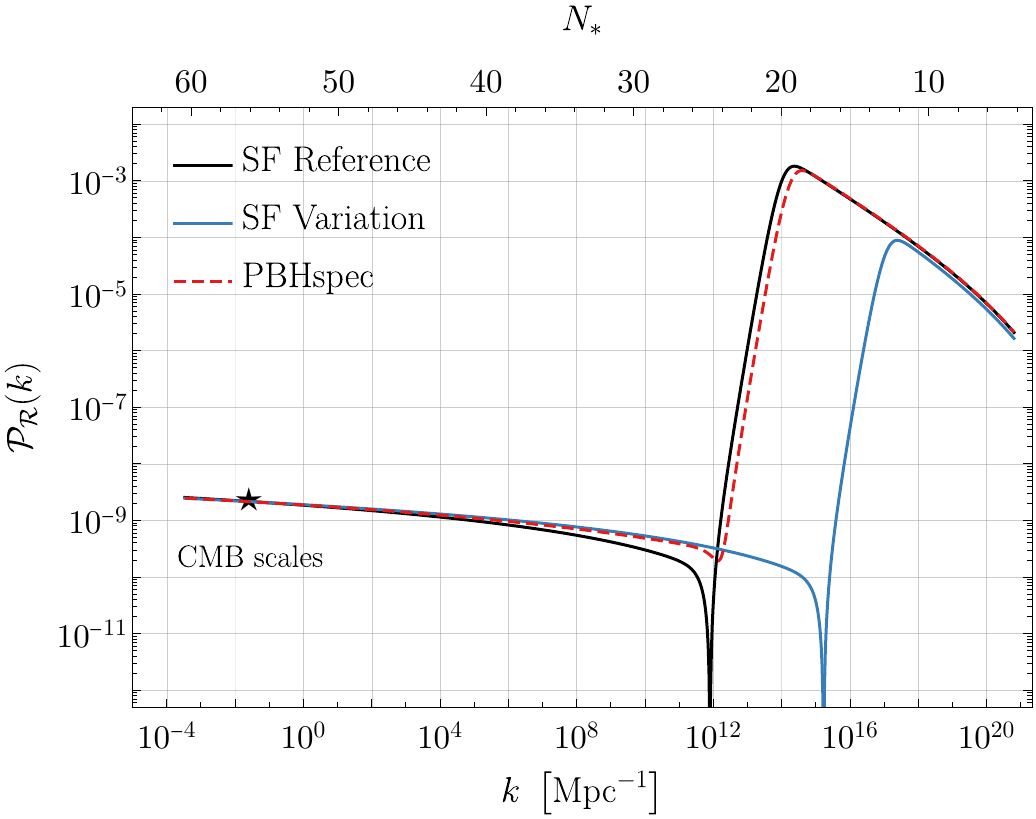}
    \caption{
    Curvature power spectrum for the $\vpbhb$ potential of \eref{eq:VPBH-B}. The parameters used in the figure are summarized in \tref{tab:PBHB-params}.
    }
    \label{fig:PBHB-PRk}
\end{figure}

The corresponding evolution of the curvature and isocurvature perturbations is illustrated in \fref{fig:PBHB-perturbations}. Here again, we find a tachyonic instability of $\mu_s^2$ during phase II: this sources a near-exponential growth of modes $\calS_k$, which then transfers power to modes $\calR_k$ at the time of the second turn in field space. The curvature power spectrum is thus enhanced, as shown in \fref{fig:PBHB-PRk}. See also \tref{tab:PBHB-obs}.

One may notice that phase II lasts longer in the PBH$B$ model than in PBH$A$, while phase III is shortened. We can expect additional modes to cross the Hubble radius during phase II and experience the tachyonic enhancement, while only a few will exit during phase III. These considerations will affect the shape of the peak of the curvature power spectrum (which is broadened in PBH$B$) but will nevertheless be in accordance with the single-field case, since the phase of ultra-slow-roll enhancement of modes $\calR_k$ will also be extended. The striking similarity between the single-field and PBHspec scenarios for $\vpbhb$ can be observed in \fref{fig:PBHB-PRk}.

\section{Discussion}
\label{sec:discussion}

Scalar fields are ubiquitous in models of high-energy physics. In the context of inflationary cosmology, scalar fields in addition to the inflaton can play the role of a {\it spectator field}, which does not couple to the inflaton directly and which contributes a subdominant fraction of the total energy density of the Universe. The phenomenology of inflation in the presence of spectator fields has been extensively studied (cf. Refs.~\cite{Chen:2018uul,Dimastrogiovanni:2016fuu,McDonough:2018xzh,Holland:2020jdh,Alexander:2018fjp}). In this work, we have extended this body of work to include spectator fields in single-field inflation models which seed primordial black holes through a phase of ultra-slow-roll (USR) evolution.

While USR models of PBH production are premised on an exponential slowdown of the inflaton and hence a sharp decrease in the slow-roll parameter $\epsilon$, a spectator field maintains a near-constant velocity at all times. This creates a floor for the combined total $\epsilon$ below which the system cannot cross, and prevents the system from entering into USR. One might naively expect this to preclude the enhancement of perturbations for which these single-field models were engineered. In this work and the companion paper~\cite{Lorenzoni:2025gni}, we have demonstrated that the enhancement survives, and moreover the spectator alleviates much of the fine-tuning of parameters required of the underlying single-field model.

The spectator qualitatively changes the dynamics of both the background and perturbations: the would-be USR phase is supplanted by a phase wherein the inflationary trajectory turns to flow along the spectator direction. Despite the spectator providing a subdominant contribution to the energy density, it briefly dominates the kinematics. This phase is characterized by tachyonic instability for isocurvature perturbations and bracketed by turns in field space, which convert power from isocurvature perturbations to curvature perturbations, following which the isocurvature modes decay. The net result is an enhancement of the amplitude of curvature perturbations that parallels that in the single-field case.

The observables of the model are in remarkable agreement with the underlying single-field models. The power spectrum inherits the same predictions for CMB observables, the same slope of the peak of the power spectrum of PBH-relevant scales, and the same decay of the peak. As we have explained in detail in Secs.~\ref{sec:spectrum} and~\ref{sec:models}, this commonality arises from the shared evolution of inflation fluctuations in the two models. 

Where the two models differ is the extent to which the feature of the inflaton potential must be set by hand: the spectator model can accommodate a much less fine-tuned feature in the inflaton potential, by compensating ${\cal O}(1)$ parameter variations of the spectator mass and initial condition. Whether this amounts  to a de-fine-tuning in a Bayesian sense requires a full statistical analysis, which is topic of future work \cite{to-appear}.

These results are illustrated in the context of a few specific single-field models---those of Refs.~\cite{Mishra:2019pzq,Garcia-Bellido:2017mdw,Germani:2017bcs}---but apply broadly to any model of single-field inflation model that is minimally coupled to gravity and which realizes USR through a feature in the potential. We have demonstrated this by undertaking our analysis for two distinct and unrelated models, finding the same results: the spectator maintains or enhances the perturbations relative to the single-field models.

There are many directions for future work. Foremost among these is a detailed exploration of parameter space to quantify the fine-tuning in the spectator versus single-field models. This exploration will enable study of other parameter regimes, such as ones in which the spectator field generates a second phase of slow-roll inflation. It will also be interesting to generalize the spectator model to other spectator potentials, to multiple spectators, or to a spectator nonminimally coupled to gravity. We leave this and other interesting directions to future work.

\vspace{0.5cm}
\section*{Acknowledgements}

The authors thank Josu Aurrekoetxea, Matteo Braglia, Chris Byrnes, Bryce Cyr, Alexandra Klipfel, J\'{e}r\^{o}me Martin, Sebastien Renaux-Petel, and Vincent Vennin for helpful discussions. 
Portions of this research were conducted in MIT's Center for Theoretical Physics---a Leinweber Institute and supported in part by the U.S.~Department of Energy under Contract No.~DE-SC0012567. 
E.M. is supported in part by a Discovery Grant from the Natural Sciences and Engineering Research Council of Canada and by a New Investigator Operating Grant from Research Manitoba. 
S.R.G. is supported by the National Science Foundation Mathematical and Physical Sciences Ascending postdoctoral fellowship, under Award No.~2317018. 
D.L.L. is supported in part by a Ph.D. Research Studentship from Research Manitoba.

\vskip -15pt
\bibliography{refs-spectators}

\begin{thebibliography}{128}%
\makeatletter
\providecommand \@ifxundefined [1]{%
 \@ifx{#1\undefined}
}%
\providecommand \@ifnum [1]{%
 \ifnum #1\expandafter \@firstoftwo
 \else \expandafter \@secondoftwo
 \fi
}%
\providecommand \@ifx [1]{%
 \ifx #1\expandafter \@firstoftwo
 \else \expandafter \@secondoftwo
 \fi
}%
\providecommand \natexlab [1]{#1}%
\providecommand \enquote  [1]{``#1''}%
\providecommand \bibnamefont  [1]{#1}%
\providecommand \bibfnamefont [1]{#1}%
\providecommand \citenamefont [1]{#1}%
\providecommand \href@noop [0]{\@secondoftwo}%
\providecommand \href [0]{\begingroup \@sanitize@url \@href}%
\providecommand \@href[1]{\@@startlink{#1}\@@href}%
\providecommand \@@href[1]{\endgroup#1\@@endlink}%
\providecommand \@sanitize@url [0]{\catcode `\\12\catcode `\$12\catcode `\&12\catcode `\#12\catcode `\^12\catcode `\_12\catcode `\%12\relax}%
\providecommand \@@startlink[1]{}%
\providecommand \@@endlink[0]{}%
\providecommand \url  [0]{\begingroup\@sanitize@url \@url }%
\providecommand \@url [1]{\endgroup\@href {#1}{\urlprefix }}%
\providecommand \urlprefix  [0]{URL }%
\providecommand \Eprint [0]{\href }%
\providecommand \doibase [0]{http://dx.doi.org/}%
\providecommand \selectlanguage [0]{\@gobble}%
\providecommand \bibinfo  [0]{\@secondoftwo}%
\providecommand \bibfield  [0]{\@secondoftwo}%
\providecommand \translation [1]{[#1]}%
\providecommand \BibitemOpen [0]{}%
\providecommand \bibitemStop [0]{}%
\providecommand \bibitemNoStop [0]{.\EOS\space}%
\providecommand \EOS [0]{\spacefactor3000\relax}%
\providecommand \BibitemShut  [1]{\csname bibitem#1\endcsname}%
\let\auto@bib@innerbib\@empty
\bibitem [{\citenamefont {Zel'dovich}(1967)}]{Zeldovich:1967lct}%
  \BibitemOpen
  \bibfield  {author} {\bibinfo {author} {\bibfnamefont {I.~D.}\ \bibnamefont {Zel'dovich}, \bibfnamefont {Ya.B.;~Novikov}},\ }\bibfield  {title} {\enquote {\bibinfo {title} {{The Hypothesis of Cores Retarded during Expansion and the Hot Cosmological Model}},}\ }\href@noop {} {\bibfield  {journal} {\bibinfo  {journal} {Soviet Astron. AJ (Engl. Transl. ),}\ }\textbf {\bibinfo {volume} {10}},\ \bibinfo {pages} {602} (\bibinfo {year} {1967})}\BibitemShut {NoStop}%
\bibitem [{\citenamefont {Hawking}(1971)}]{Hawking:1971ei}%
  \BibitemOpen
  \bibfield  {author} {\bibinfo {author} {\bibfnamefont {Stephen}\ \bibnamefont {Hawking}},\ }\bibfield  {title} {\enquote {\bibinfo {title} {{Gravitationally collapsed objects of very low mass}},}\ }\href@noop {} {\bibfield  {journal} {\bibinfo  {journal} {Mon. Not. Roy. Astron. Soc.}\ }\textbf {\bibinfo {volume} {152}},\ \bibinfo {pages} {75} (\bibinfo {year} {1971})}\BibitemShut {NoStop}%
\bibitem [{\citenamefont {Carr}\ and\ \citenamefont {Hawking}(1974)}]{Carr:1974nx}%
  \BibitemOpen
  \bibfield  {author} {\bibinfo {author} {\bibfnamefont {Bernard~J.}\ \bibnamefont {Carr}}\ and\ \bibinfo {author} {\bibfnamefont {S.~W.}\ \bibnamefont {Hawking}},\ }\bibfield  {title} {\enquote {\bibinfo {title} {{Black holes in the early Universe}},}\ }\href {\doibase 10.1093/mnras/168.2.399} {\bibfield  {journal} {\bibinfo  {journal} {Mon. Not. Roy. Astron. Soc.}\ }\textbf {\bibinfo {volume} {168}},\ \bibinfo {pages} {399--415} (\bibinfo {year} {1974})}\BibitemShut {NoStop}%
\bibitem [{\citenamefont {Meszaros}(1974)}]{Meszaros:1974tb}%
  \BibitemOpen
  \bibfield  {author} {\bibinfo {author} {\bibfnamefont {P.}~\bibnamefont {Meszaros}},\ }\bibfield  {title} {\enquote {\bibinfo {title} {{The behaviour of point masses in an expanding cosmological substratum}},}\ }\href@noop {} {\bibfield  {journal} {\bibinfo  {journal} {Astron. Astrophys.}\ }\textbf {\bibinfo {volume} {37}},\ \bibinfo {pages} {225--228} (\bibinfo {year} {1974})}\BibitemShut {NoStop}%
\bibitem [{\citenamefont {Carr}(1975)}]{Carr:1975qj}%
  \BibitemOpen
  \bibfield  {author} {\bibinfo {author} {\bibfnamefont {Bernard~J.}\ \bibnamefont {Carr}},\ }\bibfield  {title} {\enquote {\bibinfo {title} {{The Primordial black hole mass spectrum}},}\ }\href {\doibase 10.1086/153853} {\bibfield  {journal} {\bibinfo  {journal} {Astrophys. J.}\ }\textbf {\bibinfo {volume} {201}},\ \bibinfo {pages} {1--19} (\bibinfo {year} {1975})}\BibitemShut {NoStop}%
\bibitem [{\citenamefont {Khlopov}\ \emph {et~al.}(1985)\citenamefont {Khlopov}, \citenamefont {Malomed},\ and\ \citenamefont {Zeldovich}}]{Khlopov:1985jw}%
  \BibitemOpen
  \bibfield  {author} {\bibinfo {author} {\bibfnamefont {M.}~\bibnamefont {Khlopov}}, \bibinfo {author} {\bibfnamefont {B.~A.}\ \bibnamefont {Malomed}}, \ and\ \bibinfo {author} {\bibfnamefont {Ia.~B.}\ \bibnamefont {Zeldovich}},\ }\bibfield  {title} {\enquote {\bibinfo {title} {{Gravitational instability of scalar fields and formation of primordial black holes}},}\ }\href@noop {} {\bibfield  {journal} {\bibinfo  {journal} {Mon. Not. Roy. Astron. Soc.}\ }\textbf {\bibinfo {volume} {215}},\ \bibinfo {pages} {575--589} (\bibinfo {year} {1985})}\BibitemShut {NoStop}%
\bibitem [{\citenamefont {Niemeyer}\ and\ \citenamefont {Jedamzik}(1999)}]{Niemeyer:1999ak}%
  \BibitemOpen
  \bibfield  {author} {\bibinfo {author} {\bibfnamefont {Jens~C.}\ \bibnamefont {Niemeyer}}\ and\ \bibinfo {author} {\bibfnamefont {K.}~\bibnamefont {Jedamzik}},\ }\bibfield  {title} {\enquote {\bibinfo {title} {{Dynamics of primordial black hole formation}},}\ }\href {\doibase 10.1103/PhysRevD.59.124013} {\bibfield  {journal} {\bibinfo  {journal} {Phys. Rev. D}\ }\textbf {\bibinfo {volume} {59}},\ \bibinfo {pages} {124013} (\bibinfo {year} {1999})},\ \Eprint {http://arxiv.org/abs/astro-ph/9901292} {arXiv:astro-ph/9901292} \BibitemShut {NoStop}%
\bibitem [{\citenamefont {Khlopov}(2010)}]{Khlopov:2008qy}%
  \BibitemOpen
  \bibfield  {author} {\bibinfo {author} {\bibfnamefont {Maxim~Yu.}\ \bibnamefont {Khlopov}},\ }\bibfield  {title} {\enquote {\bibinfo {title} {{Primordial Black Holes}},}\ }\href {\doibase 10.1088/1674-4527/10/6/001} {\bibfield  {journal} {\bibinfo  {journal} {Res. Astron. Astrophys.}\ }\textbf {\bibinfo {volume} {10}},\ \bibinfo {pages} {495--528} (\bibinfo {year} {2010})},\ \Eprint {http://arxiv.org/abs/0801.0116} {arXiv:0801.0116 [astro-ph]} \BibitemShut {NoStop}%
\bibitem [{\citenamefont {Carr}\ \emph {et~al.}(2010)\citenamefont {Carr}, \citenamefont {Kohri}, \citenamefont {Sendouda},\ and\ \citenamefont {Yokoyama}}]{Carr:2009jm}%
  \BibitemOpen
  \bibfield  {author} {\bibinfo {author} {\bibfnamefont {B.~J.}\ \bibnamefont {Carr}}, \bibinfo {author} {\bibfnamefont {Kazunori}\ \bibnamefont {Kohri}}, \bibinfo {author} {\bibfnamefont {Yuuiti}\ \bibnamefont {Sendouda}}, \ and\ \bibinfo {author} {\bibfnamefont {Jun'ichi}\ \bibnamefont {Yokoyama}},\ }\bibfield  {title} {\enquote {\bibinfo {title} {{New cosmological constraints on primordial black holes}},}\ }\href {\doibase 10.1103/PhysRevD.81.104019} {\bibfield  {journal} {\bibinfo  {journal} {Phys. Rev. D}\ }\textbf {\bibinfo {volume} {81}},\ \bibinfo {pages} {104019} (\bibinfo {year} {2010})},\ \Eprint {http://arxiv.org/abs/0912.5297} {arXiv:0912.5297 [astro-ph.CO]} \BibitemShut {NoStop}%
\bibitem [{\citenamefont {Sasaki}\ \emph {et~al.}(2018)\citenamefont {Sasaki}, \citenamefont {Suyama}, \citenamefont {Tanaka},\ and\ \citenamefont {Yokoyama}}]{Sasaki:2018dmp}%
  \BibitemOpen
  \bibfield  {author} {\bibinfo {author} {\bibfnamefont {Misao}\ \bibnamefont {Sasaki}}, \bibinfo {author} {\bibfnamefont {Teruaki}\ \bibnamefont {Suyama}}, \bibinfo {author} {\bibfnamefont {Takahiro}\ \bibnamefont {Tanaka}}, \ and\ \bibinfo {author} {\bibfnamefont {Shuichiro}\ \bibnamefont {Yokoyama}},\ }\bibfield  {title} {\enquote {\bibinfo {title} {{Primordial black holes\textemdash{}perspectives in gravitational wave astronomy}},}\ }\href {\doibase 10.1088/1361-6382/aaa7b4} {\bibfield  {journal} {\bibinfo  {journal} {Class. Quant. Grav.}\ }\textbf {\bibinfo {volume} {35}},\ \bibinfo {pages} {063001} (\bibinfo {year} {2018})},\ \Eprint {http://arxiv.org/abs/1801.05235} {arXiv:1801.05235 [astro-ph.CO]} \BibitemShut {NoStop}%
\bibitem [{\citenamefont {Carr}\ \emph {et~al.}(2021)\citenamefont {Carr}, \citenamefont {Kohri}, \citenamefont {Sendouda},\ and\ \citenamefont {Yokoyama}}]{Carr:2020gox}%
  \BibitemOpen
  \bibfield  {author} {\bibinfo {author} {\bibfnamefont {Bernard}\ \bibnamefont {Carr}}, \bibinfo {author} {\bibfnamefont {Kazunori}\ \bibnamefont {Kohri}}, \bibinfo {author} {\bibfnamefont {Yuuiti}\ \bibnamefont {Sendouda}}, \ and\ \bibinfo {author} {\bibfnamefont {Jun'ichi}\ \bibnamefont {Yokoyama}},\ }\bibfield  {title} {\enquote {\bibinfo {title} {{Constraints on primordial black holes}},}\ }\href {\doibase 10.1088/1361-6633/ac1e31} {\bibfield  {journal} {\bibinfo  {journal} {Rept. Prog. Phys.}\ }\textbf {\bibinfo {volume} {84}},\ \bibinfo {pages} {116902} (\bibinfo {year} {2021})},\ \Eprint {http://arxiv.org/abs/2002.12778} {arXiv:2002.12778 [astro-ph.CO]} \BibitemShut {NoStop}%
\bibitem [{\citenamefont {{Carr}}\ and\ \citenamefont {{K\"{u}hnel}}(2020)}]{Carr:2020xqk}%
  \BibitemOpen
  \bibfield  {author} {\bibinfo {author} {\bibfnamefont {Bernard}\ \bibnamefont {{Carr}}}\ and\ \bibinfo {author} {\bibfnamefont {Florian}\ \bibnamefont {{K\"{u}hnel}}},\ }\bibfield  {title} {\enquote {\bibinfo {title} {{Primordial Black Holes as Dark Matter: Recent Developments}},}\ }\href {\doibase 10.1146/annurev-nucl-050520-125911} {\bibfield  {journal} {\bibinfo  {journal} {Ann. Rev. Nucl. Part. Sci.}\ }\textbf {\bibinfo {volume} {70}},\ \bibinfo {pages} {355--394} (\bibinfo {year} {2020})},\ \Eprint {http://arxiv.org/abs/2006.02838} {arXiv:2006.02838 [astro-ph.CO]} \BibitemShut {NoStop}%
\bibitem [{\citenamefont {Green}\ and\ \citenamefont {Kavanagh}(2021)}]{Green:2020jor}%
  \BibitemOpen
  \bibfield  {author} {\bibinfo {author} {\bibfnamefont {Anne~M.}\ \bibnamefont {Green}}\ and\ \bibinfo {author} {\bibfnamefont {Bradley~J.}\ \bibnamefont {Kavanagh}},\ }\bibfield  {title} {\enquote {\bibinfo {title} {{Primordial Black Holes as a dark matter candidate}},}\ }\href {\doibase 10.1088/1361-6471/abc534} {\bibfield  {journal} {\bibinfo  {journal} {J. Phys. G}\ }\textbf {\bibinfo {volume} {48}},\ \bibinfo {pages} {043001} (\bibinfo {year} {2021})},\ \Eprint {http://arxiv.org/abs/2007.10722} {arXiv:2007.10722 [astro-ph.CO]} \BibitemShut {NoStop}%
\bibitem [{\citenamefont {Escriv\`a}(2022)}]{Escriva:2021aeh}%
  \BibitemOpen
  \bibfield  {author} {\bibinfo {author} {\bibfnamefont {Albert}\ \bibnamefont {Escriv\`a}},\ }\bibfield  {title} {\enquote {\bibinfo {title} {{PBH Formation from Spherically Symmetric Hydrodynamical Perturbations: A Review}},}\ }\href {\doibase 10.3390/universe8020066} {\bibfield  {journal} {\bibinfo  {journal} {Universe}\ }\textbf {\bibinfo {volume} {8}},\ \bibinfo {pages} {66} (\bibinfo {year} {2022})},\ \Eprint {http://arxiv.org/abs/2111.12693} {arXiv:2111.12693 [gr-qc]} \BibitemShut {NoStop}%
\bibitem [{\citenamefont {Villanueva-Domingo}\ \emph {et~al.}(2021)\citenamefont {Villanueva-Domingo}, \citenamefont {Mena},\ and\ \citenamefont {Palomares-Ruiz}}]{Villanueva-Domingo:2021spv}%
  \BibitemOpen
  \bibfield  {author} {\bibinfo {author} {\bibfnamefont {Pablo}\ \bibnamefont {Villanueva-Domingo}}, \bibinfo {author} {\bibfnamefont {Olga}\ \bibnamefont {Mena}}, \ and\ \bibinfo {author} {\bibfnamefont {Sergio}\ \bibnamefont {Palomares-Ruiz}},\ }\bibfield  {title} {\enquote {\bibinfo {title} {{A brief review on primordial black holes as dark matter}},}\ }\href {\doibase 10.3389/fspas.2021.681084} {\bibfield  {journal} {\bibinfo  {journal} {Front. Astron. Space Sci.}\ }\textbf {\bibinfo {volume} {8}},\ \bibinfo {pages} {87} (\bibinfo {year} {2021})},\ \Eprint {http://arxiv.org/abs/2103.12087} {arXiv:2103.12087 [astro-ph.CO]} \BibitemShut {NoStop}%
\bibitem [{\citenamefont {Escriv{\`a}}\ \emph {et~al.}(2022)\citenamefont {Escriv{\`a}}, \citenamefont {Kuhnel},\ and\ \citenamefont {Tada}}]{Escriva:2022duf}%
  \BibitemOpen
  \bibfield  {author} {\bibinfo {author} {\bibfnamefont {Albert}\ \bibnamefont {Escriv{\`a}}}, \bibinfo {author} {\bibfnamefont {Florian}\ \bibnamefont {Kuhnel}}, \ and\ \bibinfo {author} {\bibfnamefont {Yuichiro}\ \bibnamefont {Tada}},\ }\bibfield  {title} {\enquote {\bibinfo {title} {{Primordial Black Holes}},}\ \ }(\bibinfo {year} {2022})\ \Eprint {http://arxiv.org/abs/2211.05767} {arXiv:2211.05767 [astro-ph.CO]} \BibitemShut {NoStop}%
\bibitem [{\citenamefont {Gorton}\ and\ \citenamefont {Green}(2024)}]{Gorton:2024cdm}%
  \BibitemOpen
  \bibfield  {author} {\bibinfo {author} {\bibfnamefont {Matthew}\ \bibnamefont {Gorton}}\ and\ \bibinfo {author} {\bibfnamefont {Anne~M.}\ \bibnamefont {Green}},\ }\bibfield  {title} {\enquote {\bibinfo {title} {{How open is the asteroid-mass primordial black hole window?}}}\ }\href {\doibase 10.21468/SciPostPhys.17.2.032} {\bibfield  {journal} {\bibinfo  {journal} {SciPost Phys.}\ }\textbf {\bibinfo {volume} {17}},\ \bibinfo {pages} {032} (\bibinfo {year} {2024})},\ \Eprint {http://arxiv.org/abs/2403.03839} {arXiv:2403.03839 [astro-ph.CO]} \BibitemShut {NoStop}%
\bibitem [{\citenamefont {Volonteri}\ \emph {et~al.}(2021)\citenamefont {Volonteri}, \citenamefont {Habouzit},\ and\ \citenamefont {Colpi}}]{Volonteri:2021sfo}%
  \BibitemOpen
  \bibfield  {author} {\bibinfo {author} {\bibfnamefont {Marta}\ \bibnamefont {Volonteri}}, \bibinfo {author} {\bibfnamefont {Melanie}\ \bibnamefont {Habouzit}}, \ and\ \bibinfo {author} {\bibfnamefont {Monica}\ \bibnamefont {Colpi}},\ }\bibfield  {title} {\enquote {\bibinfo {title} {{The origins of massive black holes}},}\ }\href {\doibase 10.1038/s42254-021-00364-9} {\bibfield  {journal} {\bibinfo  {journal} {Nature Rev. Phys.}\ }\textbf {\bibinfo {volume} {3}},\ \bibinfo {pages} {732--743} (\bibinfo {year} {2021})},\ \Eprint {http://arxiv.org/abs/2110.10175} {arXiv:2110.10175 [astro-ph.GA]} \BibitemShut {NoStop}%
\bibitem [{\citenamefont {Dayal}(2024)}]{Dayal:2024zwq}%
  \BibitemOpen
  \bibfield  {author} {\bibinfo {author} {\bibfnamefont {Pratika}\ \bibnamefont {Dayal}},\ }\bibfield  {title} {\enquote {\bibinfo {title} {{Exploring a primordial solution for early black holes detected with JWST}},}\ }\href {\doibase 10.1051/0004-6361/202451481} {\bibfield  {journal} {\bibinfo  {journal} {Astron. Astrophys.}\ }\textbf {\bibinfo {volume} {690}},\ \bibinfo {pages} {A182} (\bibinfo {year} {2024})},\ \Eprint {http://arxiv.org/abs/2407.07162} {arXiv:2407.07162 [astro-ph.GA]} \BibitemShut {NoStop}%
\bibitem [{\citenamefont {{Huang, Hai-Long and Wang, Yu-Tong and Piao, Yun-Song}}({2024})}]{Hai-LongHuang:2024vvz}%
  \BibitemOpen
  \bibfield  {author} {\bibinfo {author} {\bibnamefont {{Huang, Hai-Long and Wang, Yu-Tong and Piao, Yun-Song}}},\ }\bibfield  {title} {\enquote {\bibinfo {title} {{Supermassive primordial black holes for the GHZ9 and UHZ1 observed by the JWST}},}\ }\href@noop {} {\  (\bibinfo {year} {{2024}})},\ \Eprint {http://arxiv.org/abs/{2410.05891}} {{arXiv}:{2410.05891} [{astro-ph.GA}]} \BibitemShut {NoStop}%
\bibitem [{\citenamefont {Huang}\ \emph {et~al.}(2024{\natexlab{a}})\citenamefont {Huang}, \citenamefont {Jiang}, \citenamefont {He}, \citenamefont {Wang},\ and\ \citenamefont {Piao}}]{Hai-LongHuang:2024gtx}%
  \BibitemOpen
  \bibfield  {author} {\bibinfo {author} {\bibfnamefont {Hai-Long}\ \bibnamefont {Huang}}, \bibinfo {author} {\bibfnamefont {Jun-Qian}\ \bibnamefont {Jiang}}, \bibinfo {author} {\bibfnamefont {Jibin}\ \bibnamefont {He}}, \bibinfo {author} {\bibfnamefont {Yu-Tong}\ \bibnamefont {Wang}}, \ and\ \bibinfo {author} {\bibfnamefont {Yun-Song}\ \bibnamefont {Piao}},\ }\bibfield  {title} {\enquote {\bibinfo {title} {{Sub-Eddington accreting supermassive primordial black holes explain Little Red Dots}},}\ }\href@noop {} {\  (\bibinfo {year} {2024}{\natexlab{a}})},\ \Eprint {http://arxiv.org/abs/2410.20663} {arXiv:2410.20663 [astro-ph.GA]} \BibitemShut {NoStop}%
\bibitem [{\citenamefont {Huang}\ \emph {et~al.}(2024{\natexlab{b}})\citenamefont {Huang}, \citenamefont {Jiang},\ and\ \citenamefont {Piao}}]{Huang:2024aog}%
  \BibitemOpen
  \bibfield  {author} {\bibinfo {author} {\bibfnamefont {Hai-Long}\ \bibnamefont {Huang}}, \bibinfo {author} {\bibfnamefont {Jun-Qian}\ \bibnamefont {Jiang}}, \ and\ \bibinfo {author} {\bibfnamefont {Yun-Song}\ \bibnamefont {Piao}},\ }\bibfield  {title} {\enquote {\bibinfo {title} {{High-redshift JWST massive galaxies and the initial clustering of supermassive primordial black holes}},}\ }\href {\doibase 10.1103/PhysRevD.110.103540} {\bibfield  {journal} {\bibinfo  {journal} {Phys. Rev. D}\ }\textbf {\bibinfo {volume} {110}},\ \bibinfo {pages} {103540} (\bibinfo {year} {2024}{\natexlab{b}})},\ \Eprint {http://arxiv.org/abs/2407.15781} {arXiv:2407.15781 [astro-ph.CO]} \BibitemShut {NoStop}%
\bibitem [{\citenamefont {Bird}\ \emph {et~al.}(2016)\citenamefont {Bird}, \citenamefont {Cholis}, \citenamefont {Mu{\~n}oz}, \citenamefont {Ali-Ha{\"\i}moud}, \citenamefont {Kamionkowski}, \citenamefont {Kovetz}, \citenamefont {Raccanelli},\ and\ \citenamefont {Riess}}]{Bird:2016dcv}%
  \BibitemOpen
  \bibfield  {author} {\bibinfo {author} {\bibfnamefont {Simeon}\ \bibnamefont {Bird}}, \bibinfo {author} {\bibfnamefont {Ilias}\ \bibnamefont {Cholis}}, \bibinfo {author} {\bibfnamefont {Julian~B.}\ \bibnamefont {Mu{\~n}oz}}, \bibinfo {author} {\bibfnamefont {Yacine}\ \bibnamefont {Ali-Ha{\"\i}moud}}, \bibinfo {author} {\bibfnamefont {Marc}\ \bibnamefont {Kamionkowski}}, \bibinfo {author} {\bibfnamefont {Ely~D.}\ \bibnamefont {Kovetz}}, \bibinfo {author} {\bibfnamefont {Alvise}\ \bibnamefont {Raccanelli}}, \ and\ \bibinfo {author} {\bibfnamefont {Adam~G.}\ \bibnamefont {Riess}},\ }\bibfield  {title} {\enquote {\bibinfo {title} {{Did LIGO detect dark matter?}}}\ }\href {\doibase 10.1103/PhysRevLett.116.201301} {\bibfield  {journal} {\bibinfo  {journal} {Phys. Rev. Lett.}\ }\textbf {\bibinfo {volume} {116}},\ \bibinfo {pages} {201301} (\bibinfo {year} {2016})},\ \Eprint {http://arxiv.org/abs/1603.00464} {arXiv:1603.00464 [astro-ph.CO]} \BibitemShut {NoStop}%
\bibitem [{\citenamefont {De~Luca}\ \emph {et~al.}(2021)\citenamefont {De~Luca}, \citenamefont {Franciolini},\ and\ \citenamefont {Riotto}}]{DeLuca:2020agl}%
  \BibitemOpen
  \bibfield  {author} {\bibinfo {author} {\bibfnamefont {V.}~\bibnamefont {De~Luca}}, \bibinfo {author} {\bibfnamefont {G.}~\bibnamefont {Franciolini}}, \ and\ \bibinfo {author} {\bibfnamefont {A.}~\bibnamefont {Riotto}},\ }\bibfield  {title} {\enquote {\bibinfo {title} {{NANOGrav Data Hints at Primordial Black Holes as Dark Matter}},}\ }\href {\doibase 10.1103/PhysRevLett.126.041303} {\bibfield  {journal} {\bibinfo  {journal} {Phys. Rev. Lett.}\ }\textbf {\bibinfo {volume} {126}},\ \bibinfo {pages} {041303} (\bibinfo {year} {2021})},\ \Eprint {http://arxiv.org/abs/2009.08268} {arXiv:2009.08268 [astro-ph.CO]} \BibitemShut {NoStop}%
\bibitem [{\citenamefont {De~Lorenci}\ \emph {et~al.}(2025)\citenamefont {De~Lorenci}, \citenamefont {Kaiser}, \citenamefont {Peter}, \citenamefont {Ruiz},\ and\ \citenamefont {Wolfe}}]{DeLorenci:2025wbn}%
  \BibitemOpen
  \bibfield  {author} {\bibinfo {author} {\bibfnamefont {Vitorio~A.}\ \bibnamefont {De~Lorenci}}, \bibinfo {author} {\bibfnamefont {David~I.}\ \bibnamefont {Kaiser}}, \bibinfo {author} {\bibfnamefont {Patrick}\ \bibnamefont {Peter}}, \bibinfo {author} {\bibfnamefont {Lucas~S.}\ \bibnamefont {Ruiz}}, \ and\ \bibinfo {author} {\bibfnamefont {Noah~E.}\ \bibnamefont {Wolfe}},\ }\bibfield  {title} {\enquote {\bibinfo {title} {{Gravitational wave signals from primordial black holes orbiting solar-type stars}},}\ }\href {\doibase 10.1103/294z-nfj4} {\bibfield  {journal} {\bibinfo  {journal} {Phys. Rev. D}\ }\textbf {\bibinfo {volume} {112}},\ \bibinfo {pages} {063063} (\bibinfo {year} {2025})},\ \Eprint {http://arxiv.org/abs/2504.07517} {arXiv:2504.07517 [gr-qc]} \BibitemShut {NoStop}%
\bibitem [{\citenamefont {Qin}\ \emph {et~al.}(2023)\citenamefont {Qin}, \citenamefont {Geller}, \citenamefont {Balaji}, \citenamefont {McDonough},\ and\ \citenamefont {Kaiser}}]{Qin:2023lgo}%
  \BibitemOpen
  \bibfield  {author} {\bibinfo {author} {\bibfnamefont {Wenzer}\ \bibnamefont {Qin}}, \bibinfo {author} {\bibfnamefont {Sarah~R.}\ \bibnamefont {Geller}}, \bibinfo {author} {\bibfnamefont {Shyam}\ \bibnamefont {Balaji}}, \bibinfo {author} {\bibfnamefont {Evan}\ \bibnamefont {McDonough}}, \ and\ \bibinfo {author} {\bibfnamefont {David~I.}\ \bibnamefont {Kaiser}},\ }\bibfield  {title} {\enquote {\bibinfo {title} {{Planck constraints and gravitational wave forecasts for primordial black hole dark matter seeded by multifield inflation}},}\ }\href {\doibase 10.1103/PhysRevD.108.043508} {\bibfield  {journal} {\bibinfo  {journal} {Phys. Rev. D}\ }\textbf {\bibinfo {volume} {108}},\ \bibinfo {pages} {043508} (\bibinfo {year} {2023})},\ \Eprint {http://arxiv.org/abs/2303.02168} {arXiv:2303.02168 [astro-ph.CO]} \BibitemShut {NoStop}%
\bibitem [{\citenamefont {Klipfel}\ and\ \citenamefont {Kaiser}(2025)}]{Klipfel:2025jql}%
  \BibitemOpen
  \bibfield  {author} {\bibinfo {author} {\bibfnamefont {Alexandra~P.}\ \bibnamefont {Klipfel}}\ and\ \bibinfo {author} {\bibfnamefont {David~I.}\ \bibnamefont {Kaiser}},\ }\bibfield  {title} {\enquote {\bibinfo {title} {{Ultrahigh-Energy Neutrinos from Primordial Black Holes}},}\ }\href {\doibase 10.1103/vnm4-7wdc} {\bibfield  {journal} {\bibinfo  {journal} {Phys. Rev. Lett.}\ }\textbf {\bibinfo {volume} {135}},\ \bibinfo {pages} {121003} (\bibinfo {year} {2025})},\ \Eprint {http://arxiv.org/abs/2503.19227} {arXiv:2503.19227 [hep-ph]} \BibitemShut {NoStop}%
\bibitem [{\citenamefont {Baker}\ \emph {et~al.}(2025)\citenamefont {Baker}, \citenamefont {Iguaz~Juan}, \citenamefont {Symons},\ and\ \citenamefont {Thamm}}]{Baker:2025cff}%
  \BibitemOpen
  \bibfield  {author} {\bibinfo {author} {\bibfnamefont {Michael~J.}\ \bibnamefont {Baker}}, \bibinfo {author} {\bibfnamefont {Joaquim}\ \bibnamefont {Iguaz~Juan}}, \bibinfo {author} {\bibfnamefont {Aidan}\ \bibnamefont {Symons}}, \ and\ \bibinfo {author} {\bibfnamefont {Andrea}\ \bibnamefont {Thamm}},\ }\bibfield  {title} {\enquote {\bibinfo {title} {{Explaining the PeV Neutrino Fluxes at KM3NeT and IceCube with Quasi-Extremal Primordial Black Holes}},}\ }\href@noop {} {\  (\bibinfo {year} {2025})},\ \Eprint {http://arxiv.org/abs/2505.22722} {arXiv:2505.22722 [hep-ph]} \BibitemShut {NoStop}%
\bibitem [{\citenamefont {Klipfel}\ \emph {et~al.}(2025)\citenamefont {Klipfel}, \citenamefont {Fisher},\ and\ \citenamefont {Kaiser}}]{Klipfel:2025bvh}%
  \BibitemOpen
  \bibfield  {author} {\bibinfo {author} {\bibfnamefont {Alexandra~P.}\ \bibnamefont {Klipfel}}, \bibinfo {author} {\bibfnamefont {Peter}\ \bibnamefont {Fisher}}, \ and\ \bibinfo {author} {\bibfnamefont {David~I.}\ \bibnamefont {Kaiser}},\ }\bibfield  {title} {\enquote {\bibinfo {title} {{Hawking radiation signatures from primordial black holes transiting the inner Solar System: Prospects for detection}},}\ }\href {\doibase 10.1103/9jyp-24sw} {\bibfield  {journal} {\bibinfo  {journal} {Phys. Rev. D}\ }\textbf {\bibinfo {volume} {112}},\ \bibinfo {pages} {103007} (\bibinfo {year} {2025})},\ \Eprint {http://arxiv.org/abs/2506.14041} {arXiv:2506.14041 [astro-ph.CO]} \BibitemShut {NoStop}%
\bibitem [{\citenamefont {Hawking}\ \emph {et~al.}(1982)\citenamefont {Hawking}, \citenamefont {Moss},\ and\ \citenamefont {Stewart}}]{Hawking:1982ga}%
  \BibitemOpen
  \bibfield  {author} {\bibinfo {author} {\bibfnamefont {S.~W.}\ \bibnamefont {Hawking}}, \bibinfo {author} {\bibfnamefont {I.~G.}\ \bibnamefont {Moss}}, \ and\ \bibinfo {author} {\bibfnamefont {J.~M.}\ \bibnamefont {Stewart}},\ }\bibfield  {title} {\enquote {\bibinfo {title} {{Bubble Collisions in the Very Early Universe}},}\ }\href {\doibase 10.1103/PhysRevD.26.2681} {\bibfield  {journal} {\bibinfo  {journal} {Phys. Rev. D}\ }\textbf {\bibinfo {volume} {26}},\ \bibinfo {pages} {2681} (\bibinfo {year} {1982})}\BibitemShut {NoStop}%
\bibitem [{\citenamefont {Deng}\ and\ \citenamefont {Vilenkin}(2017)}]{Deng:2017uwc}%
  \BibitemOpen
  \bibfield  {author} {\bibinfo {author} {\bibfnamefont {Heling}\ \bibnamefont {Deng}}\ and\ \bibinfo {author} {\bibfnamefont {Alexander}\ \bibnamefont {Vilenkin}},\ }\bibfield  {title} {\enquote {\bibinfo {title} {{Primordial black hole formation by vacuum bubbles}},}\ }\href {\doibase 10.1088/1475-7516/2017/12/044} {\bibfield  {journal} {\bibinfo  {journal} {JCAP}\ }\textbf {\bibinfo {volume} {12}},\ \bibinfo {pages} {044} (\bibinfo {year} {2017})},\ \Eprint {http://arxiv.org/abs/1710.02865} {arXiv:1710.02865 [gr-qc]} \BibitemShut {NoStop}%
\bibitem [{\citenamefont {Hawking}(1989)}]{Hawking:1987bn}%
  \BibitemOpen
  \bibfield  {author} {\bibinfo {author} {\bibfnamefont {S.~W.}\ \bibnamefont {Hawking}},\ }\bibfield  {title} {\enquote {\bibinfo {title} {{Black Holes From Cosmic Strings}},}\ }\href {\doibase 10.1016/0370-2693(89)90206-2} {\bibfield  {journal} {\bibinfo  {journal} {Phys. Lett. B}\ }\textbf {\bibinfo {volume} {231}},\ \bibinfo {pages} {237--239} (\bibinfo {year} {1989})}\BibitemShut {NoStop}%
\bibitem [{\citenamefont {Polnarev}\ and\ \citenamefont {Zembowicz}(1991)}]{Polnarev:1988dh}%
  \BibitemOpen
  \bibfield  {author} {\bibinfo {author} {\bibfnamefont {Alexander}\ \bibnamefont {Polnarev}}\ and\ \bibinfo {author} {\bibfnamefont {Robert}\ \bibnamefont {Zembowicz}},\ }\bibfield  {title} {\enquote {\bibinfo {title} {{Formation of Primordial Black Holes by Cosmic Strings}},}\ }\href {\doibase 10.1103/PhysRevD.43.1106} {\bibfield  {journal} {\bibinfo  {journal} {Phys. Rev. D}\ }\textbf {\bibinfo {volume} {43}},\ \bibinfo {pages} {1106--1109} (\bibinfo {year} {1991})}\BibitemShut {NoStop}%
\bibitem [{\citenamefont {Rubin}\ \emph {et~al.}(2001)\citenamefont {Rubin}, \citenamefont {Sakharov},\ and\ \citenamefont {Khlopov}}]{Rubin:2001yw}%
  \BibitemOpen
  \bibfield  {author} {\bibinfo {author} {\bibfnamefont {Sergei~G.}\ \bibnamefont {Rubin}}, \bibinfo {author} {\bibfnamefont {Alexander~S.}\ \bibnamefont {Sakharov}}, \ and\ \bibinfo {author} {\bibfnamefont {Maxim~Yu.}\ \bibnamefont {Khlopov}},\ }\bibfield  {title} {\enquote {\bibinfo {title} {{The Formation of primary galactic nuclei during phase transitions in the early universe}},}\ }\href {\doibase 10.1134/1.1385631} {\bibfield  {journal} {\bibinfo  {journal} {J. Exp. Theor. Phys.}\ }\textbf {\bibinfo {volume} {91}},\ \bibinfo {pages} {921--929} (\bibinfo {year} {2001})},\ \Eprint {http://arxiv.org/abs/hep-ph/0106187} {arXiv:hep-ph/0106187} \BibitemShut {NoStop}%
\bibitem [{\citenamefont {Dvali}\ \emph {et~al.}(2021)\citenamefont {Dvali}, \citenamefont {K{\"u}hnel},\ and\ \citenamefont {Zantedeschi}}]{Dvali:2021byy}%
  \BibitemOpen
  \bibfield  {author} {\bibinfo {author} {\bibfnamefont {Gia}\ \bibnamefont {Dvali}}, \bibinfo {author} {\bibfnamefont {Florian}\ \bibnamefont {K{\"u}hnel}}, \ and\ \bibinfo {author} {\bibfnamefont {Michael}\ \bibnamefont {Zantedeschi}},\ }\bibfield  {title} {\enquote {\bibinfo {title} {{Primordial black holes from confinement}},}\ }\href {\doibase 10.1103/PhysRevD.104.123507} {\bibfield  {journal} {\bibinfo  {journal} {Phys. Rev. D}\ }\textbf {\bibinfo {volume} {104}},\ \bibinfo {pages} {123507} (\bibinfo {year} {2021})},\ \Eprint {http://arxiv.org/abs/2108.09471} {arXiv:2108.09471 [hep-ph]} \BibitemShut {NoStop}%
\bibitem [{\citenamefont {Kinney}(2005)}]{Kinney:2005vj}%
  \BibitemOpen
  \bibfield  {author} {\bibinfo {author} {\bibfnamefont {William~H.}\ \bibnamefont {Kinney}},\ }\bibfield  {title} {\enquote {\bibinfo {title} {{Horizon crossing and inflation with large eta}},}\ }\href {\doibase 10.1103/PhysRevD.72.023515} {\bibfield  {journal} {\bibinfo  {journal} {Phys. Rev. D}\ }\textbf {\bibinfo {volume} {72}},\ \bibinfo {pages} {023515} (\bibinfo {year} {2005})},\ \Eprint {http://arxiv.org/abs/gr-qc/0503017} {arXiv:gr-qc/0503017} \BibitemShut {NoStop}%
\bibitem [{\citenamefont {Martin}\ \emph {et~al.}(2013)\citenamefont {Martin}, \citenamefont {Motohashi},\ and\ \citenamefont {Suyama}}]{Martin:2012pe}%
  \BibitemOpen
  \bibfield  {author} {\bibinfo {author} {\bibfnamefont {Jerome}\ \bibnamefont {Martin}}, \bibinfo {author} {\bibfnamefont {Hayato}\ \bibnamefont {Motohashi}}, \ and\ \bibinfo {author} {\bibfnamefont {Teruaki}\ \bibnamefont {Suyama}},\ }\bibfield  {title} {\enquote {\bibinfo {title} {{Ultra Slow-Roll Inflation and the non-Gaussianity Consistency Relation}},}\ }\href {\doibase 10.1103/PhysRevD.87.023514} {\bibfield  {journal} {\bibinfo  {journal} {Phys. Rev. D}\ }\textbf {\bibinfo {volume} {87}},\ \bibinfo {pages} {023514} (\bibinfo {year} {2013})},\ \Eprint {http://arxiv.org/abs/1211.0083} {arXiv:1211.0083 [astro-ph.CO]} \BibitemShut {NoStop}%
\bibitem [{\citenamefont {Ezquiaga}\ \emph {et~al.}(2018)\citenamefont {Ezquiaga}, \citenamefont {Garcia-Bellido},\ and\ \citenamefont {Ruiz~Morales}}]{Ezquiaga:2017fvi}%
  \BibitemOpen
  \bibfield  {author} {\bibinfo {author} {\bibfnamefont {Jose~Maria}\ \bibnamefont {Ezquiaga}}, \bibinfo {author} {\bibfnamefont {Juan}\ \bibnamefont {Garcia-Bellido}}, \ and\ \bibinfo {author} {\bibfnamefont {Ester}\ \bibnamefont {Ruiz~Morales}},\ }\bibfield  {title} {\enquote {\bibinfo {title} {{Primordial Black Hole production in Critical Higgs Inflation}},}\ }\href {\doibase 10.1016/j.physletb.2017.11.039} {\bibfield  {journal} {\bibinfo  {journal} {Phys. Lett. B}\ }\textbf {\bibinfo {volume} {776}},\ \bibinfo {pages} {345--349} (\bibinfo {year} {2018})},\ \Eprint {http://arxiv.org/abs/1705.04861} {arXiv:1705.04861 [astro-ph.CO]} \BibitemShut {NoStop}%
\bibitem [{\citenamefont {Garcia-Bellido}\ and\ \citenamefont {Ruiz~Morales}(2017)}]{Garcia-Bellido:2017mdw}%
  \BibitemOpen
  \bibfield  {author} {\bibinfo {author} {\bibfnamefont {Juan}\ \bibnamefont {Garcia-Bellido}}\ and\ \bibinfo {author} {\bibfnamefont {Ester}\ \bibnamefont {Ruiz~Morales}},\ }\bibfield  {title} {\enquote {\bibinfo {title} {{Primordial black holes from single field models of inflation}},}\ }\href {\doibase 10.1016/j.dark.2017.09.007} {\bibfield  {journal} {\bibinfo  {journal} {Phys. Dark Univ.}\ }\textbf {\bibinfo {volume} {18}},\ \bibinfo {pages} {47--54} (\bibinfo {year} {2017})},\ \Eprint {http://arxiv.org/abs/1702.03901} {arXiv:1702.03901 [astro-ph.CO]} \BibitemShut {NoStop}%
\bibitem [{\citenamefont {Germani}\ and\ \citenamefont {Prokopec}(2017)}]{Germani:2017bcs}%
  \BibitemOpen
  \bibfield  {author} {\bibinfo {author} {\bibfnamefont {Cristiano}\ \bibnamefont {Germani}}\ and\ \bibinfo {author} {\bibfnamefont {Tomislav}\ \bibnamefont {Prokopec}},\ }\bibfield  {title} {\enquote {\bibinfo {title} {{On primordial black holes from an inflection point}},}\ }\href {\doibase 10.1016/j.dark.2017.09.001} {\bibfield  {journal} {\bibinfo  {journal} {Phys. Dark Univ.}\ }\textbf {\bibinfo {volume} {18}},\ \bibinfo {pages} {6--10} (\bibinfo {year} {2017})},\ \Eprint {http://arxiv.org/abs/1706.04226} {arXiv:1706.04226 [astro-ph.CO]} \BibitemShut {NoStop}%
\bibitem [{\citenamefont {Kannike}\ \emph {et~al.}(2017)\citenamefont {Kannike}, \citenamefont {Marzola}, \citenamefont {Raidal},\ and\ \citenamefont {Veerm\"ae}}]{Kannike:2017bxn}%
  \BibitemOpen
  \bibfield  {author} {\bibinfo {author} {\bibfnamefont {Kristjan}\ \bibnamefont {Kannike}}, \bibinfo {author} {\bibfnamefont {Luca}\ \bibnamefont {Marzola}}, \bibinfo {author} {\bibfnamefont {Martti}\ \bibnamefont {Raidal}}, \ and\ \bibinfo {author} {\bibfnamefont {Hardi}\ \bibnamefont {Veerm\"ae}},\ }\bibfield  {title} {\enquote {\bibinfo {title} {{Single Field Double Inflation and Primordial Black Holes}},}\ }\href {\doibase 10.1088/1475-7516/2017/09/020} {\bibfield  {journal} {\bibinfo  {journal} {JCAP}\ }\textbf {\bibinfo {volume} {09}},\ \bibinfo {pages} {020} (\bibinfo {year} {2017})},\ \Eprint {http://arxiv.org/abs/1705.06225} {arXiv:1705.06225 [astro-ph.CO]} \BibitemShut {NoStop}%
\bibitem [{\citenamefont {Motohashi}\ and\ \citenamefont {Hu}(2017)}]{Motohashi:2017kbs}%
  \BibitemOpen
  \bibfield  {author} {\bibinfo {author} {\bibfnamefont {Hayato}\ \bibnamefont {Motohashi}}\ and\ \bibinfo {author} {\bibfnamefont {Wayne}\ \bibnamefont {Hu}},\ }\bibfield  {title} {\enquote {\bibinfo {title} {{Primordial Black Holes and Slow-Roll Violation}},}\ }\href {\doibase 10.1103/PhysRevD.96.063503} {\bibfield  {journal} {\bibinfo  {journal} {Phys. Rev. D}\ }\textbf {\bibinfo {volume} {96}},\ \bibinfo {pages} {063503} (\bibinfo {year} {2017})},\ \Eprint {http://arxiv.org/abs/1706.06784} {arXiv:1706.06784 [astro-ph.CO]} \BibitemShut {NoStop}%
\bibitem [{\citenamefont {Di}\ and\ \citenamefont {Gong}(2018)}]{Di:2017ndc}%
  \BibitemOpen
  \bibfield  {author} {\bibinfo {author} {\bibfnamefont {Haoran}\ \bibnamefont {Di}}\ and\ \bibinfo {author} {\bibfnamefont {Yungui}\ \bibnamefont {Gong}},\ }\bibfield  {title} {\enquote {\bibinfo {title} {{Primordial black holes and second order gravitational waves from ultra-slow-roll inflation}},}\ }\href {\doibase 10.1088/1475-7516/2018/07/007} {\bibfield  {journal} {\bibinfo  {journal} {JCAP}\ }\textbf {\bibinfo {volume} {07}},\ \bibinfo {pages} {007} (\bibinfo {year} {2018})},\ \Eprint {http://arxiv.org/abs/1707.09578} {arXiv:1707.09578 [astro-ph.CO]} \BibitemShut {NoStop}%
\bibitem [{\citenamefont {Ballesteros}\ and\ \citenamefont {Taoso}(2018)}]{Ballesteros:2017fsr}%
  \BibitemOpen
  \bibfield  {author} {\bibinfo {author} {\bibfnamefont {Guillermo}\ \bibnamefont {Ballesteros}}\ and\ \bibinfo {author} {\bibfnamefont {Marco}\ \bibnamefont {Taoso}},\ }\bibfield  {title} {\enquote {\bibinfo {title} {{Primordial black hole dark matter from single field inflation}},}\ }\href {\doibase 10.1103/PhysRevD.97.023501} {\bibfield  {journal} {\bibinfo  {journal} {Phys. Rev. D}\ }\textbf {\bibinfo {volume} {97}},\ \bibinfo {pages} {023501} (\bibinfo {year} {2018})},\ \Eprint {http://arxiv.org/abs/1709.05565} {arXiv:1709.05565 [hep-ph]} \BibitemShut {NoStop}%
\bibitem [{\citenamefont {Pattison}\ \emph {et~al.}(2017)\citenamefont {Pattison}, \citenamefont {Vennin}, \citenamefont {Assadullahi},\ and\ \citenamefont {Wands}}]{Pattison:2017mbe}%
  \BibitemOpen
  \bibfield  {author} {\bibinfo {author} {\bibfnamefont {Chris}\ \bibnamefont {Pattison}}, \bibinfo {author} {\bibfnamefont {Vincent}\ \bibnamefont {Vennin}}, \bibinfo {author} {\bibfnamefont {Hooshyar}\ \bibnamefont {Assadullahi}}, \ and\ \bibinfo {author} {\bibfnamefont {David}\ \bibnamefont {Wands}},\ }\bibfield  {title} {\enquote {\bibinfo {title} {{Quantum diffusion during inflation and primordial black holes}},}\ }\href {\doibase 10.1088/1475-7516/2017/10/046} {\bibfield  {journal} {\bibinfo  {journal} {JCAP}\ }\textbf {\bibinfo {volume} {10}},\ \bibinfo {pages} {046} (\bibinfo {year} {2017})},\ \Eprint {http://arxiv.org/abs/1707.00537} {arXiv:1707.00537 [hep-th]} \BibitemShut {NoStop}%
\bibitem [{\citenamefont {Passaglia}\ \emph {et~al.}(2019)\citenamefont {Passaglia}, \citenamefont {Hu},\ and\ \citenamefont {Motohashi}}]{Passaglia:2018ixg}%
  \BibitemOpen
  \bibfield  {author} {\bibinfo {author} {\bibfnamefont {Samuel}\ \bibnamefont {Passaglia}}, \bibinfo {author} {\bibfnamefont {Wayne}\ \bibnamefont {Hu}}, \ and\ \bibinfo {author} {\bibfnamefont {Hayato}\ \bibnamefont {Motohashi}},\ }\bibfield  {title} {\enquote {\bibinfo {title} {{Primordial black holes and local non-Gaussianity in canonical inflation}},}\ }\href {\doibase 10.1103/PhysRevD.99.043536} {\bibfield  {journal} {\bibinfo  {journal} {Phys. Rev. D}\ }\textbf {\bibinfo {volume} {99}},\ \bibinfo {pages} {043536} (\bibinfo {year} {2019})},\ \Eprint {http://arxiv.org/abs/1812.08243} {arXiv:1812.08243 [astro-ph.CO]} \BibitemShut {NoStop}%
\bibitem [{\citenamefont {Biagetti}\ \emph {et~al.}(2018)\citenamefont {Biagetti}, \citenamefont {Franciolini}, \citenamefont {Kehagias},\ and\ \citenamefont {Riotto}}]{Biagetti:2018pjj}%
  \BibitemOpen
  \bibfield  {author} {\bibinfo {author} {\bibfnamefont {Matteo}\ \bibnamefont {Biagetti}}, \bibinfo {author} {\bibfnamefont {Gabriele}\ \bibnamefont {Franciolini}}, \bibinfo {author} {\bibfnamefont {Alex}\ \bibnamefont {Kehagias}}, \ and\ \bibinfo {author} {\bibfnamefont {Antonio}\ \bibnamefont {Riotto}},\ }\bibfield  {title} {\enquote {\bibinfo {title} {{Primordial Black Holes from Inflation and Quantum Diffusion}},}\ }\href {\doibase 10.1088/1475-7516/2018/07/032} {\bibfield  {journal} {\bibinfo  {journal} {JCAP}\ }\textbf {\bibinfo {volume} {07}},\ \bibinfo {pages} {032} (\bibinfo {year} {2018})},\ \Eprint {http://arxiv.org/abs/1804.07124} {arXiv:1804.07124 [astro-ph.CO]} \BibitemShut {NoStop}%
\bibitem [{\citenamefont {Mishra}\ and\ \citenamefont {Sahni}(2020)}]{Mishra:2019pzq}%
  \BibitemOpen
  \bibfield  {author} {\bibinfo {author} {\bibfnamefont {Swagat~S.}\ \bibnamefont {Mishra}}\ and\ \bibinfo {author} {\bibfnamefont {Varun}\ \bibnamefont {Sahni}},\ }\bibfield  {title} {\enquote {\bibinfo {title} {{Primordial Black Holes from a tiny bump/dip in the Inflaton potential}},}\ }\href {\doibase 10.1088/1475-7516/2020/04/007} {\bibfield  {journal} {\bibinfo  {journal} {JCAP}\ }\textbf {\bibinfo {volume} {04}},\ \bibinfo {pages} {007} (\bibinfo {year} {2020})},\ \Eprint {http://arxiv.org/abs/1911.00057} {arXiv:1911.00057 [gr-qc]} \BibitemShut {NoStop}%
\bibitem [{\citenamefont {Figueroa}\ \emph {et~al.}(2021)\citenamefont {Figueroa}, \citenamefont {Raatikainen}, \citenamefont {Rasanen},\ and\ \citenamefont {Tomberg}}]{Figueroa:2020jkf}%
  \BibitemOpen
  \bibfield  {author} {\bibinfo {author} {\bibfnamefont {Daniel~G.}\ \bibnamefont {Figueroa}}, \bibinfo {author} {\bibfnamefont {Sami}\ \bibnamefont {Raatikainen}}, \bibinfo {author} {\bibfnamefont {Syksy}\ \bibnamefont {Rasanen}}, \ and\ \bibinfo {author} {\bibfnamefont {Eemeli}\ \bibnamefont {Tomberg}},\ }\bibfield  {title} {\enquote {\bibinfo {title} {{Non-Gaussian Tail of the Curvature Perturbation in Stochastic Ultraslow-Roll Inflation: Implications for Primordial Black Hole Production}},}\ }\href {\doibase 10.1103/PhysRevLett.127.101302} {\bibfield  {journal} {\bibinfo  {journal} {Phys. Rev. Lett.}\ }\textbf {\bibinfo {volume} {127}},\ \bibinfo {pages} {101302} (\bibinfo {year} {2021})},\ \Eprint {http://arxiv.org/abs/2012.06551} {arXiv:2012.06551 [astro-ph.CO]} \BibitemShut {NoStop}%
\bibitem [{\citenamefont {Karam}\ \emph {et~al.}(2023)\citenamefont {Karam}, \citenamefont {Koivunen}, \citenamefont {Tomberg}, \citenamefont {Vaskonen},\ and\ \citenamefont {Veerm\"ae}}]{Karam:2022nym}%
  \BibitemOpen
  \bibfield  {author} {\bibinfo {author} {\bibfnamefont {Alexandros}\ \bibnamefont {Karam}}, \bibinfo {author} {\bibfnamefont {Niko}\ \bibnamefont {Koivunen}}, \bibinfo {author} {\bibfnamefont {Eemeli}\ \bibnamefont {Tomberg}}, \bibinfo {author} {\bibfnamefont {Ville}\ \bibnamefont {Vaskonen}}, \ and\ \bibinfo {author} {\bibfnamefont {Hardi}\ \bibnamefont {Veerm\"ae}},\ }\bibfield  {title} {\enquote {\bibinfo {title} {{Anatomy of single-field inflationary models for primordial black holes}},}\ }\href {\doibase 10.1088/1475-7516/2023/03/013} {\bibfield  {journal} {\bibinfo  {journal} {JCAP}\ }\textbf {\bibinfo {volume} {03}},\ \bibinfo {pages} {013} (\bibinfo {year} {2023})},\ \Eprint {http://arxiv.org/abs/2205.13540} {arXiv:2205.13540 [astro-ph.CO]} \BibitemShut {NoStop}%
\bibitem [{\citenamefont {\"Ozsoy}\ and\ \citenamefont {Tasinato}(2023)}]{Ozsoy:2023ryl}%
  \BibitemOpen
  \bibfield  {author} {\bibinfo {author} {\bibfnamefont {Ogan}\ \bibnamefont {\"Ozsoy}}\ and\ \bibinfo {author} {\bibfnamefont {Gianmassimo}\ \bibnamefont {Tasinato}},\ }\bibfield  {title} {\enquote {\bibinfo {title} {{Inflation and Primordial Black Holes}},}\ }\href {\doibase 10.3390/universe9050203} {\bibfield  {journal} {\bibinfo  {journal} {Universe}\ }\textbf {\bibinfo {volume} {9}},\ \bibinfo {pages} {203} (\bibinfo {year} {2023})},\ \Eprint {http://arxiv.org/abs/2301.03600} {arXiv:2301.03600 [astro-ph.CO]} \BibitemShut {NoStop}%
\bibitem [{\citenamefont {Cole}\ \emph {et~al.}(2023)\citenamefont {Cole}, \citenamefont {Gow}, \citenamefont {Byrnes},\ and\ \citenamefont {Patil}}]{Cole:2023wyx}%
  \BibitemOpen
  \bibfield  {author} {\bibinfo {author} {\bibfnamefont {Philippa~S.}\ \bibnamefont {Cole}}, \bibinfo {author} {\bibfnamefont {Andrew~D.}\ \bibnamefont {Gow}}, \bibinfo {author} {\bibfnamefont {Christian~T.}\ \bibnamefont {Byrnes}}, \ and\ \bibinfo {author} {\bibfnamefont {Subodh~P.}\ \bibnamefont {Patil}},\ }\bibfield  {title} {\enquote {\bibinfo {title} {{Primordial black holes from single-field inflation: a fine-tuning audit}},}\ }\href {\doibase 10.1088/1475-7516/2023/08/031} {\bibfield  {journal} {\bibinfo  {journal} {JCAP}\ }\textbf {\bibinfo {volume} {08}},\ \bibinfo {pages} {031} (\bibinfo {year} {2023})},\ \Eprint {http://arxiv.org/abs/2304.01997} {arXiv:2304.01997 [astro-ph.CO]} \BibitemShut {NoStop}%
\bibitem [{\citenamefont {Cicoli}\ \emph {et~al.}(2018)\citenamefont {Cicoli}, \citenamefont {Diaz},\ and\ \citenamefont {Pedro}}]{Cicoli:2018asa}%
  \BibitemOpen
  \bibfield  {author} {\bibinfo {author} {\bibfnamefont {Michele}\ \bibnamefont {Cicoli}}, \bibinfo {author} {\bibfnamefont {Victor~A.}\ \bibnamefont {Diaz}}, \ and\ \bibinfo {author} {\bibfnamefont {Francisco~G.}\ \bibnamefont {Pedro}},\ }\bibfield  {title} {\enquote {\bibinfo {title} {{Primordial Black Holes from String Inflation}},}\ }\href {\doibase 10.1088/1475-7516/2018/06/034} {\bibfield  {journal} {\bibinfo  {journal} {JCAP}\ }\textbf {\bibinfo {volume} {06}},\ \bibinfo {pages} {034} (\bibinfo {year} {2018})},\ \Eprint {http://arxiv.org/abs/1803.02837} {arXiv:1803.02837 [hep-th]} \BibitemShut {NoStop}%
\bibitem [{\citenamefont {Cicoli}\ \emph {et~al.}(2022)\citenamefont {Cicoli}, \citenamefont {Pedro},\ and\ \citenamefont {Pedron}}]{Cicoli:2022sih}%
  \BibitemOpen
  \bibfield  {author} {\bibinfo {author} {\bibfnamefont {Michele}\ \bibnamefont {Cicoli}}, \bibinfo {author} {\bibfnamefont {Francisco~G.}\ \bibnamefont {Pedro}}, \ and\ \bibinfo {author} {\bibfnamefont {Nicola}\ \bibnamefont {Pedron}},\ }\bibfield  {title} {\enquote {\bibinfo {title} {{Secondary GWs and PBHs in string inflation: formation and detectability}},}\ }\href {\doibase 10.1088/1475-7516/2022/08/030} {\bibfield  {journal} {\bibinfo  {journal} {JCAP}\ }\textbf {\bibinfo {volume} {08}},\ \bibinfo {pages} {030} (\bibinfo {year} {2022})},\ \Eprint {http://arxiv.org/abs/2203.00021} {arXiv:2203.00021 [hep-th]} \BibitemShut {NoStop}%
\bibitem [{\citenamefont {Cai}\ \emph {et~al.}(2022)\citenamefont {Cai}, \citenamefont {Ma}, \citenamefont {Sasaki}, \citenamefont {Wang},\ and\ \citenamefont {Zhou}}]{Cai:2022erk}%
  \BibitemOpen
  \bibfield  {author} {\bibinfo {author} {\bibfnamefont {Yi-Fu}\ \bibnamefont {Cai}}, \bibinfo {author} {\bibfnamefont {Xiao-Han}\ \bibnamefont {Ma}}, \bibinfo {author} {\bibfnamefont {Misao}\ \bibnamefont {Sasaki}}, \bibinfo {author} {\bibfnamefont {Dong-Gang}\ \bibnamefont {Wang}}, \ and\ \bibinfo {author} {\bibfnamefont {Zihan}\ \bibnamefont {Zhou}},\ }\bibfield  {title} {\enquote {\bibinfo {title} {{Highly non-Gaussian tails and primordial black holes from single-field inflation}},}\ }\href {\doibase 10.1088/1475-7516/2022/12/034} {\bibfield  {journal} {\bibinfo  {journal} {JCAP}\ }\textbf {\bibinfo {volume} {12}},\ \bibinfo {pages} {034} (\bibinfo {year} {2022})},\ \Eprint {http://arxiv.org/abs/2207.11910} {arXiv:2207.11910 [astro-ph.CO]} \BibitemShut {NoStop}%
\bibitem [{\citenamefont {Inomata}\ \emph {et~al.}(2021)\citenamefont {Inomata}, \citenamefont {McDonough},\ and\ \citenamefont {Hu}}]{Inomata:2021uqj}%
  \BibitemOpen
  \bibfield  {author} {\bibinfo {author} {\bibfnamefont {Keisuke}\ \bibnamefont {Inomata}}, \bibinfo {author} {\bibfnamefont {Evan}\ \bibnamefont {McDonough}}, \ and\ \bibinfo {author} {\bibfnamefont {Wayne}\ \bibnamefont {Hu}},\ }\bibfield  {title} {\enquote {\bibinfo {title} {{Primordial black holes arise when the inflaton falls}},}\ }\href {\doibase 10.1103/PhysRevD.104.123553} {\bibfield  {journal} {\bibinfo  {journal} {Phys. Rev. D}\ }\textbf {\bibinfo {volume} {104}},\ \bibinfo {pages} {123553} (\bibinfo {year} {2021})},\ \Eprint {http://arxiv.org/abs/2104.03972} {arXiv:2104.03972 [astro-ph.CO]} \BibitemShut {NoStop}%
\bibitem [{\citenamefont {Inomata}\ \emph {et~al.}(2022)\citenamefont {Inomata}, \citenamefont {McDonough},\ and\ \citenamefont {Hu}}]{Inomata:2021tpx}%
  \BibitemOpen
  \bibfield  {author} {\bibinfo {author} {\bibfnamefont {Keisuke}\ \bibnamefont {Inomata}}, \bibinfo {author} {\bibfnamefont {Evan}\ \bibnamefont {McDonough}}, \ and\ \bibinfo {author} {\bibfnamefont {Wayne}\ \bibnamefont {Hu}},\ }\bibfield  {title} {\enquote {\bibinfo {title} {{Amplification of primordial perturbations from the rise or fall of the inflaton}},}\ }\href {\doibase 10.1088/1475-7516/2022/02/031} {\bibfield  {journal} {\bibinfo  {journal} {JCAP}\ }\textbf {\bibinfo {volume} {02}},\ \bibinfo {pages} {031} (\bibinfo {year} {2022})},\ \Eprint {http://arxiv.org/abs/2110.14641} {arXiv:2110.14641 [astro-ph.CO]} \BibitemShut {NoStop}%
\bibitem [{\citenamefont {Bhaumik}\ and\ \citenamefont {Jain}(2020)}]{Bhaumik:2019tvl}%
  \BibitemOpen
  \bibfield  {author} {\bibinfo {author} {\bibfnamefont {Nilanjandev}\ \bibnamefont {Bhaumik}}\ and\ \bibinfo {author} {\bibfnamefont {Rajeev~Kumar}\ \bibnamefont {Jain}},\ }\bibfield  {title} {\enquote {\bibinfo {title} {{Primordial black holes dark matter from inflection point models of inflation and the effects of reheating}},}\ }\href {\doibase 10.1088/1475-7516/2020/01/037} {\bibfield  {journal} {\bibinfo  {journal} {JCAP}\ }\textbf {\bibinfo {volume} {01}},\ \bibinfo {pages} {037} (\bibinfo {year} {2020})},\ \Eprint {http://arxiv.org/abs/1907.04125} {arXiv:1907.04125 [astro-ph.CO]} \BibitemShut {NoStop}%
\bibitem [{\citenamefont {Pi}\ and\ \citenamefont {Sasaki}(2023)}]{Pi:2022ysn}%
  \BibitemOpen
  \bibfield  {author} {\bibinfo {author} {\bibfnamefont {Shi}\ \bibnamefont {Pi}}\ and\ \bibinfo {author} {\bibfnamefont {Misao}\ \bibnamefont {Sasaki}},\ }\bibfield  {title} {\enquote {\bibinfo {title} {{Logarithmic Duality of the Curvature Perturbation}},}\ }\href {\doibase 10.1103/PhysRevLett.131.011002} {\bibfield  {journal} {\bibinfo  {journal} {Phys. Rev. Lett.}\ }\textbf {\bibinfo {volume} {131}},\ \bibinfo {pages} {011002} (\bibinfo {year} {2023})},\ \Eprint {http://arxiv.org/abs/2211.13932} {arXiv:2211.13932 [astro-ph.CO]} \BibitemShut {NoStop}%
\bibitem [{\citenamefont {Choudhury}\ \emph {et~al.}(2024)\citenamefont {Choudhury}, \citenamefont {Karde}, \citenamefont {Panda},\ and\ \citenamefont {Sami}}]{Choudhury:2024one}%
  \BibitemOpen
  \bibfield  {author} {\bibinfo {author} {\bibfnamefont {Sayantan}\ \bibnamefont {Choudhury}}, \bibinfo {author} {\bibfnamefont {Ahaskar}\ \bibnamefont {Karde}}, \bibinfo {author} {\bibfnamefont {Sudhakar}\ \bibnamefont {Panda}}, \ and\ \bibinfo {author} {\bibfnamefont {M.}~\bibnamefont {Sami}},\ }\bibfield  {title} {\enquote {\bibinfo {title} {{Realisation of the ultra-slow roll phase in Galileon inflation and PBH overproduction}},}\ }\href {\doibase 10.1088/1475-7516/2024/07/034} {\bibfield  {journal} {\bibinfo  {journal} {JCAP}\ }\textbf {\bibinfo {volume} {07}},\ \bibinfo {pages} {034} (\bibinfo {year} {2024})},\ \Eprint {http://arxiv.org/abs/2401.10925} {arXiv:2401.10925 [astro-ph.CO]} \BibitemShut {NoStop}%
\bibitem [{\citenamefont {Akrami}\ \emph {et~al.}(2020{\natexlab{a}})\citenamefont {Akrami} \emph {et~al.}}]{Planck:2019kim}%
  \BibitemOpen
  \bibfield  {author} {\bibinfo {author} {\bibfnamefont {Y.}~\bibnamefont {Akrami}} \emph {et~al.} (\bibinfo {collaboration} {Planck}),\ }\bibfield  {title} {\enquote {\bibinfo {title} {{Planck 2018 results. IX. Constraints on primordial non-Gaussianity}},}\ }\href {\doibase 10.1051/0004-6361/201935891} {\bibfield  {journal} {\bibinfo  {journal} {Astron. Astrophys.}\ }\textbf {\bibinfo {volume} {641}},\ \bibinfo {pages} {A9} (\bibinfo {year} {2020}{\natexlab{a}})},\ \Eprint {http://arxiv.org/abs/1905.05697} {arXiv:1905.05697 [astro-ph.CO]} \BibitemShut {NoStop}%
\bibitem [{\citenamefont {Aghanim}\ \emph {et~al.}(2020)\citenamefont {Aghanim} \emph {et~al.}}]{Planck:2018vyg}%
  \BibitemOpen
  \bibfield  {author} {\bibinfo {author} {\bibfnamefont {N.}~\bibnamefont {Aghanim}} \emph {et~al.} (\bibinfo {collaboration} {Planck}),\ }\bibfield  {title} {\enquote {\bibinfo {title} {{Planck 2018 results. VI. Cosmological parameters}},}\ }\href {\doibase 10.1051/0004-6361/201833910} {\bibfield  {journal} {\bibinfo  {journal} {Astron. Astrophys.}\ }\textbf {\bibinfo {volume} {641}},\ \bibinfo {pages} {A6} (\bibinfo {year} {2020})},\ \bibinfo {note} {[Erratum: Astron.Astrophys. 652, C4 (2021)]},\ \Eprint {http://arxiv.org/abs/1807.06209} {arXiv:1807.06209 [astro-ph.CO]} \BibitemShut {NoStop}%
\bibitem [{\citenamefont {Akrami}\ \emph {et~al.}(2020{\natexlab{b}})\citenamefont {Akrami} \emph {et~al.}}]{Planck:2018jri}%
  \BibitemOpen
  \bibfield  {author} {\bibinfo {author} {\bibfnamefont {Y.}~\bibnamefont {Akrami}} \emph {et~al.} (\bibinfo {collaboration} {Planck}),\ }\bibfield  {title} {\enquote {\bibinfo {title} {{Planck 2018 results. X. Constraints on inflation}},}\ }\href {\doibase 10.1051/0004-6361/201833887} {\bibfield  {journal} {\bibinfo  {journal} {Astron. Astrophys.}\ }\textbf {\bibinfo {volume} {641}},\ \bibinfo {pages} {A10} (\bibinfo {year} {2020}{\natexlab{b}})},\ \Eprint {http://arxiv.org/abs/1807.06211} {arXiv:1807.06211 [astro-ph.CO]} \BibitemShut {NoStop}%
\bibitem [{\citenamefont {Ade}\ \emph {et~al.}(2021)\citenamefont {Ade} \emph {et~al.}}]{BICEP:2021xfz}%
  \BibitemOpen
  \bibfield  {author} {\bibinfo {author} {\bibfnamefont {P.~A.~R.}\ \bibnamefont {Ade}} \emph {et~al.} (\bibinfo {collaboration} {BICEP, Keck}),\ }\bibfield  {title} {\enquote {\bibinfo {title} {{Improved Constraints on Primordial Gravitational Waves using Planck, WMAP, and BICEP/Keck Observations through the 2018 Observing Season}},}\ }\href {\doibase 10.1103/PhysRevLett.127.151301} {\bibfield  {journal} {\bibinfo  {journal} {Phys. Rev. Lett.}\ }\textbf {\bibinfo {volume} {127}},\ \bibinfo {pages} {151301} (\bibinfo {year} {2021})},\ \Eprint {http://arxiv.org/abs/2110.00483} {arXiv:2110.00483 [astro-ph.CO]} \BibitemShut {NoStop}%
\bibitem [{\citenamefont {Branco}\ \emph {et~al.}(2012)\citenamefont {Branco}, \citenamefont {Ferreira}, \citenamefont {Lavoura}, \citenamefont {Rebelo}, \citenamefont {Sher},\ and\ \citenamefont {Silva}}]{Branco:2011iw}%
  \BibitemOpen
  \bibfield  {author} {\bibinfo {author} {\bibfnamefont {G.~C.}\ \bibnamefont {Branco}}, \bibinfo {author} {\bibfnamefont {P.~M.}\ \bibnamefont {Ferreira}}, \bibinfo {author} {\bibfnamefont {L.}~\bibnamefont {Lavoura}}, \bibinfo {author} {\bibfnamefont {M.~N.}\ \bibnamefont {Rebelo}}, \bibinfo {author} {\bibfnamefont {Marc}\ \bibnamefont {Sher}}, \ and\ \bibinfo {author} {\bibfnamefont {Joao~P.}\ \bibnamefont {Silva}},\ }\bibfield  {title} {\enquote {\bibinfo {title} {{Theory and phenomenology of two-Higgs-doublet models}},}\ }\href {\doibase 10.1016/j.physrep.2012.02.002} {\bibfield  {journal} {\bibinfo  {journal} {Phys. Rept.}\ }\textbf {\bibinfo {volume} {516}},\ \bibinfo {pages} {1--102} (\bibinfo {year} {2012})},\ \Eprint {http://arxiv.org/abs/1106.0034} {arXiv:1106.0034 [hep-ph]} \BibitemShut {NoStop}%
\bibitem [{\citenamefont {Lyth}\ and\ \citenamefont {Riotto}(1999)}]{Lyth:1998xn}%
  \BibitemOpen
  \bibfield  {author} {\bibinfo {author} {\bibfnamefont {David~H.}\ \bibnamefont {Lyth}}\ and\ \bibinfo {author} {\bibfnamefont {Antonio}\ \bibnamefont {Riotto}},\ }\bibfield  {title} {\enquote {\bibinfo {title} {{Particle physics models of inflation and the cosmological density perturbation}},}\ }\href {\doibase 10.1016/S0370-1573(98)00128-8} {\bibfield  {journal} {\bibinfo  {journal} {Phys. Rept.}\ }\textbf {\bibinfo {volume} {314}},\ \bibinfo {pages} {1--146} (\bibinfo {year} {1999})},\ \Eprint {http://arxiv.org/abs/hep-ph/9807278} {arXiv:hep-ph/9807278} \BibitemShut {NoStop}%
\bibitem [{\citenamefont {Mazumdar}\ and\ \citenamefont {Rocher}(2011)}]{Mazumdar:2010sa}%
  \BibitemOpen
  \bibfield  {author} {\bibinfo {author} {\bibfnamefont {Anupam}\ \bibnamefont {Mazumdar}}\ and\ \bibinfo {author} {\bibfnamefont {Jonathan}\ \bibnamefont {Rocher}},\ }\bibfield  {title} {\enquote {\bibinfo {title} {{Particle physics models of inflation and curvaton scenarios}},}\ }\href {\doibase 10.1016/j.physrep.2010.08.001} {\bibfield  {journal} {\bibinfo  {journal} {Phys. Rept.}\ }\textbf {\bibinfo {volume} {497}},\ \bibinfo {pages} {85--215} (\bibinfo {year} {2011})},\ \Eprint {http://arxiv.org/abs/1001.0993} {arXiv:1001.0993 [hep-ph]} \BibitemShut {NoStop}%
\bibitem [{\citenamefont {Fallon}\ \emph {et~al.}(2025)\citenamefont {Fallon}, \citenamefont {Halverson}, \citenamefont {McAllister},\ and\ \citenamefont {Zhu}}]{Fallon:2025lvn}%
  \BibitemOpen
  \bibfield  {author} {\bibinfo {author} {\bibfnamefont {Sebastian Vander~Ploeg}\ \bibnamefont {Fallon}}, \bibinfo {author} {\bibfnamefont {James}\ \bibnamefont {Halverson}}, \bibinfo {author} {\bibfnamefont {Liam}\ \bibnamefont {McAllister}}, \ and\ \bibinfo {author} {\bibfnamefont {Yunhao}\ \bibnamefont {Zhu}},\ }\bibfield  {title} {\enquote {\bibinfo {title} {{F-theory Axiverse}},}\ }\href@noop {} {\  (\bibinfo {year} {2025})},\ \Eprint {http://arxiv.org/abs/2511.20458} {arXiv:2511.20458 [hep-th]} \BibitemShut {NoStop}%
\bibitem [{\citenamefont {McDonough}\ \emph {et~al.}(2020)\citenamefont {McDonough}, \citenamefont {Guth},\ and\ \citenamefont {Kaiser}}]{McDonough:2020gmn}%
  \BibitemOpen
  \bibfield  {author} {\bibinfo {author} {\bibfnamefont {Evan}\ \bibnamefont {McDonough}}, \bibinfo {author} {\bibfnamefont {Alan~H.}\ \bibnamefont {Guth}}, \ and\ \bibinfo {author} {\bibfnamefont {David~I.}\ \bibnamefont {Kaiser}},\ }\bibfield  {title} {\enquote {\bibinfo {title} {{Nonminimal Couplings and the Forgotten Field of Axion Inflation}},}\ }\href@noop {} {\  (\bibinfo {year} {2020})},\ \Eprint {http://arxiv.org/abs/2010.04179} {arXiv:2010.04179 [hep-th]} \BibitemShut {NoStop}%
\bibitem [{\citenamefont {Maleknejad}\ and\ \citenamefont {McDonough}(2022)}]{Maleknejad:2022gyf}%
  \BibitemOpen
  \bibfield  {author} {\bibinfo {author} {\bibfnamefont {Azadeh}\ \bibnamefont {Maleknejad}}\ and\ \bibinfo {author} {\bibfnamefont {Evan}\ \bibnamefont {McDonough}},\ }\bibfield  {title} {\enquote {\bibinfo {title} {{Ultralight pion and superheavy baryon dark matter}},}\ }\href {\doibase 10.1103/PhysRevD.106.095011} {\bibfield  {journal} {\bibinfo  {journal} {Phys. Rev. D}\ }\textbf {\bibinfo {volume} {106}},\ \bibinfo {pages} {095011} (\bibinfo {year} {2022})},\ \Eprint {http://arxiv.org/abs/2205.12983} {arXiv:2205.12983 [hep-ph]} \BibitemShut {NoStop}%
\bibitem [{\citenamefont {Alexander}\ \emph {et~al.}(2023)\citenamefont {Alexander}, \citenamefont {Gilmer}, \citenamefont {Manton},\ and\ \citenamefont {McDonough}}]{Alexander:2023wgk}%
  \BibitemOpen
  \bibfield  {author} {\bibinfo {author} {\bibfnamefont {Stephon}\ \bibnamefont {Alexander}}, \bibinfo {author} {\bibfnamefont {Humberto}\ \bibnamefont {Gilmer}}, \bibinfo {author} {\bibfnamefont {Tucker}\ \bibnamefont {Manton}}, \ and\ \bibinfo {author} {\bibfnamefont {Evan}\ \bibnamefont {McDonough}},\ }\bibfield  {title} {\enquote {\bibinfo {title} {{\ensuremath{\pi}-axion and \ensuremath{\pi}-axiverse of dark QCD}},}\ }\href {\doibase 10.1103/PhysRevD.108.123014} {\bibfield  {journal} {\bibinfo  {journal} {Phys. Rev. D}\ }\textbf {\bibinfo {volume} {108}},\ \bibinfo {pages} {123014} (\bibinfo {year} {2023})},\ \Eprint {http://arxiv.org/abs/2304.11176} {arXiv:2304.11176 [hep-ph]} \BibitemShut {NoStop}%
\bibitem [{\citenamefont {Alexander}\ \emph {et~al.}(2024)\citenamefont {Alexander}, \citenamefont {Manton},\ and\ \citenamefont {McDonough}}]{Alexander:2024nvi}%
  \BibitemOpen
  \bibfield  {author} {\bibinfo {author} {\bibfnamefont {Stephon}\ \bibnamefont {Alexander}}, \bibinfo {author} {\bibfnamefont {Tucker}\ \bibnamefont {Manton}}, \ and\ \bibinfo {author} {\bibfnamefont {Evan}\ \bibnamefont {McDonough}},\ }\bibfield  {title} {\enquote {\bibinfo {title} {{Field theory axiverse}},}\ }\href {\doibase 10.1103/PhysRevD.109.116019} {\bibfield  {journal} {\bibinfo  {journal} {Phys. Rev. D}\ }\textbf {\bibinfo {volume} {109}},\ \bibinfo {pages} {116019} (\bibinfo {year} {2024})},\ \Eprint {http://arxiv.org/abs/2404.11642} {arXiv:2404.11642 [hep-ph]} \BibitemShut {NoStop}%
\bibitem [{\citenamefont {Randall}\ \emph {et~al.}(1996)\citenamefont {Randall}, \citenamefont {Soljacic},\ and\ \citenamefont {Guth}}]{Randall:1995dj}%
  \BibitemOpen
  \bibfield  {author} {\bibinfo {author} {\bibfnamefont {Lisa}\ \bibnamefont {Randall}}, \bibinfo {author} {\bibfnamefont {Marin}\ \bibnamefont {Soljacic}}, \ and\ \bibinfo {author} {\bibfnamefont {Alan~H.}\ \bibnamefont {Guth}},\ }\bibfield  {title} {\enquote {\bibinfo {title} {{Supernatural inflation: Inflation from supersymmetry with no (very) small parameters}},}\ }\href {\doibase 10.1016/0550-3213(96)00174-5} {\bibfield  {journal} {\bibinfo  {journal} {Nucl. Phys. B}\ }\textbf {\bibinfo {volume} {472}},\ \bibinfo {pages} {377--408} (\bibinfo {year} {1996})},\ \Eprint {http://arxiv.org/abs/hep-ph/9512439} {arXiv:hep-ph/9512439} \BibitemShut {NoStop}%
\bibitem [{\citenamefont {Garcia-Bellido}\ \emph {et~al.}(1996)\citenamefont {Garcia-Bellido}, \citenamefont {Linde},\ and\ \citenamefont {Wands}}]{Garcia-Bellido:1996mdl}%
  \BibitemOpen
  \bibfield  {author} {\bibinfo {author} {\bibfnamefont {Juan}\ \bibnamefont {Garcia-Bellido}}, \bibinfo {author} {\bibfnamefont {Andrei~D.}\ \bibnamefont {Linde}}, \ and\ \bibinfo {author} {\bibfnamefont {David}\ \bibnamefont {Wands}},\ }\bibfield  {title} {\enquote {\bibinfo {title} {{Density perturbations and black hole formation in hybrid inflation}},}\ }\href {\doibase 10.1103/PhysRevD.54.6040} {\bibfield  {journal} {\bibinfo  {journal} {Phys. Rev. D}\ }\textbf {\bibinfo {volume} {54}},\ \bibinfo {pages} {6040--6058} (\bibinfo {year} {1996})},\ \Eprint {http://arxiv.org/abs/astro-ph/9605094} {arXiv:astro-ph/9605094} \BibitemShut {NoStop}%
\bibitem [{\citenamefont {Lyth}(2011)}]{Lyth:2010zq}%
  \BibitemOpen
  \bibfield  {author} {\bibinfo {author} {\bibfnamefont {David~H.}\ \bibnamefont {Lyth}},\ }\bibfield  {title} {\enquote {\bibinfo {title} {{Contribution of the hybrid inflation waterfall to the primordial curvature perturbation}},}\ }\href {\doibase 10.1088/1475-7516/2011/07/035} {\bibfield  {journal} {\bibinfo  {journal} {JCAP}\ }\textbf {\bibinfo {volume} {07}},\ \bibinfo {pages} {035} (\bibinfo {year} {2011})},\ \Eprint {http://arxiv.org/abs/1012.4617} {arXiv:1012.4617 [astro-ph.CO]} \BibitemShut {NoStop}%
\bibitem [{\citenamefont {Bugaev}\ and\ \citenamefont {Klimai}(2012)}]{Bugaev:2011wy}%
  \BibitemOpen
  \bibfield  {author} {\bibinfo {author} {\bibfnamefont {Edgar}\ \bibnamefont {Bugaev}}\ and\ \bibinfo {author} {\bibfnamefont {Peter}\ \bibnamefont {Klimai}},\ }\bibfield  {title} {\enquote {\bibinfo {title} {{Formation of primordial black holes from non-Gaussian perturbations produced in a waterfall transition}},}\ }\href {\doibase 10.1103/PhysRevD.85.103504} {\bibfield  {journal} {\bibinfo  {journal} {Phys. Rev. D}\ }\textbf {\bibinfo {volume} {85}},\ \bibinfo {pages} {103504} (\bibinfo {year} {2012})},\ \Eprint {http://arxiv.org/abs/1112.5601} {arXiv:1112.5601 [astro-ph.CO]} \BibitemShut {NoStop}%
\bibitem [{\citenamefont {Halpern}\ \emph {et~al.}(2015)\citenamefont {Halpern}, \citenamefont {Hertzberg}, \citenamefont {Joss},\ and\ \citenamefont {Sfakianakis}}]{Halpern:2014mca}%
  \BibitemOpen
  \bibfield  {author} {\bibinfo {author} {\bibfnamefont {Illan~F.}\ \bibnamefont {Halpern}}, \bibinfo {author} {\bibfnamefont {Mark~P.}\ \bibnamefont {Hertzberg}}, \bibinfo {author} {\bibfnamefont {Matthew~A.}\ \bibnamefont {Joss}}, \ and\ \bibinfo {author} {\bibfnamefont {Evangelos~I.}\ \bibnamefont {Sfakianakis}},\ }\bibfield  {title} {\enquote {\bibinfo {title} {{A Density Spike on Astrophysical Scales from an N-Field Waterfall Transition}},}\ }\href {\doibase 10.1016/j.physletb.2015.06.076} {\bibfield  {journal} {\bibinfo  {journal} {Phys. Lett. B}\ }\textbf {\bibinfo {volume} {748}},\ \bibinfo {pages} {132--143} (\bibinfo {year} {2015})},\ \Eprint {http://arxiv.org/abs/1410.1878} {arXiv:1410.1878 [astro-ph.CO]} \BibitemShut {NoStop}%
\bibitem [{\citenamefont {Clesse}\ and\ \citenamefont {Garc\'\i{}a-Bellido}(2015)}]{Clesse:2015wea}%
  \BibitemOpen
  \bibfield  {author} {\bibinfo {author} {\bibfnamefont {S\'ebastien}\ \bibnamefont {Clesse}}\ and\ \bibinfo {author} {\bibfnamefont {Juan}\ \bibnamefont {Garc\'\i{}a-Bellido}},\ }\bibfield  {title} {\enquote {\bibinfo {title} {{Massive Primordial Black Holes from Hybrid Inflation as Dark Matter and the seeds of Galaxies}},}\ }\href {\doibase 10.1103/PhysRevD.92.023524} {\bibfield  {journal} {\bibinfo  {journal} {Phys. Rev. D}\ }\textbf {\bibinfo {volume} {92}},\ \bibinfo {pages} {023524} (\bibinfo {year} {2015})},\ \Eprint {http://arxiv.org/abs/1501.07565} {arXiv:1501.07565 [astro-ph.CO]} \BibitemShut {NoStop}%
\bibitem [{\citenamefont {Kawasaki}\ and\ \citenamefont {Tada}(2016)}]{Kawasaki:2015ppx}%
  \BibitemOpen
  \bibfield  {author} {\bibinfo {author} {\bibfnamefont {Masahiro}\ \bibnamefont {Kawasaki}}\ and\ \bibinfo {author} {\bibfnamefont {Yuichiro}\ \bibnamefont {Tada}},\ }\bibfield  {title} {\enquote {\bibinfo {title} {{Can massive primordial black holes be produced in mild waterfall hybrid inflation?}}}\ }\href {\doibase 10.1088/1475-7516/2016/08/041} {\bibfield  {journal} {\bibinfo  {journal} {JCAP}\ }\textbf {\bibinfo {volume} {08}},\ \bibinfo {pages} {041} (\bibinfo {year} {2016})},\ \Eprint {http://arxiv.org/abs/1512.03515} {arXiv:1512.03515 [astro-ph.CO]} \BibitemShut {NoStop}%
\bibitem [{\citenamefont {Braglia}\ \emph {et~al.}(2023)\citenamefont {Braglia}, \citenamefont {Linde}, \citenamefont {Kallosh},\ and\ \citenamefont {Finelli}}]{Braglia:2022phb}%
  \BibitemOpen
  \bibfield  {author} {\bibinfo {author} {\bibfnamefont {Matteo}\ \bibnamefont {Braglia}}, \bibinfo {author} {\bibfnamefont {Andrei}\ \bibnamefont {Linde}}, \bibinfo {author} {\bibfnamefont {Renata}\ \bibnamefont {Kallosh}}, \ and\ \bibinfo {author} {\bibfnamefont {Fabio}\ \bibnamefont {Finelli}},\ }\bibfield  {title} {\enquote {\bibinfo {title} {{Hybrid \ensuremath{\alpha}-attractors, primordial black holes and gravitational wave backgrounds}},}\ }\href {\doibase 10.1088/1475-7516/2023/04/033} {\bibfield  {journal} {\bibinfo  {journal} {JCAP}\ }\textbf {\bibinfo {volume} {04}},\ \bibinfo {pages} {033} (\bibinfo {year} {2023})},\ \Eprint {http://arxiv.org/abs/2211.14262} {arXiv:2211.14262 [astro-ph.CO]} \BibitemShut {NoStop}%
\bibitem [{\citenamefont {Fumagalli}\ \emph {et~al.}(2023)\citenamefont {Fumagalli}, \citenamefont {Renaux-Petel}, \citenamefont {Ronayne},\ and\ \citenamefont {Witkowski}}]{Fumagalli:2020adf}%
  \BibitemOpen
  \bibfield  {author} {\bibinfo {author} {\bibfnamefont {Jacopo}\ \bibnamefont {Fumagalli}}, \bibinfo {author} {\bibfnamefont {S\'ebastien}\ \bibnamefont {Renaux-Petel}}, \bibinfo {author} {\bibfnamefont {John~W.}\ \bibnamefont {Ronayne}}, \ and\ \bibinfo {author} {\bibfnamefont {Lukas~T.}\ \bibnamefont {Witkowski}},\ }\bibfield  {title} {\enquote {\bibinfo {title} {{Turning in the landscape: A new mechanism for generating primordial black holes}},}\ }\href {\doibase 10.1016/j.physletb.2023.137921} {\bibfield  {journal} {\bibinfo  {journal} {Phys. Lett. B}\ }\textbf {\bibinfo {volume} {841}},\ \bibinfo {pages} {137921} (\bibinfo {year} {2023})},\ \Eprint {http://arxiv.org/abs/2004.08369} {arXiv:2004.08369 [hep-th]} \BibitemShut {NoStop}%
\bibitem [{\citenamefont {Braglia}\ \emph {et~al.}(2020)\citenamefont {Braglia}, \citenamefont {Hazra}, \citenamefont {Finelli}, \citenamefont {Smoot}, \citenamefont {Sriramkumar},\ and\ \citenamefont {Starobinsky}}]{Braglia:2020eai}%
  \BibitemOpen
  \bibfield  {author} {\bibinfo {author} {\bibfnamefont {Matteo}\ \bibnamefont {Braglia}}, \bibinfo {author} {\bibfnamefont {Dhiraj~Kumar}\ \bibnamefont {Hazra}}, \bibinfo {author} {\bibfnamefont {Fabio}\ \bibnamefont {Finelli}}, \bibinfo {author} {\bibfnamefont {George~F.}\ \bibnamefont {Smoot}}, \bibinfo {author} {\bibfnamefont {L.}~\bibnamefont {Sriramkumar}}, \ and\ \bibinfo {author} {\bibfnamefont {Alexei~A.}\ \bibnamefont {Starobinsky}},\ }\bibfield  {title} {\enquote {\bibinfo {title} {{Generating PBHs and small-scale GWs in two-field models of inflation}},}\ }\href {\doibase 10.1088/1475-7516/2020/08/001} {\bibfield  {journal} {\bibinfo  {journal} {JCAP}\ }\textbf {\bibinfo {volume} {08}},\ \bibinfo {pages} {001} (\bibinfo {year} {2020})},\ \Eprint {http://arxiv.org/abs/2005.02895} {arXiv:2005.02895 [astro-ph.CO]} \BibitemShut {NoStop}%
\bibitem [{\citenamefont {Palma}\ \emph {et~al.}(2020)\citenamefont {Palma}, \citenamefont {Sypsas},\ and\ \citenamefont {Zenteno}}]{Palma:2020ejf}%
  \BibitemOpen
  \bibfield  {author} {\bibinfo {author} {\bibfnamefont {Gonzalo~A.}\ \bibnamefont {Palma}}, \bibinfo {author} {\bibfnamefont {Spyros}\ \bibnamefont {Sypsas}}, \ and\ \bibinfo {author} {\bibfnamefont {Cristobal}\ \bibnamefont {Zenteno}},\ }\bibfield  {title} {\enquote {\bibinfo {title} {{Seeding primordial black holes in multifield inflation}},}\ }\href {\doibase 10.1103/PhysRevLett.125.121301} {\bibfield  {journal} {\bibinfo  {journal} {Phys. Rev. Lett.}\ }\textbf {\bibinfo {volume} {125}},\ \bibinfo {pages} {121301} (\bibinfo {year} {2020})},\ \Eprint {http://arxiv.org/abs/2004.06106} {arXiv:2004.06106 [astro-ph.CO]} \BibitemShut {NoStop}%
\bibitem [{\citenamefont {Geller}\ \emph {et~al.}(2022)\citenamefont {Geller}, \citenamefont {Qin}, \citenamefont {McDonough},\ and\ \citenamefont {Kaiser}}]{Geller:2022nkr}%
  \BibitemOpen
  \bibfield  {author} {\bibinfo {author} {\bibfnamefont {Sarah~R.}\ \bibnamefont {Geller}}, \bibinfo {author} {\bibfnamefont {Wenzer}\ \bibnamefont {Qin}}, \bibinfo {author} {\bibfnamefont {Evan}\ \bibnamefont {McDonough}}, \ and\ \bibinfo {author} {\bibfnamefont {David~I.}\ \bibnamefont {Kaiser}},\ }\bibfield  {title} {\enquote {\bibinfo {title} {{Primordial black holes from multifield inflation with nonminimal couplings}},}\ }\href {\doibase 10.1103/PhysRevD.106.063535} {\bibfield  {journal} {\bibinfo  {journal} {Phys. Rev. D}\ }\textbf {\bibinfo {volume} {106}},\ \bibinfo {pages} {063535} (\bibinfo {year} {2022})},\ \Eprint {http://arxiv.org/abs/2205.04471} {arXiv:2205.04471 [hep-th]} \BibitemShut {NoStop}%
\bibitem [{\citenamefont {Ferraz}\ and\ \citenamefont {Rosa}(2025)}]{Ferraz:2024bvd}%
  \BibitemOpen
  \bibfield  {author} {\bibinfo {author} {\bibfnamefont {Paulo~B.}\ \bibnamefont {Ferraz}}\ and\ \bibinfo {author} {\bibfnamefont {Jo{\~a}o~G.}\ \bibnamefont {Rosa}},\ }\bibfield  {title} {\enquote {\bibinfo {title} {{The inflation trilogy and primordial black holes}},}\ }\href {\doibase 10.1088/1475-7516/2025/03/040} {\bibfield  {journal} {\bibinfo  {journal} {JCAP}\ }\textbf {\bibinfo {volume} {03}},\ \bibinfo {pages} {040} (\bibinfo {year} {2025})},\ \Eprint {http://arxiv.org/abs/2410.10996} {arXiv:2410.10996 [hep-ph]} \BibitemShut {NoStop}%
\bibitem [{\citenamefont {Bhattacharya}\ and\ \citenamefont {Zavala}(2023)}]{Bhattacharya:2022fze}%
  \BibitemOpen
  \bibfield  {author} {\bibinfo {author} {\bibfnamefont {Sukannya}\ \bibnamefont {Bhattacharya}}\ and\ \bibinfo {author} {\bibfnamefont {Ivonne}\ \bibnamefont {Zavala}},\ }\bibfield  {title} {\enquote {\bibinfo {title} {{Sharp turns in axion monodromy: primordial black holes and gravitational waves}},}\ }\href {\doibase 10.1088/1475-7516/2023/04/065} {\bibfield  {journal} {\bibinfo  {journal} {JCAP}\ }\textbf {\bibinfo {volume} {04}},\ \bibinfo {pages} {065} (\bibinfo {year} {2023})},\ \Eprint {http://arxiv.org/abs/2205.06065} {arXiv:2205.06065 [astro-ph.CO]} \BibitemShut {NoStop}%
\bibitem [{\citenamefont {Cai}\ \emph {et~al.}(2021)\citenamefont {Cai}, \citenamefont {Chen},\ and\ \citenamefont {Fu}}]{Cai:2021wzd}%
  \BibitemOpen
  \bibfield  {author} {\bibinfo {author} {\bibfnamefont {Rong-Gen}\ \bibnamefont {Cai}}, \bibinfo {author} {\bibfnamefont {Chao}\ \bibnamefont {Chen}}, \ and\ \bibinfo {author} {\bibfnamefont {Chengjie}\ \bibnamefont {Fu}},\ }\bibfield  {title} {\enquote {\bibinfo {title} {{Primordial black holes and stochastic gravitational wave background from inflation with a noncanonical spectator field}},}\ }\href {\doibase 10.1103/PhysRevD.104.083537} {\bibfield  {journal} {\bibinfo  {journal} {Phys. Rev. D}\ }\textbf {\bibinfo {volume} {104}},\ \bibinfo {pages} {083537} (\bibinfo {year} {2021})},\ \Eprint {http://arxiv.org/abs/2108.03422} {arXiv:2108.03422 [astro-ph.CO]} \BibitemShut {NoStop}%
\bibitem [{\citenamefont {Pi}\ \emph {et~al.}(2018)\citenamefont {Pi}, \citenamefont {Zhang}, \citenamefont {Huang},\ and\ \citenamefont {Sasaki}}]{Pi:2017gih}%
  \BibitemOpen
  \bibfield  {author} {\bibinfo {author} {\bibfnamefont {Shi}\ \bibnamefont {Pi}}, \bibinfo {author} {\bibfnamefont {Ying-li}\ \bibnamefont {Zhang}}, \bibinfo {author} {\bibfnamefont {Qing-Guo}\ \bibnamefont {Huang}}, \ and\ \bibinfo {author} {\bibfnamefont {Misao}\ \bibnamefont {Sasaki}},\ }\bibfield  {title} {\enquote {\bibinfo {title} {{Scalaron from $R^2$-gravity as a heavy field}},}\ }\href {\doibase 10.1088/1475-7516/2018/05/042} {\bibfield  {journal} {\bibinfo  {journal} {JCAP}\ }\textbf {\bibinfo {volume} {05}},\ \bibinfo {pages} {042} (\bibinfo {year} {2018})},\ \Eprint {http://arxiv.org/abs/1712.09896} {arXiv:1712.09896 [astro-ph.CO]} \BibitemShut {NoStop}%
\bibitem [{\citenamefont {Chen}\ \emph {et~al.}(2018)\citenamefont {Chen}, \citenamefont {Palma}, \citenamefont {Riquelme}, \citenamefont {Scheihing~Hitschfeld},\ and\ \citenamefont {Sypsas}}]{Chen:2018uul}%
  \BibitemOpen
  \bibfield  {author} {\bibinfo {author} {\bibfnamefont {Xingang}\ \bibnamefont {Chen}}, \bibinfo {author} {\bibfnamefont {Gonzalo~A.}\ \bibnamefont {Palma}}, \bibinfo {author} {\bibfnamefont {Walter}\ \bibnamefont {Riquelme}}, \bibinfo {author} {\bibfnamefont {Bruno}\ \bibnamefont {Scheihing~Hitschfeld}}, \ and\ \bibinfo {author} {\bibfnamefont {Spyros}\ \bibnamefont {Sypsas}},\ }\bibfield  {title} {\enquote {\bibinfo {title} {{Landscape tomography through primordial non-Gaussianity}},}\ }\href {\doibase 10.1103/PhysRevD.98.083528} {\bibfield  {journal} {\bibinfo  {journal} {Phys. Rev. D}\ }\textbf {\bibinfo {volume} {98}},\ \bibinfo {pages} {083528} (\bibinfo {year} {2018})},\ \Eprint {http://arxiv.org/abs/1804.07315} {arXiv:1804.07315 [hep-th]} \BibitemShut {NoStop}%
\bibitem [{\citenamefont {Dimastrogiovanni}\ \emph {et~al.}(2017)\citenamefont {Dimastrogiovanni}, \citenamefont {Fasiello},\ and\ \citenamefont {Fujita}}]{Dimastrogiovanni:2016fuu}%
  \BibitemOpen
  \bibfield  {author} {\bibinfo {author} {\bibfnamefont {Emanuela}\ \bibnamefont {Dimastrogiovanni}}, \bibinfo {author} {\bibfnamefont {Matteo}\ \bibnamefont {Fasiello}}, \ and\ \bibinfo {author} {\bibfnamefont {Tomohiro}\ \bibnamefont {Fujita}},\ }\bibfield  {title} {\enquote {\bibinfo {title} {{Primordial Gravitational Waves from Axion-Gauge Fields Dynamics}},}\ }\href {\doibase 10.1088/1475-7516/2017/01/019} {\bibfield  {journal} {\bibinfo  {journal} {JCAP}\ }\textbf {\bibinfo {volume} {01}},\ \bibinfo {pages} {019} (\bibinfo {year} {2017})},\ \Eprint {http://arxiv.org/abs/1608.04216} {arXiv:1608.04216 [astro-ph.CO]} \BibitemShut {NoStop}%
\bibitem [{\citenamefont {McDonough}\ and\ \citenamefont {Alexander}(2018)}]{McDonough:2018xzh}%
  \BibitemOpen
  \bibfield  {author} {\bibinfo {author} {\bibfnamefont {Evan}\ \bibnamefont {McDonough}}\ and\ \bibinfo {author} {\bibfnamefont {Stephon}\ \bibnamefont {Alexander}},\ }\bibfield  {title} {\enquote {\bibinfo {title} {{Observable Chiral Gravitational Waves from Inflation in String Theory}},}\ }\href {\doibase 10.1088/1475-7516/2018/11/030} {\bibfield  {journal} {\bibinfo  {journal} {JCAP}\ }\textbf {\bibinfo {volume} {11}},\ \bibinfo {pages} {030} (\bibinfo {year} {2018})},\ \Eprint {http://arxiv.org/abs/1806.05684} {arXiv:1806.05684 [hep-th]} \BibitemShut {NoStop}%
\bibitem [{\citenamefont {Holland}\ \emph {et~al.}(2020)\citenamefont {Holland}, \citenamefont {Zavala},\ and\ \citenamefont {Tasinato}}]{Holland:2020jdh}%
  \BibitemOpen
  \bibfield  {author} {\bibinfo {author} {\bibfnamefont {Jonathan}\ \bibnamefont {Holland}}, \bibinfo {author} {\bibfnamefont {Ivonne}\ \bibnamefont {Zavala}}, \ and\ \bibinfo {author} {\bibfnamefont {Gianmassimo}\ \bibnamefont {Tasinato}},\ }\bibfield  {title} {\enquote {\bibinfo {title} {{On chromonatural inflation in string theory}},}\ }\href {\doibase 10.1088/1475-7516/2020/12/026} {\bibfield  {journal} {\bibinfo  {journal} {JCAP}\ }\textbf {\bibinfo {volume} {12}},\ \bibinfo {pages} {026} (\bibinfo {year} {2020})},\ \Eprint {http://arxiv.org/abs/2009.00653} {arXiv:2009.00653 [hep-th]} \BibitemShut {NoStop}%
\bibitem [{\citenamefont {Alexander}\ \emph {et~al.}(2018)\citenamefont {Alexander}, \citenamefont {McDonough},\ and\ \citenamefont {Spergel}}]{Alexander:2018fjp}%
  \BibitemOpen
  \bibfield  {author} {\bibinfo {author} {\bibfnamefont {Stephon}\ \bibnamefont {Alexander}}, \bibinfo {author} {\bibfnamefont {Evan}\ \bibnamefont {McDonough}}, \ and\ \bibinfo {author} {\bibfnamefont {David~N.}\ \bibnamefont {Spergel}},\ }\bibfield  {title} {\enquote {\bibinfo {title} {{Chiral Gravitational Waves and Baryon Superfluid Dark Matter}},}\ }\href {\doibase 10.1088/1475-7516/2018/05/003} {\bibfield  {journal} {\bibinfo  {journal} {JCAP}\ }\textbf {\bibinfo {volume} {05}},\ \bibinfo {pages} {003} (\bibinfo {year} {2018})},\ \Eprint {http://arxiv.org/abs/1801.07255} {arXiv:1801.07255 [hep-th]} \BibitemShut {NoStop}%
\bibitem [{\citenamefont {Lorenzoni}\ \emph {et~al.}(2026)\citenamefont {Lorenzoni}, \citenamefont {Geller}, \citenamefont {Ireland}, \citenamefont {Kaiser}, \citenamefont {McDonough},\ and\ \citenamefont {Wittmeier}}]{Lorenzoni:2025gni}%
  \BibitemOpen
  \bibfield  {author} {\bibinfo {author} {\bibfnamefont {Dario~L.}\ \bibnamefont {Lorenzoni}}, \bibinfo {author} {\bibfnamefont {Sarah~R.}\ \bibnamefont {Geller}}, \bibinfo {author} {\bibfnamefont {Zachary}\ \bibnamefont {Ireland}}, \bibinfo {author} {\bibfnamefont {David~I.}\ \bibnamefont {Kaiser}}, \bibinfo {author} {\bibfnamefont {Evan}\ \bibnamefont {McDonough}}, \ and\ \bibinfo {author} {\bibfnamefont {Kyle~A.}\ \bibnamefont {Wittmeier}},\ }\bibfield  {title} {\enquote {\bibinfo {title} {{Resilience of primordial black hole formation from inflation with a light spectator field}},}\ }\href {\doibase 10.1103/gwfm-jx9r} {\bibfield  {journal} {\bibinfo  {journal} {Phys. Rev. D}\ }\textbf {\bibinfo {volume} {114}},\ \bibinfo {pages} {083528} (\bibinfo {year} {2026})},\ \Eprint {http://arxiv.org/abs/2504.13251} {arXiv:2504.13251 [astro-ph.CO]} \BibitemShut {NoStop}%
\bibitem [{\citenamefont {Kachru}\ \emph {et~al.}(2003)\citenamefont {Kachru}, \citenamefont {Kallosh}, \citenamefont {Linde},\ and\ \citenamefont {Trivedi}}]{Kachru:2003aw}%
  \BibitemOpen
  \bibfield  {author} {\bibinfo {author} {\bibfnamefont {Shamit}\ \bibnamefont {Kachru}}, \bibinfo {author} {\bibfnamefont {Renata}\ \bibnamefont {Kallosh}}, \bibinfo {author} {\bibfnamefont {Andrei~D.}\ \bibnamefont {Linde}}, \ and\ \bibinfo {author} {\bibfnamefont {Sandip~P.}\ \bibnamefont {Trivedi}},\ }\bibfield  {title} {\enquote {\bibinfo {title} {{De Sitter vacua in string theory}},}\ }\href {\doibase 10.1103/PhysRevD.68.046005} {\bibfield  {journal} {\bibinfo  {journal} {Phys. Rev. D}\ }\textbf {\bibinfo {volume} {68}},\ \bibinfo {pages} {046005} (\bibinfo {year} {2003})},\ \Eprint {http://arxiv.org/abs/hep-th/0301240} {arXiv:hep-th/0301240} \BibitemShut {NoStop}%
\bibitem [{\citenamefont {Sasaki}\ and\ \citenamefont {Stewart}(1996)}]{Sasaki:1995aw}%
  \BibitemOpen
  \bibfield  {author} {\bibinfo {author} {\bibfnamefont {Misao}\ \bibnamefont {Sasaki}}\ and\ \bibinfo {author} {\bibfnamefont {Ewan~D.}\ \bibnamefont {Stewart}},\ }\bibfield  {title} {\enquote {\bibinfo {title} {{A General analytic formula for the spectral index of the density perturbations produced during inflation}},}\ }\href {\doibase 10.1143/PTP.95.71} {\bibfield  {journal} {\bibinfo  {journal} {Prog. Theor. Phys.}\ }\textbf {\bibinfo {volume} {95}},\ \bibinfo {pages} {71--78} (\bibinfo {year} {1996})},\ \Eprint {http://arxiv.org/abs/astro-ph/9507001} {arXiv:astro-ph/9507001} \BibitemShut {NoStop}%
\bibitem [{\citenamefont {Gordon}\ \emph {et~al.}(2000)\citenamefont {Gordon}, \citenamefont {Wands}, \citenamefont {Bassett},\ and\ \citenamefont {Maartens}}]{Gordon:2000hv}%
  \BibitemOpen
  \bibfield  {author} {\bibinfo {author} {\bibfnamefont {Christopher}\ \bibnamefont {Gordon}}, \bibinfo {author} {\bibfnamefont {David}\ \bibnamefont {Wands}}, \bibinfo {author} {\bibfnamefont {Bruce~A.}\ \bibnamefont {Bassett}}, \ and\ \bibinfo {author} {\bibfnamefont {Roy}\ \bibnamefont {Maartens}},\ }\bibfield  {title} {\enquote {\bibinfo {title} {{Adiabatic and entropy perturbations from inflation}},}\ }\href {\doibase 10.1103/PhysRevD.63.023506} {\bibfield  {journal} {\bibinfo  {journal} {Phys. Rev. D}\ }\textbf {\bibinfo {volume} {63}},\ \bibinfo {pages} {023506} (\bibinfo {year} {2000})},\ \Eprint {http://arxiv.org/abs/astro-ph/0009131} {arXiv:astro-ph/0009131} \BibitemShut {NoStop}%
\bibitem [{\citenamefont {Wands}\ \emph {et~al.}(2002)\citenamefont {Wands}, \citenamefont {Bartolo}, \citenamefont {Matarrese},\ and\ \citenamefont {Riotto}}]{Wands:2002bn}%
  \BibitemOpen
  \bibfield  {author} {\bibinfo {author} {\bibfnamefont {David}\ \bibnamefont {Wands}}, \bibinfo {author} {\bibfnamefont {Nicola}\ \bibnamefont {Bartolo}}, \bibinfo {author} {\bibfnamefont {Sabino}\ \bibnamefont {Matarrese}}, \ and\ \bibinfo {author} {\bibfnamefont {Antonio}\ \bibnamefont {Riotto}},\ }\bibfield  {title} {\enquote {\bibinfo {title} {{An Observational test of two-field inflation}},}\ }\href {\doibase 10.1103/PhysRevD.66.043520} {\bibfield  {journal} {\bibinfo  {journal} {Phys. Rev. D}\ }\textbf {\bibinfo {volume} {66}},\ \bibinfo {pages} {043520} (\bibinfo {year} {2002})},\ \Eprint {http://arxiv.org/abs/astro-ph/0205253} {arXiv:astro-ph/0205253} \BibitemShut {NoStop}%
\bibitem [{\citenamefont {Langlois}\ and\ \citenamefont {Renaux-Petel}(2008)}]{Langlois:2008mn}%
  \BibitemOpen
  \bibfield  {author} {\bibinfo {author} {\bibfnamefont {David}\ \bibnamefont {Langlois}}\ and\ \bibinfo {author} {\bibfnamefont {Sebastien}\ \bibnamefont {Renaux-Petel}},\ }\bibfield  {title} {\enquote {\bibinfo {title} {{Perturbations in generalized multi-field inflation}},}\ }\href {\doibase 10.1088/1475-7516/2008/04/017} {\bibfield  {journal} {\bibinfo  {journal} {JCAP}\ }\textbf {\bibinfo {volume} {04}},\ \bibinfo {pages} {017} (\bibinfo {year} {2008})},\ \Eprint {http://arxiv.org/abs/0801.1085} {arXiv:0801.1085 [hep-th]} \BibitemShut {NoStop}%
\bibitem [{\citenamefont {Peterson}\ and\ \citenamefont {Tegmark}(2011)}]{Peterson:2010np}%
  \BibitemOpen
  \bibfield  {author} {\bibinfo {author} {\bibfnamefont {Courtney~M.}\ \bibnamefont {Peterson}}\ and\ \bibinfo {author} {\bibfnamefont {Max}\ \bibnamefont {Tegmark}},\ }\bibfield  {title} {\enquote {\bibinfo {title} {{Testing Two-Field Inflation}},}\ }\href {\doibase 10.1103/PhysRevD.83.023522} {\bibfield  {journal} {\bibinfo  {journal} {Phys. Rev. D}\ }\textbf {\bibinfo {volume} {83}},\ \bibinfo {pages} {023522} (\bibinfo {year} {2011})},\ \Eprint {http://arxiv.org/abs/1005.4056} {arXiv:1005.4056 [astro-ph.CO]} \BibitemShut {NoStop}%
\bibitem [{\citenamefont {Gong}\ and\ \citenamefont {Tanaka}(2011)}]{Gong:2011uw}%
  \BibitemOpen
  \bibfield  {author} {\bibinfo {author} {\bibfnamefont {Jinn-Ouk}\ \bibnamefont {Gong}}\ and\ \bibinfo {author} {\bibfnamefont {Takahiro}\ \bibnamefont {Tanaka}},\ }\bibfield  {title} {\enquote {\bibinfo {title} {{A covariant approach to general field space metric in multi-field inflation}},}\ }\href {\doibase 10.1088/1475-7516/2012/02/E01} {\bibfield  {journal} {\bibinfo  {journal} {JCAP}\ }\textbf {\bibinfo {volume} {03}},\ \bibinfo {pages} {015} (\bibinfo {year} {2011})},\ \bibinfo {note} {[Erratum: JCAP 02, E01 (2012)]},\ \Eprint {http://arxiv.org/abs/1101.4809} {arXiv:1101.4809 [astro-ph.CO]} \BibitemShut {NoStop}%
\bibitem [{\citenamefont {Kaiser}\ \emph {et~al.}(2013)\citenamefont {Kaiser}, \citenamefont {Mazenc},\ and\ \citenamefont {Sfakianakis}}]{Kaiser:2012ak}%
  \BibitemOpen
  \bibfield  {author} {\bibinfo {author} {\bibfnamefont {David~I.}\ \bibnamefont {Kaiser}}, \bibinfo {author} {\bibfnamefont {Edward~A.}\ \bibnamefont {Mazenc}}, \ and\ \bibinfo {author} {\bibfnamefont {Evangelos~I.}\ \bibnamefont {Sfakianakis}},\ }\bibfield  {title} {\enquote {\bibinfo {title} {{Primordial Bispectrum from Multifield Inflation with Nonminimal Couplings}},}\ }\href {\doibase 10.1103/PhysRevD.87.064004} {\bibfield  {journal} {\bibinfo  {journal} {Phys. Rev. D}\ }\textbf {\bibinfo {volume} {87}},\ \bibinfo {pages} {064004} (\bibinfo {year} {2013})},\ \Eprint {http://arxiv.org/abs/1210.7487} {arXiv:1210.7487 [astro-ph.CO]} \BibitemShut {NoStop}%
\bibitem [{\citenamefont {Gong}(2016)}]{Gong:2016qmq}%
  \BibitemOpen
  \bibfield  {author} {\bibinfo {author} {\bibfnamefont {Jinn-Ouk}\ \bibnamefont {Gong}},\ }\bibfield  {title} {\enquote {\bibinfo {title} {{Multi-field inflation and cosmological perturbations}},}\ }\href {\doibase 10.1142/S021827181740003X} {\bibfield  {journal} {\bibinfo  {journal} {Int. J. Mod. Phys. D}\ }\textbf {\bibinfo {volume} {26}},\ \bibinfo {pages} {1740003} (\bibinfo {year} {2016})},\ \Eprint {http://arxiv.org/abs/1606.06971} {arXiv:1606.06971 [gr-qc]} \BibitemShut {NoStop}%
\bibitem [{\citenamefont {Kaiser}\ and\ \citenamefont {Sfakianakis}(2014)}]{Kaiser:2013sna}%
  \BibitemOpen
  \bibfield  {author} {\bibinfo {author} {\bibfnamefont {David~I.}\ \bibnamefont {Kaiser}}\ and\ \bibinfo {author} {\bibfnamefont {Evangelos~I.}\ \bibnamefont {Sfakianakis}},\ }\bibfield  {title} {\enquote {\bibinfo {title} {{Multifield Inflation after Planck: The Case for Nonminimal Couplings}},}\ }\href {\doibase 10.1103/PhysRevLett.112.011302} {\bibfield  {journal} {\bibinfo  {journal} {Phys. Rev. Lett.}\ }\textbf {\bibinfo {volume} {112}},\ \bibinfo {pages} {011302} (\bibinfo {year} {2014})},\ \Eprint {http://arxiv.org/abs/1304.0363} {arXiv:1304.0363 [astro-ph.CO]} \BibitemShut {NoStop}%
\bibitem [{\citenamefont {Schutz}\ \emph {et~al.}(2014)\citenamefont {Schutz}, \citenamefont {Sfakianakis},\ and\ \citenamefont {Kaiser}}]{Schutz:2013fua}%
  \BibitemOpen
  \bibfield  {author} {\bibinfo {author} {\bibfnamefont {Katelin}\ \bibnamefont {Schutz}}, \bibinfo {author} {\bibfnamefont {Evangelos~I.}\ \bibnamefont {Sfakianakis}}, \ and\ \bibinfo {author} {\bibfnamefont {David~I.}\ \bibnamefont {Kaiser}},\ }\bibfield  {title} {\enquote {\bibinfo {title} {{Multifield Inflation after Planck: Isocurvature Modes from Nonminimal Couplings}},}\ }\href {\doibase 10.1103/PhysRevD.89.064044} {\bibfield  {journal} {\bibinfo  {journal} {Phys. Rev. D}\ }\textbf {\bibinfo {volume} {89}},\ \bibinfo {pages} {064044} (\bibinfo {year} {2014})},\ \Eprint {http://arxiv.org/abs/1310.8285} {arXiv:1310.8285 [astro-ph.CO]} \BibitemShut {NoStop}%
\bibitem [{COD()}]{CODEnote}%
  \BibitemOpen
  \href@noop {} {}\bibinfo {note} {Code available at the public GitHub repository \href{https://github.com/dario-lorenzoni/PBHspec}{https://github.com/dario-lorenzoni/PBHspec} (Dario L. Lorenzoni, 2025)}\BibitemShut {NoStop}%
\bibitem [{\citenamefont {Dias}\ \emph {et~al.}(2016)\citenamefont {Dias}, \citenamefont {Frazer}, \citenamefont {Mulryne},\ and\ \citenamefont {Seery}}]{Dias:2016rjq}%
  \BibitemOpen
  \bibfield  {author} {\bibinfo {author} {\bibfnamefont {Mafalda}\ \bibnamefont {Dias}}, \bibinfo {author} {\bibfnamefont {Jonathan}\ \bibnamefont {Frazer}}, \bibinfo {author} {\bibfnamefont {David~J.}\ \bibnamefont {Mulryne}}, \ and\ \bibinfo {author} {\bibfnamefont {David}\ \bibnamefont {Seery}},\ }\bibfield  {title} {\enquote {\bibinfo {title} {{Numerical evaluation of the bispectrum in multiple field inflation\textemdash{}the transport approach with code}},}\ }\href {\doibase 10.1088/1475-7516/2016/12/033} {\bibfield  {journal} {\bibinfo  {journal} {JCAP}\ }\textbf {\bibinfo {volume} {12}},\ \bibinfo {pages} {033} (\bibinfo {year} {2016})},\ \Eprint {http://arxiv.org/abs/1609.00379} {arXiv:1609.00379 [astro-ph.CO]} \BibitemShut {NoStop}%
\bibitem [{\citenamefont {Mulryne}\ and\ \citenamefont {Ronayne}(2018)}]{Mulryne:2016mzv}%
  \BibitemOpen
  \bibfield  {author} {\bibinfo {author} {\bibfnamefont {David~J.}\ \bibnamefont {Mulryne}}\ and\ \bibinfo {author} {\bibfnamefont {John~W.}\ \bibnamefont {Ronayne}},\ }\bibfield  {title} {\enquote {\bibinfo {title} {{PyTransport: A Python package for the calculation of inflationary correlation functions}},}\ }\href {\doibase 10.21105/joss.00494} {\bibfield  {journal} {\bibinfo  {journal} {J. Open Source Softw.}\ }\textbf {\bibinfo {volume} {3}},\ \bibinfo {pages} {494} (\bibinfo {year} {2018})},\ \Eprint {http://arxiv.org/abs/1609.00381} {arXiv:1609.00381 [astro-ph.CO]} \BibitemShut {NoStop}%
\bibitem [{\citenamefont {Dias}\ \emph {et~al.}(2015)\citenamefont {Dias}, \citenamefont {Frazer},\ and\ \citenamefont {Seery}}]{Dias:2015rca}%
  \BibitemOpen
  \bibfield  {author} {\bibinfo {author} {\bibfnamefont {Mafalda}\ \bibnamefont {Dias}}, \bibinfo {author} {\bibfnamefont {Jonathan}\ \bibnamefont {Frazer}}, \ and\ \bibinfo {author} {\bibfnamefont {David}\ \bibnamefont {Seery}},\ }\bibfield  {title} {\enquote {\bibinfo {title} {{Computing observables in curved multifield models of inflation{\textemdash}A guide (with code) to the transport method}},}\ }\href {\doibase 10.1088/1475-7516/2015/12/030} {\bibfield  {journal} {\bibinfo  {journal} {JCAP}\ }\textbf {\bibinfo {volume} {12}},\ \bibinfo {pages} {030} (\bibinfo {year} {2015})},\ \Eprint {http://arxiv.org/abs/1502.03125} {arXiv:1502.03125 [astro-ph.CO]} \BibitemShut {NoStop}%
\bibitem [{\citenamefont {Achucarro}\ \emph {et~al.}(2011)\citenamefont {Achucarro}, \citenamefont {Gong}, \citenamefont {Hardeman}, \citenamefont {Palma},\ and\ \citenamefont {Patil}}]{Achucarro:2010da}%
  \BibitemOpen
  \bibfield  {author} {\bibinfo {author} {\bibfnamefont {Ana}\ \bibnamefont {Achucarro}}, \bibinfo {author} {\bibfnamefont {Jinn-Ouk}\ \bibnamefont {Gong}}, \bibinfo {author} {\bibfnamefont {Sjoerd}\ \bibnamefont {Hardeman}}, \bibinfo {author} {\bibfnamefont {Gonzalo~A.}\ \bibnamefont {Palma}}, \ and\ \bibinfo {author} {\bibfnamefont {Subodh~P.}\ \bibnamefont {Patil}},\ }\bibfield  {title} {\enquote {\bibinfo {title} {{Features of heavy physics in the CMB power spectrum}},}\ }\href {\doibase 10.1088/1475-7516/2011/01/030} {\bibfield  {journal} {\bibinfo  {journal} {JCAP}\ }\textbf {\bibinfo {volume} {01}},\ \bibinfo {pages} {030} (\bibinfo {year} {2011})},\ \Eprint {http://arxiv.org/abs/1010.3693} {arXiv:1010.3693 [hep-ph]} \BibitemShut {NoStop}%
\bibitem [{\citenamefont {Byrnes}\ \emph {et~al.}(2019)\citenamefont {Byrnes}, \citenamefont {Cole},\ and\ \citenamefont {Patil}}]{Byrnes:2018txb}%
  \BibitemOpen
  \bibfield  {author} {\bibinfo {author} {\bibfnamefont {Christian~T.}\ \bibnamefont {Byrnes}}, \bibinfo {author} {\bibfnamefont {Philippa~S.}\ \bibnamefont {Cole}}, \ and\ \bibinfo {author} {\bibfnamefont {Subodh~P.}\ \bibnamefont {Patil}},\ }\bibfield  {title} {\enquote {\bibinfo {title} {{Steepest growth of the power spectrum and primordial black holes}},}\ }\href {\doibase 10.1088/1475-7516/2019/06/028} {\bibfield  {journal} {\bibinfo  {journal} {JCAP}\ }\textbf {\bibinfo {volume} {06}},\ \bibinfo {pages} {028} (\bibinfo {year} {2019})},\ \Eprint {http://arxiv.org/abs/1811.11158} {arXiv:1811.11158 [astro-ph.CO]} \BibitemShut {NoStop}%
\bibitem [{\citenamefont {Young}\ \emph {et~al.}(2019)\citenamefont {Young}, \citenamefont {Musco},\ and\ \citenamefont {Byrnes}}]{Young:2019yug}%
  \BibitemOpen
  \bibfield  {author} {\bibinfo {author} {\bibfnamefont {Sam}\ \bibnamefont {Young}}, \bibinfo {author} {\bibfnamefont {Ilia}\ \bibnamefont {Musco}}, \ and\ \bibinfo {author} {\bibfnamefont {Christian~T.}\ \bibnamefont {Byrnes}},\ }\bibfield  {title} {\enquote {\bibinfo {title} {{Primordial black hole formation and abundance: contribution from the non-linear relation between the density and curvature perturbation}},}\ }\href {\doibase 10.1088/1475-7516/2019/11/012} {\bibfield  {journal} {\bibinfo  {journal} {JCAP}\ }\textbf {\bibinfo {volume} {11}},\ \bibinfo {pages} {012} (\bibinfo {year} {2019})},\ \Eprint {http://arxiv.org/abs/1904.00984} {arXiv:1904.00984 [astro-ph.CO]} \BibitemShut {NoStop}%
\bibitem [{\citenamefont {Kehagias}\ \emph {et~al.}(2019)\citenamefont {Kehagias}, \citenamefont {Musco},\ and\ \citenamefont {Riotto}}]{Kehagias:2019eil}%
  \BibitemOpen
  \bibfield  {author} {\bibinfo {author} {\bibfnamefont {Alex}\ \bibnamefont {Kehagias}}, \bibinfo {author} {\bibfnamefont {Ilia}\ \bibnamefont {Musco}}, \ and\ \bibinfo {author} {\bibfnamefont {Antonio}\ \bibnamefont {Riotto}},\ }\bibfield  {title} {\enquote {\bibinfo {title} {{Non-Gaussian Formation of Primordial Black Holes: Effects on the Threshold}},}\ }\href {\doibase 10.1088/1475-7516/2019/12/029} {\bibfield  {journal} {\bibinfo  {journal} {JCAP}\ }\textbf {\bibinfo {volume} {12}},\ \bibinfo {pages} {029} (\bibinfo {year} {2019})},\ \Eprint {http://arxiv.org/abs/1906.07135} {arXiv:1906.07135 [astro-ph.CO]} \BibitemShut {NoStop}%
\bibitem [{\citenamefont {Escriv\`a}\ \emph {et~al.}(2020)\citenamefont {Escriv\`a}, \citenamefont {Germani},\ and\ \citenamefont {Sheth}}]{Escriva:2019phb}%
  \BibitemOpen
  \bibfield  {author} {\bibinfo {author} {\bibfnamefont {Albert}\ \bibnamefont {Escriv\`a}}, \bibinfo {author} {\bibfnamefont {Cristiano}\ \bibnamefont {Germani}}, \ and\ \bibinfo {author} {\bibfnamefont {Ravi~K.}\ \bibnamefont {Sheth}},\ }\bibfield  {title} {\enquote {\bibinfo {title} {{Universal threshold for primordial black hole formation}},}\ }\href {\doibase 10.1103/PhysRevD.101.044022} {\bibfield  {journal} {\bibinfo  {journal} {Phys. Rev. D}\ }\textbf {\bibinfo {volume} {101}},\ \bibinfo {pages} {044022} (\bibinfo {year} {2020})},\ \Eprint {http://arxiv.org/abs/1907.13311} {arXiv:1907.13311 [gr-qc]} \BibitemShut {NoStop}%
\bibitem [{\citenamefont {De~Luca}\ \emph {et~al.}(2020)\citenamefont {De~Luca}, \citenamefont {Franciolini},\ and\ \citenamefont {Riotto}}]{DeLuca:2020ioi}%
  \BibitemOpen
  \bibfield  {author} {\bibinfo {author} {\bibfnamefont {V.}~\bibnamefont {De~Luca}}, \bibinfo {author} {\bibfnamefont {G.}~\bibnamefont {Franciolini}}, \ and\ \bibinfo {author} {\bibfnamefont {A.}~\bibnamefont {Riotto}},\ }\bibfield  {title} {\enquote {\bibinfo {title} {{On the Primordial Black Hole Mass Function for Broad Spectra}},}\ }\href {\doibase 10.1016/j.physletb.2020.135550} {\bibfield  {journal} {\bibinfo  {journal} {Phys. Lett. B}\ }\textbf {\bibinfo {volume} {807}},\ \bibinfo {pages} {135550} (\bibinfo {year} {2020})},\ \Eprint {http://arxiv.org/abs/2001.04371} {arXiv:2001.04371 [astro-ph.CO]} \BibitemShut {NoStop}%
\bibitem [{\citenamefont {Musco}\ \emph {et~al.}(2021)\citenamefont {Musco}, \citenamefont {De~Luca}, \citenamefont {Franciolini},\ and\ \citenamefont {Riotto}}]{Musco:2020jjb}%
  \BibitemOpen
  \bibfield  {author} {\bibinfo {author} {\bibfnamefont {Ilia}\ \bibnamefont {Musco}}, \bibinfo {author} {\bibfnamefont {Valerio}\ \bibnamefont {De~Luca}}, \bibinfo {author} {\bibfnamefont {Gabriele}\ \bibnamefont {Franciolini}}, \ and\ \bibinfo {author} {\bibfnamefont {Antonio}\ \bibnamefont {Riotto}},\ }\bibfield  {title} {\enquote {\bibinfo {title} {{Threshold for primordial black holes. II. A simple analytic prescription}},}\ }\href {\doibase 10.1103/PhysRevD.103.063538} {\bibfield  {journal} {\bibinfo  {journal} {Phys. Rev. D}\ }\textbf {\bibinfo {volume} {103}},\ \bibinfo {pages} {063538} (\bibinfo {year} {2021})},\ \Eprint {http://arxiv.org/abs/2011.03014} {arXiv:2011.03014 [astro-ph.CO]} \BibitemShut {NoStop}%
\bibitem [{\citenamefont {Dodelson}\ and\ \citenamefont {Hui}(2003)}]{Dodelson:2003vq}%
  \BibitemOpen
  \bibfield  {author} {\bibinfo {author} {\bibfnamefont {Scott}\ \bibnamefont {Dodelson}}\ and\ \bibinfo {author} {\bibfnamefont {Lam}\ \bibnamefont {Hui}},\ }\bibfield  {title} {\enquote {\bibinfo {title} {{A Horizon ratio bound for inflationary fluctuations}},}\ }\href {\doibase 10.1103/PhysRevLett.91.131301} {\bibfield  {journal} {\bibinfo  {journal} {Phys. Rev. Lett.}\ }\textbf {\bibinfo {volume} {91}},\ \bibinfo {pages} {131301} (\bibinfo {year} {2003})},\ \Eprint {http://arxiv.org/abs/astro-ph/0305113} {arXiv:astro-ph/0305113} \BibitemShut {NoStop}%
\bibitem [{\citenamefont {Liddle}\ and\ \citenamefont {Leach}(2003)}]{Liddle:2003as}%
  \BibitemOpen
  \bibfield  {author} {\bibinfo {author} {\bibfnamefont {Andrew~R}\ \bibnamefont {Liddle}}\ and\ \bibinfo {author} {\bibfnamefont {Samuel~M}\ \bibnamefont {Leach}},\ }\bibfield  {title} {\enquote {\bibinfo {title} {{How long before the end of inflation were observable perturbations produced?}}}\ }\href {\doibase 10.1103/PhysRevD.68.103503} {\bibfield  {journal} {\bibinfo  {journal} {Phys. Rev. D}\ }\textbf {\bibinfo {volume} {68}},\ \bibinfo {pages} {103503} (\bibinfo {year} {2003})},\ \Eprint {http://arxiv.org/abs/astro-ph/0305263} {arXiv:astro-ph/0305263} \BibitemShut {NoStop}%
\bibitem [{\citenamefont {Amin}\ \emph {et~al.}(2014)\citenamefont {Amin}, \citenamefont {Hertzberg}, \citenamefont {Kaiser},\ and\ \citenamefont {Karouby}}]{Amin:2014eta}%
  \BibitemOpen
  \bibfield  {author} {\bibinfo {author} {\bibfnamefont {Mustafa~A.}\ \bibnamefont {Amin}}, \bibinfo {author} {\bibfnamefont {Mark~P.}\ \bibnamefont {Hertzberg}}, \bibinfo {author} {\bibfnamefont {David~I.}\ \bibnamefont {Kaiser}}, \ and\ \bibinfo {author} {\bibfnamefont {Johanna}\ \bibnamefont {Karouby}},\ }\bibfield  {title} {\enquote {\bibinfo {title} {{Nonperturbative Dynamics Of Reheating After Inflation: A Review}},}\ }\href {\doibase 10.1142/S0218271815300037} {\bibfield  {journal} {\bibinfo  {journal} {Int. J. Mod. Phys. D}\ }\textbf {\bibinfo {volume} {24}},\ \bibinfo {pages} {1530003} (\bibinfo {year} {2014})},\ \Eprint {http://arxiv.org/abs/1410.3808} {arXiv:1410.3808 [hep-ph]} \BibitemShut {NoStop}%
\bibitem [{\citenamefont {Cook}\ \emph {et~al.}(2015)\citenamefont {Cook}, \citenamefont {Dimastrogiovanni}, \citenamefont {Easson},\ and\ \citenamefont {Krauss}}]{Cook:2015vqa}%
  \BibitemOpen
  \bibfield  {author} {\bibinfo {author} {\bibfnamefont {Jessica~L.}\ \bibnamefont {Cook}}, \bibinfo {author} {\bibfnamefont {Emanuela}\ \bibnamefont {Dimastrogiovanni}}, \bibinfo {author} {\bibfnamefont {Damien~A.}\ \bibnamefont {Easson}}, \ and\ \bibinfo {author} {\bibfnamefont {Lawrence~M.}\ \bibnamefont {Krauss}},\ }\bibfield  {title} {\enquote {\bibinfo {title} {{Reheating predictions in single field inflation}},}\ }\href {\doibase 10.1088/1475-7516/2015/04/047} {\bibfield  {journal} {\bibinfo  {journal} {JCAP}\ }\textbf {\bibinfo {volume} {04}},\ \bibinfo {pages} {047} (\bibinfo {year} {2015})},\ \Eprint {http://arxiv.org/abs/1502.04673} {arXiv:1502.04673 [astro-ph.CO]} \BibitemShut {NoStop}%
\bibitem [{\citenamefont {Martin}\ \emph {et~al.}(2016)\citenamefont {Martin}, \citenamefont {Ringeval},\ and\ \citenamefont {Vennin}}]{Martin:2016oyk}%
  \BibitemOpen
  \bibfield  {author} {\bibinfo {author} {\bibfnamefont {Jerome}\ \bibnamefont {Martin}}, \bibinfo {author} {\bibfnamefont {Christophe}\ \bibnamefont {Ringeval}}, \ and\ \bibinfo {author} {\bibfnamefont {Vincent}\ \bibnamefont {Vennin}},\ }\bibfield  {title} {\enquote {\bibinfo {title} {{Information Gain on Reheating: the One Bit Milestone}},}\ }\href {\doibase 10.1103/PhysRevD.93.103532} {\bibfield  {journal} {\bibinfo  {journal} {Phys. Rev. D}\ }\textbf {\bibinfo {volume} {93}},\ \bibinfo {pages} {103532} (\bibinfo {year} {2016})},\ \Eprint {http://arxiv.org/abs/1603.02606} {arXiv:1603.02606 [astro-ph.CO]} \BibitemShut {NoStop}%
\bibitem [{\citenamefont {Allahverdi}\ \emph {et~al.}(2020)\citenamefont {Allahverdi} \emph {et~al.}}]{Allahverdi:2020bys}%
  \BibitemOpen
  \bibfield  {author} {\bibinfo {author} {\bibfnamefont {Rouzbeh}\ \bibnamefont {Allahverdi}} \emph {et~al.},\ }\bibfield  {title} {\enquote {\bibinfo {title} {{The First Three Seconds: a Review of Possible Expansion Histories of the Early Universe}},}\ }\href {\doibase 10.21105/astro.2006.16182} {\bibfield  {journal} {\bibinfo  {journal} {Open J. Astrophys.}\ }\textbf {\bibinfo {volume} {4}} (\bibinfo {year} {2020}),\ 10.21105/astro.2006.16182},\ \Eprint {http://arxiv.org/abs/2006.16182} {arXiv:2006.16182 [astro-ph.CO]} \BibitemShut {NoStop}%
\bibitem [{\citenamefont {Ade}\ \emph {et~al.}(2016)\citenamefont {Ade} \emph {et~al.}}]{Planck:2015sxf}%
  \BibitemOpen
  \bibfield  {author} {\bibinfo {author} {\bibfnamefont {P.~A.~R.}\ \bibnamefont {Ade}} \emph {et~al.} (\bibinfo {collaboration} {Planck}),\ }\bibfield  {title} {\enquote {\bibinfo {title} {{Planck 2015 results. XX. Constraints on inflation}},}\ }\href {\doibase 10.1051/0004-6361/201525898} {\bibfield  {journal} {\bibinfo  {journal} {Astron. Astrophys.}\ }\textbf {\bibinfo {volume} {594}},\ \bibinfo {pages} {A20} (\bibinfo {year} {2016})},\ \Eprint {http://arxiv.org/abs/1502.02114} {arXiv:1502.02114 [astro-ph.CO]} \BibitemShut {NoStop}%
\bibitem [{\citenamefont {Komatsu}\ and\ \citenamefont {Spergel}(2001)}]{Komatsu:2001rj}%
  \BibitemOpen
  \bibfield  {author} {\bibinfo {author} {\bibfnamefont {Eiichiro}\ \bibnamefont {Komatsu}}\ and\ \bibinfo {author} {\bibfnamefont {David~N.}\ \bibnamefont {Spergel}},\ }\bibfield  {title} {\enquote {\bibinfo {title} {{Acoustic signatures in the primary microwave background bispectrum}},}\ }\href {\doibase 10.1103/PhysRevD.63.063002} {\bibfield  {journal} {\bibinfo  {journal} {Phys. Rev. D}\ }\textbf {\bibinfo {volume} {63}},\ \bibinfo {pages} {063002} (\bibinfo {year} {2001})},\ \Eprint {http://arxiv.org/abs/astro-ph/0005036} {arXiv:astro-ph/0005036} \BibitemShut {NoStop}%
\bibitem [{\citenamefont {Fergusson}\ and\ \citenamefont {Shellard}(2007)}]{Fergusson:2006pr}%
  \BibitemOpen
  \bibfield  {author} {\bibinfo {author} {\bibfnamefont {J.~R.}\ \bibnamefont {Fergusson}}\ and\ \bibinfo {author} {\bibfnamefont {Edward P.~S.}\ \bibnamefont {Shellard}},\ }\bibfield  {title} {\enquote {\bibinfo {title} {{Primordial non-Gaussianity and the CMB bispectrum}},}\ }\href {\doibase 10.1103/PhysRevD.76.083523} {\bibfield  {journal} {\bibinfo  {journal} {Phys. Rev. D}\ }\textbf {\bibinfo {volume} {76}},\ \bibinfo {pages} {083523} (\bibinfo {year} {2007})},\ \Eprint {http://arxiv.org/abs/astro-ph/0612713} {arXiv:astro-ph/0612713} \BibitemShut {NoStop}%
\bibitem [{\citenamefont {Rigopoulos}\ \emph {et~al.}(2005)\citenamefont {Rigopoulos}, \citenamefont {Shellard},\ and\ \citenamefont {van Tent}}]{Rigopoulos:2004ba}%
  \BibitemOpen
  \bibfield  {author} {\bibinfo {author} {\bibfnamefont {G.~I.}\ \bibnamefont {Rigopoulos}}, \bibinfo {author} {\bibfnamefont {E.~P.~S.}\ \bibnamefont {Shellard}}, \ and\ \bibinfo {author} {\bibfnamefont {B.~J.~W.}\ \bibnamefont {van Tent}},\ }\bibfield  {title} {\enquote {\bibinfo {title} {{A Simple route to non-Gaussianity in inflation}},}\ }\href {\doibase 10.1103/PhysRevD.72.083507} {\bibfield  {journal} {\bibinfo  {journal} {Phys. Rev. D}\ }\textbf {\bibinfo {volume} {72}},\ \bibinfo {pages} {083507} (\bibinfo {year} {2005})},\ \Eprint {http://arxiv.org/abs/astro-ph/0410486} {arXiv:astro-ph/0410486} \BibitemShut {NoStop}%
\bibitem [{\citenamefont {Upadyaya}\ \emph {et~al.}()\citenamefont {Upadyaya}, \citenamefont {Lorenzoni}, \citenamefont {Geller}, \citenamefont {Kaiser},\ and\ \citenamefont {McDonough}}]{to-appear}%
  \BibitemOpen
  \bibfield  {author} {\bibinfo {author} {\bibfnamefont {Param}\ \bibnamefont {Upadyaya}}, \bibinfo {author} {\bibfnamefont {Dario~L.}\ \bibnamefont {Lorenzoni}}, \bibinfo {author} {\bibfnamefont {Sarah~R.}\ \bibnamefont {Geller}}, \bibinfo {author} {\bibfnamefont {David~I.}\ \bibnamefont {Kaiser}}, \ and\ \bibinfo {author} {\bibfnamefont {Evan}\ \bibnamefont {McDonough}},\ }\href@noop {} {\bibinfo  {journal} {{In prep}}\ }\BibitemShut {NoStop}%
\bibitem [{\citenamefont {Bezrukov}\ and\ \citenamefont {Shaposhnikov}(2008)}]{Bezrukov:2007ep}%
  \BibitemOpen
\bibfield  {journal} {  }\bibfield  {author} {\bibinfo {author} {\bibfnamefont {Fedor~L.}\ \bibnamefont {Bezrukov}}\ and\ \bibinfo {author} {\bibfnamefont {Mikhail}\ \bibnamefont {Shaposhnikov}},\ }\bibfield  {title} {\enquote {\bibinfo {title} {{The Standard Model Higgs boson as the inflaton}},}\ }\href {\doibase 10.1016/j.physletb.2007.11.072} {\bibfield  {journal} {\bibinfo  {journal} {Phys. Lett. B}\ }\textbf {\bibinfo {volume} {659}},\ \bibinfo {pages} {703--706} (\bibinfo {year} {2008})},\ \Eprint {http://arxiv.org/abs/0710.3755} {arXiv:0710.3755 [hep-th]} \BibitemShut {NoStop}%
\end{thebibliography}%

\clearpage
\appendix
\setcounter{equation}{0}
\setcounter{table}{0}
\setcounter{figure}{0}
\renewcommand{\thetable}{A\Roman{table}}
\renewcommand{\thefigure}{A\arabic{figure}}
\renewcommand{\theequation}{A\arabic{equation}}

\end{document}